\documentclass[11pt]{article}
\usepackage{amsthm,amsfonts,
}
\swapnumbers
\theoremstyle{plain}
\newtheorem{thm}{THEOREM}[section]
\newtheorem{lm}[thm]{LEMMA}
\newtheorem{cl}[thm]{COROLLARY}

\theoremstyle{definition}

\theoremstyle{remark}

\newcommand{\upchi}{\raise1pt\hbox{$\chi$}}
\newcommand{\R}{{\mathord{\mathbb R}}}
\newcommand{\C}{{\mathord{\mathbb C}}}
\newcommand{\Z}{{\mathord{\mathbb Z}}}
\newcommand{\N}{{\mathord{\mathbb N}}}
\newcommand{\supp}{{\mathop{\rm supp\ }}}
\newcommand{\mfr}[2]{{\textstyle\frac{#1}{#2}}}
\newcommand{\const}{C}
\newcommand{\iint}{\mathop{\displaystyle\int\!\!\int}}
\newcommand{\hw}{{\mathord{\widehat{w}}^{\phantom{*}}}}
\newcommand{\hwp}{{\mathord{\widehat{w}'}}}
\newcommand{\hn}{{\mathord{\widehat{n}}}}
\newcommand{\hnu}{{\mathord{\widehat{\nu}}}}
\newcommand{\an}{{\mathord{a}}^{\phantom{*}}}
\newcommand{\bn}{{\mathord{b}}^{\phantom{*}}}
\newcommand{\dn}{{\mathord{d}}^{\phantom{*}}}
\newcommand{\cn}{{\mathord{c}}^{\phantom{*}}}

\newcommand{\cA}{{\mathord{\cal A}}}
\newcommand{\cB}{{\mathord{\cal B}}}
\newcommand{\cP}{{\mathord{\cal P}}}
\newcommand{\cK}{{\mathord{\cal K}}}
\newcommand{\cH}{{\mathord{\cal H}}}
\newcommand{\cN}{{\mathord{\cal N}}}
\newcommand{\wH}{{\mathord{\widetilde H}}}
\newcommand{\gamep}{\gamma_{\varepsilon,t}}
\begin{document}
\title{GROUND STATE ENERGY OF THE TWO-COMPONENT CHARGED BOSE GAS}
\author{
  \begin{tabular}{ccc}
    Elliott H. Lieb\thanks{Work partially supported by U.S. National Science
      Foundation grant PHY01 39984-A01.} &\hspace{2cm}& Jan Philip
    Solovej\thanks{Work partially supported by NSF grant DMS-0111298, by EU grant HPRN-CT-2002-00277,
      by MaPhySto -- A Network in Mathematical Physics and
      Stochastics, funded by The Danish National Research Foundation, and by grants from the Danish research
      council.
    }\\
    \normalsize Departments of Physics and Mathematics &&
    \normalsize School of Mathematics\thanks{On leave from Department of Mathematics,
      University of Copenhagen, Universitetsparken 5, DK-2100 Copenhagen, DENMARK.
      \newline \copyright 2003 by the authors. This article may be
      reproduced in its entirety for non-commercial purposes.\hfill\newline
    To appear in {\it Communications in  Mathematical Physics. } }\\
    \normalsize Jadwin Hall, Princeton University &&
    \normalsize Institute for Advanced Study\\
    \normalsize PO Box 708 &&  \normalsize 1 Einstein Drive \normalsize \\
    \normalsize  Princeton, N.J. 08544-0708 & & \normalsize Princeton, N.J. 08540
    \\
    \normalsize {\it e-mail\/}: lieb@princeton.edu &&
    \normalsize {\it e-mail\/}: solovej@math.ku.dk
  \end{tabular}
  \bigskip
  \date{Jun 3, 2004 \\
    $ {\phantom x}  $  \\
    Dedicated to Freeman J. Dyson on the occasion of his 80th birthday }}

                                \maketitle

\begin{abstract}
  We continue the study of the two-component charged Bose gas
  initiated by Dyson in 1967. He showed that the ground state energy
  for $N$ particles is at least as negative as $-CN^{7/5}$ for large
  $N$ and this power law was verified by a lower bound found by Conlon, Lieb and Yau in
1988. Dyson  conjectured that the exact constant $C$ was given by a
  mean-field minimization problem that used, as input, Foldy's
  calculation (using Bogolubov's 1947 formalism) for the one-component
  gas. Earlier we showed that Foldy's calculation is exact insofar as
  a lower bound of his form was obtained. In this paper we do the same
  thing for Dyson's conjecture. The two-component case is considerably
  more difficult because the gas is very non-homogeneous in its ground
  state.
\end{abstract}

\section{Introduction}

In 1967 Dyson \cite{D} showed that a system composed of
non-relativistic, charged bosons is unstable in the sense that the
ground state energy of $N$ particles is at least as negative as
$-CN^{7/5}$ instead of $-CN$, where $C$ is some constant. A
lower bound of the form $-C'N^{7/5}$ was derived later \cite{CLY}, thereby
establishing the correctness of the exponent $7/5$, but not the
constant $C$.

In an earlier, parallel development, in 1961 Foldy \cite{F} considered the
problem of the one-component Bose gas (``jellium'') in which charged
particles (all of the same charge) move in a uniformly charged,
neutralizing background. Using Bogolubov's 1947 theory \cite{B}, Foldy
``derived'' the high density asymptotics for the ground state energy
of this problem as proportional to $-I_0N \rho^{1/4}$ where $I_0$ is
defined in Eq.\ (\ref{eq:I0def}) below. The correctness of this
$\rho^{1/4}$ law, but not the coefficient $I_0$,  was also proved in
(\cite{CLY}).

Dyson was motivated by Foldy's work, for he realized that if one
treated one of the two components (say, the positive one) as a
background for the other, and if one allowed the density to be
variable, one would easily arrive --- heuristically, at least --- at
the $N^{7/5}$ law. This will be explained below.

Two obvious questions arise from this earlier work. What is the
correct coefficient for the $\rho^{1/4}$ law at high density and what
is the correct coefficient for the $N^{7/5}$ law? The former question
was resolved by us in \cite{LSo}, where we showed that Foldy's $I_0$
is, indeed, correct at high density as a lower bound. Foldy's
calculation is ``essentially'' an upper bound, but some technical
issues must be clarified.  The proof that Foldy's calculation, indeed,
gives an upper bound can be found in \cite{S}.

In \cite{D} Dyson derives a rigorous upper bound for the $N^{7/5}$ law,
but with a coefficient $C$ that is clearly too small. He conjectures a
``correct'' coefficient, however, and in the present paper we shall
show that Dyson's conjectured coefficient gives a correct lower bound
(asymptotically as $N\to \infty$). An asymptotically correct upper
bound for the two-component gas energy is also given in \cite{S}.

Actually, our lower bound is slightly more general than just the case
of $N/2$ particles of each charge. We prove the lower bound for the
case in which the total number is $N$ without restriction to the
$N/2,\ N/2 $ case.

In order to understand the reason that the proof for the two-component
case is more difficult than that for jellium, presented in
\cite{LSo}, it is necessary to
recapitulate Dyson's argument briefly. His picture is that there is a
local density of particles $\rho(x)$, which has a local energy density
given by Foldy's formula, i.e., $I_0 \rho(x)^{5/4} $. One might
question whether the jellium energy can be simply taken over to the
two-component situation, but it is correct to do so, as our lower
bound shows. There is a good reason for this within the Bogolubov
theory, as we shall explain in Sect.~\ref{sec:heuristics}, but let us
continue with Dyson's picture now.  In addition to the local energy
there is also a kinetic energy caused by the variation in $\rho$,
namely $\int |\nabla \sqrt{\rho(x)}|^2dx$. Such an ``envelope'' energy
is familiar from Thomas-Fermi-Weizs\"acker and Gross-Pitaevskii
theories, for example.

If this total energy is minimized with respect to $\rho$ we are led to
a differential equation in $\sqrt{\rho}$ with the side condition that
$\int \rho =N$, but the basic features are clear. The scale length of
$\rho(x)$ will be of the order $N^{-1/5}$, the amplitude of $\rho$
will be $N^{8/5}$ and the energy will be $N^{7/5}$. Indeed, if we
define $\Phi(x)=N^{-4/5}\sqrt{\rho(N^{-1/5}x)}$ such that
$\int\Phi^2=1$ then the energy as a function of $\rho$ that we have to
minimize can be rewritten as
\begin{equation}\label{minprob}
\mfr{1}{2}\int_{\R^3}|\nabla\sqrt{\rho}|^2-I_0\int_{\R^3}\rho^{5/4}=
N^{7/5}\left(\mfr{1}{2}\int_{\R^3}|\nabla\Phi|^2-I_0\int_{\R^3}
\Phi^{5/2}\right),
\end{equation}
which makes the scaling explicit.  We then have to minimize the right
side with respect to $\Phi$ under the condition $\int\Phi^2=1$.  The
finiteness of this minimum energy is an easy consequence of the
Sobolev inequality.  The existence of a minimizing $\Phi$ is a little
harder and follows with the help of rearrangement inequalities.  It
satisfies a Lane-Emden differential equation for some $\mu>0$ and
all $x\in \R^3$
\begin{equation}
-\Delta \Phi(x) - \mfr{5}{2}I_0 \ \Phi(x)^{3/2} + \mu \Phi(x)=0.
\end{equation}
The uniqueness of $\Phi$, up to translations, is harder still. See
\cite{Be,K,MS,Z}.

Dyson's conjecture is contained in \cite[eqs.\ 104,\ 105]{D}. The
normalization convention and units employed there are not completely
clear, but the heuristics leading to (104) is clear. This will be clarified
in Sect.~\ref{sec:heuristics}.

The last topic to discuss in this introduction is the essential
difficulty inherent in the two-component problem. As in the jellium
proof in \cite{LSo} we decompose space into suitable small boxes and
impose Neumann boundary conditions on each. The scale length of these
boxes is $N^{-2/5 +\varepsilon}$.  Using the ``sliding argument'' of
\cite{CLY} we can ignore the Coulomb interaction between different
boxes. We must then distribute the $N$ particles in the boxes and what
we must prove is that the number in the various boxes has a coarse
grained density given by the solution $\rho$ to (\ref{minprob}). Since
the energy is super concave (i.e., $-N^{7/5}$), the lowest energy is
obtained by putting all the $N$ particles in one box. What prevents
this from happening is that the boxes are not really totally
independent, thanks to the kinetic energy operator. In other words, we
must somehow save a little bit of the kinetic energy operator to
prevent wild variations in particle density between neighboring boxes.

The conundrum is that the mean-field energy in (\ref{minprob}) uses
all the kinetic energy, not just some of it. Likewise, to get the
intrabox energy (the second term in (\ref{minprob})) we also need the
full kinetic energy. The resolution is to split the kinetic energy
operator $-\Delta$ into a high-momentum part for use in calculating the
intrabox energy and a low-momentum part for use in reproducing the
first term in (\ref{minprob}). Naturally, error terms will arise and
the chief difference between this paper and our earlier jellium paper
is centered on the definition of the splitting and the management of
the induced error terms.

The proof of the main theorem starts in Sect.~\ref{sec:loc} where we
show how to localize the problem into large boxes of size $L\gg
N^{-1/5}$ with Dirichlet boundary conditions.  This size is larger
than the expected size of the bound complex.  The decomposition of the
kinetic energy into large and small momentum is carried out in an
appendix. It is used in Sect.~\ref{sec:lochml}, which localizes
further into really small boxes of size $\ell \gg N^{-2/5}$.  For the
relevance of the scale $N^{-2/5}$ the reader is referred to the
heuristic discussion in Sect.~\ref{sec:heuristics}. The control of the
electrostatics using sliding is also discussed in both of these
sections.  Sect.~\ref{sec:cutoff} discusses the ultraviolet and
infrared cutoffs of the interaction potential with control of the
errors they introduce. In contrast to the treatment in \cite{LSo} we
now need to use the kinetic energy to control the errors caused by the
ultraviolet and infrared cutoffs in the potential.
Sect.~\ref{sec:estimates} controls all the unimportant parts of the
localized Hamiltonian and reduces the problem to Bogolubov's
Hamiltonian, which is analyzed in Sect.~\ref{sec:quadratic}.

Sect.~\ref{sec:simple} gives the first,
simple bound on kinetic energy, local non-neutrality, and an estimate
on the local condensation. Sect.~\ref{sec:localizingexcitations}
improves the estimate on condensation with the help of the method of
``localizing large matrices'' in \cite{LSo} (and which is reviewed in
an appendix). (Recall that in \cite{LSo} we had to reach the final
estimate on the various energies by a succession of finer and finer
error bounds, each taking the previous bound as input.) In
Sect.~\ref{sec:energybound}, we give the final bound on the energy in
each small box. We have to treat boxes with few particles as well as
many particles separately. In Sect.~\ref{sec:lattice} we show how the
kinetic energy estimate in the appendix (in which the low momentum
kinetic energy {\it between} boxes leads to a difference energy on a
lattice) leads, in turn, to the term $\int |\nabla \sqrt{\rho}|^2 $ in
the energy functional (\ref{minprob}). In the final Sect.~\ref{sec:final} all the
estimates are put together and we show how to choose the various parameters to
get the desired minimization problem for the lower bound.

We thank Y.Y.~Li and M.I. Weinstein for pointing out the references \cite{K,MS,Z} to us.

\section{Basic definitions and main theorem}

We consider $N$ particles with charges $e_i=\pm1$, $i=1,\ldots,N$. The
Hamiltonian describing the system is
\begin{equation}
  H_N=\sum_{i=1}^N-\mfr{1}{2}\Delta_i+\sum_{1\leq i<j\leq N}\frac{e_ie_j}{|x_i-x_j|}
\end{equation}
acting on
$L^2(\R^{3N})$. We shall not specify the number of positive or
negative particles, but simply consider the smallest possible energy
$$
E(N)=\inf\{\inf\hbox{\rm spec}_{L^2(\R^{3N})} H_N\ |\ e_i=\pm1,\ i=1,\ldots,N\}.
$$
Instead of considering $H_N$ depending on the parameters
$e_1,\ldots,e_N$ we may consider it as {\it one} operator on the enlarged
space $L^2\left((\R^{3}\times\{1,-1\})^N\right)$, where the set
$\{1,-1\}$ contains the values of the charge variables.
Then
$$
E(N)=\inf\hbox{\rm spec}_{L^2\left((\R^{3}\times\{1,-1\})^N\right)} H_N.
$$
Thus $E(N)$ is the infimum of
$\langle\Psi,H_N\Psi\rangle$, over normalized functions $\Psi$ in
$L^2\left((\R^{3}\times\{1,-1\})^N\right)$.
Here we may restrict $\Psi$ to be non-negative
and thus moreover also to be symmetric under the interchange of
particles. Thus $E(N)$ is the energy of a charged Bose gas.

Our main result in this paper is the asymptotic {\it lower} bound on
$E(N)$ conjectured by Dyson in \cite{D}. The corresponding asymptotic
{\it upper} bound is given in \cite{S}. Together these results prove
Dyson's formula.

\begin{thm}[Dyson's formula]\label{thm:main} As $N\to\infty$ we have
$$
E(N)= -AN^{7/5}+o(N^{7/5}),
$$
where $A$ is the positive constant determined by the variational principle
\begin{equation}\label{eq:Dysonformula}
  -A=\inf\biggl\{\mfr{1}{2}\int_{\R^3}|\nabla\Phi|^2-I_0\int_{\R^3} \Phi^{5/2}\
  \biggr|\ 0\leq \Phi,\ \int_{\R^3}\Phi^2\leq 1\biggr\}
\end{equation}
with
\begin{equation}\label{eq:I0def}
  I_0=(2/\pi)^{3/4}\int_0^\infty\left(1+x^4-x^2\left(x^4+2\right)^{1/2}\right)\,dx=
\frac{2^{3/2}\Gamma(3/4)}{5\pi^{1/4}\Gamma(5/4)} .
\end{equation}
\end{thm}

It will be clear from the proof (see the discussion in
Sect.~\ref{sec:final}) that the error in Dyson's formula could have
been written in the form $N^{7/5-\xi}$ for some $\xi>0$. Although this
is in principle straightforward we have not attempted to optimize the
error term to determine the exact exponent $\xi$.

\section{Heuristic derivation of the energy}\label{sec:heuristics}

In this section we  give the heuristic derivation that leads to
the local energy density $I_0\rho^{5/4}$ in the second term in
(\ref{minprob}) (see also (\ref{eq:Dysonformula})).  The fact that the
constant for the two-component gas is the same as $I_0$, the constant
for the one-component gas, is somewhat mysterious. After all, there is
not only the negative $+-$ interaction energy, but there are also the
$++$ and $--$ energies.  Moreover, the $\rho$ in the two-component
case refers to the total number of particles, which is twice the
number of each charge (in the neutral case), whereas $\rho$ refers
only to the movable particles of one sign in the one-component case.

Another problem is that the usual thermodynamic limit exists for the
one-component gas only because we constrain the background to have
uniform density and we do not allow it to contract to a high density
ball, as it would if one merely minimizes the energy. The
two-component gas does not have the usual thermodynamic limit because,
as Dyson showed, its energy goes as $-N^{7/5}$.

Nevertheless, we shall describe the two-component charged gas on a
small local scale $\ell$ as a gas of uniform density $\rho$. For such
a uniform gas we shall imitate Foldy's calculation \cite{F} to arrive
at the energy $-I_0\rho^{5/4}\ell^3$. We shall assume that
$\ell\ll N^{-1/5}$ (the scale on which $\rho(x)$ varies), but
$\ell\gg\rho^{-1/4}$ which, as we shall see, is the relevant scale for
the uniform gas. Note that we expect $\rho\sim N^{8/5}$ and hence
$\rho^{-1/4}\sim N^{-2/5}$.
In any event, we are being ``heuristic''
in this section and the reader is welcome to ignore this
``derivation'' if it is not pleasing.

We shall use periodic boundary conditions, as usual, and write the
Hamiltonian in second quantized form in the manner of Bogolubov/Foldy
as
\begin{eqnarray}
  H_{\rm F}&=&\sum_{k} \mfr{1}{2}k^2 \left(a^*_{k+}\an_{k+}+a^*_{k-}\an_{k-}\right)\nonumber\\&&{}+
  \frac{1}{2\ell^3}\sum_{k\ne0}
  \sum_{pq}\frac{4\pi}{k^2}\Bigl(a^*_{p+}a^*_{q+}\an_{(q-k)+}\an_{(p+k)+}
  +a^*_{p-}a^*_{q-}\an_{(q-k)-}\an_{(p+k)-}\nonumber\\&&{}
  \phantom{+\frac{1}{2\ell^3}\sum_{k\ne0}
   \sum_{pq}\frac{4\pi}{k^2}\Bigl(}
   -2a^*_{p+}a^*_{q-}\an_{(q-k)-}\an_{(p+k)+}\Bigr),\label{heur}
\end{eqnarray}
where the sums are over momenta in the set $\frac{2\pi}{\ell} \Z^3$.
Here $a^*_{p\pm}$ creates a state with momentum $p$ and charge $\pm1$.

In the second term we have excluded $k=0$, which is also what Foldy
does.  It would of course be meaningless to include this term and
leave the Fourier transform of the potential equal to
$\frac{4\pi}{k^2}$. If, instead, we defined the Fourier transform of
the potential for $k=0$ to be proportional to $\ell^2$ (which is the
maximal value of the Fourier transform for $k\ne0$), the contribution from the term $k=0$
would be proportional to $-(\rho \ell^3) \ell^{-1}$ for a neutral
system. If, as we assume, $\ell\gg\rho^{-1/4}$ then $(\rho \ell^3) \ell^{-1}\ll\rho^{5/4}\ell^3$.
Hence we may ignore the $k=0$ contribution.
In Foldy's situation the corresponding term would not
contribute to the thermodynamic limit.

The next step in the heuristic derivation is to exclude those terms in
the second sum above that do not contain precisely two creation or
annihilation operators of particles of momentum zero. Subsequently
these zero momentum creation or annihilation operators $a_{0\pm}^*$
and $\an_{0\pm}$ are replaced by the square root of half the particle
number, namely $\sqrt{\rho\ell^3/2}$. We then arrive at
\begin{eqnarray*}
  &&\sum_{k} \mfr{1}{4}k^2 \left(
    a_{k+}^*\an_{k+}+a_{-k+}^*\an_{-k+}
    +a_{k-}^*\an_{k-}+a_{-k-}^*\an_{-k-}\right)\nonumber\\
  &&+
  \frac{\rho}{4}\sum_{k\ne0}
  \frac{4\pi}{k^2}\Bigl[a^*_{k+}\an_{k+}+a^*_{-k+}\an_{-k+}+a^*_{k+}a^*_{-k+}+
  \an_{k+}\an_{-k+}
  \nonumber\\&&
  \phantom{+\frac{\rho}{2}\sum_{k\ne0}\frac{4\pi}{k^2}\Bigl[}
  +a^*_{k-}\an_{k-}+a^*_{-k-}\an_{-k-}
  +a^*_{k-}a^*_{-k-}+\an_{k-}\an_{-k-}
  \nonumber\\&&
  \phantom{+\frac{\rho}{2}\sum_{k\ne0}\frac{4\pi}{k^2}\Bigl[}
  -(a^*_{k+}\an_{k-}+a^*_{-k+}\an_{-k-}
  +a^*_{k-}\an_{k+}+a^*_{-k-}\an_{-k+}
  \nonumber\\&&
  \phantom{+\frac{\rho}{2}\sum_{k\ne0}\frac{4\pi}{k^2}\Bigl[}
  a^*_{k+}a^*_{-k-}+a^*_{-k+}a^*_{k-}+\an_{k+}\an_{-k-}
  +\an_{-k+}\an_{k-})
  \Bigr].
\end{eqnarray*}
This expression should be compared with Lemma~\ref{lm:hQ}, where we
arrive at a similar expression as a rigorous lower bound on part of
the full Hamiltonian. (In comparing with the lemma one should replace
$\bn_k\to\an_k$, $\nu^+,\nu^-\to\rho\ell^3/2$, $\hat
V_{r,R}(k)\to\frac{4\pi}{|k|^2}$, $\gamep\to1$, $(\ell t^6)^{-2}\to0$,
and of course also $\frac{\ell^3}{(2\pi)^3}\int\,dk\to\sum_k$.)

The final step is to recognize that the resulting quadratic
Hamiltonian has the following property, as the reader can easily
check: The operators $\an_{k\pm}$ always appear in the potential
energy term (the last sum above) in the combination $\dn_k= (\an_{k+}
- \an_{k-})/\sqrt2$.  This is a normal mode since $[\dn_k, d^*_q] =
\delta_{k,q}$. The other normal mode $\cn_k= (\an_{k+} +
\an_{k-})/\sqrt2$ appears only in the kinetic energy term (the first
sum above), i.e., the kinetic energy is $a^*_{k+}\an_{k+}
+a^*_{k-}\an_{k-} = c^*_k\cn_k +d^*_k\dn_k$. The ground state is
achieved by having no $\cn_k$ excitations, which leaves us just with
the term $d^*_k \dn_k$ in the kinetic energy term.

The conclusion is that the quadratic Hamiltonian is now exactly the
same as Foldy's (but with $\dn_k $ in place of $\an_k$) and,
therefore, the ground state energy is $-I_0 \rho^{5/4} \ell^3$. This
conclusion could also have been arrived at by an explicit
diagonalization of the total quadratic Hamiltonian. (See
Sect.~\ref{sec:quadratic}, in particular Theorem~\ref{thm:bogolubov},
for comparison.)

The detailed diagonalization analysis shows that the relevant momenta
k are of magnitude $\rho^{1/4}$ and hence as mentioned above the
relevant length scale is $\rho^{-1/4}$. The assumption
$\ell\gg\rho^{-1/4}$ allows one to replace sums over the lattice
$\frac{2\pi}{\ell}\Z^3$ by integrals
$\frac{\ell^3}{(2\pi)^3}\int_{\R^3}dk$.  This is how the integral in
(\ref{eq:I0def}) appears.

\section{Localization}\label{sec:loc}
If Dyson's conjecture is correct then the size of the boson cloud is
proportional to $N^{-1/5}$. As a first step we shall  localize
the problem into cubes of size $L$, where we choose $L$ as a function
of $N$ in such a way that $N^{1/5}L\to\infty$ as $N\to\infty$. Exactly
how $N^{1/5}L\to\infty$ will be determined at the end of the analysis
in Sect.~\ref{sec:final}. As a consequence of our results we shall see
that essentially all particles will concentrate within one of these
cubes of size $L$.

We shall do the localization in such a way that the cubes do not
interact and the analysis can be done in each cube independently. The
only thing to bear in mind is that the total number of particles in
all cubes is $N$.

In analyzing the individual cubes we shall perform a further
localization into smaller cells of a size $\ell< L$ depending on $N$ in
such a way that $N^{2/5}\ell\to\infty$ as $N\to\infty$ (precisely how
will again be determined in Sect.~\ref{sec:final}).

We shall first describe how we control the electrostatic
interaction between the different regions into which we localize.
We do this in a manner very
similar to what was done in \cite{LSo} using the sliding technique
of \cite{CLY}, and  shall use this technique both for the localization
into the large cubes and again when we localize into the smaller cells.

Let $t$, with $0<t<1/2$, be a parameter which will be chosen later in Sect.~\ref{sec:final} to
depend on $N$ in such a way that $t\to0$ as $N\to\infty$.

Let $\Theta,\theta\in C^\infty_0(\R^3)$ satisfy
\begin{enumerate}
\item $0\leq \Theta,\theta\leq1$, $\theta(x)=\theta(-x)$, and $\Theta(x)=\Theta(-x)$.
\item $\supp\theta\subset\left[(-1+t)/2,(1-t)/2\right]^3$,  $\supp\Theta\subset\left[(-1-t)/2,(1+t)/2\right]^3$.
\item $\theta(x)=1$ for $x\in\left[(-1+2t)/2,(1-2t)/2\right]^3$, and
$\Theta(x)=1$ for $x\in\left[(-1+t)/2,(1-t)/2\right]^3$,
\item All derivatives of order $m$ for $m\leq
3$ of the functions $\theta, \sqrt{1-\theta^2}, \Theta$ are uniformly bounded by $\const t^{-m}$, where
$\const$  is some universal constant.
\item For all $x\in\R^3$ we have $\sum\limits_{k\in\Z^3}\Theta(x-k)^2=1$.
\end{enumerate}
We introduce the two constants $\gamma$, $\widetilde\gamma$ such
that $\gamma\int\theta(y)^2dy=1$ and
$\widetilde\gamma\int\Theta^4(y)^2dy=1$.
Then
\begin{equation}\label{eq:gammaprop}
1\leq \gamma\leq (1-2t)^{-3},\quad (1+t)^{-3}\leq\widetilde\gamma\leq(1-t)^{-3}.
\end{equation}

We also
introduce the Yukawa potential $Y_{m}(x)=|x|^{-1}e^{-m|x|}$ for
$m\geq0$. For $m=0$ this is, of course, the Coulomb potential.

\begin{lm}[Electrostatic
  decoupling of boxes using sliding]\label{lm:sliding}\hfill\\
  There exists a function of the form $\omega(t)=\const t^{-4}$ (we
  assume that $\omega(t)\geq1$ for $t<1/2$) such that
  for all $x_1,x_2, \ldots,x_N\in\R^3$, all
  $e_1,e_2,\ldots,e_N$, with $|e_i|=1$ for $i=1,2,\ldots,N$, all
  $m\geq0$, and all $\lambda>0$ we have
  \begin{equation}\label{eq:thetasliding}
    \sum_{1\leq i< j\leq N}{e_ie_j}Y_m(x_i-x_j) \geq\gamma\int\limits_{\R^3}
    \sum_{1\leq i<j\leq N}{e_ie_j}
    \theta\left(\frac{x_i}{\lambda}-z\right)Y_{m+\frac{\omega(t)}{\lambda}}(x_i-x_j)
    \theta\left(\frac{x_j}{\lambda}-z\right)dz-\frac{N\omega(t)}{2\lambda}
  \end{equation}
  and likewise
  \begin{equation}\label{eq:Thetasliding}
    \sum_{1\leq i< j\leq N}{e_ie_j}Y_m(x_i-x_j) \geq\widetilde\gamma\int\limits_{\R^3}
    \sum_{1\leq i<j\leq N}{e_ie_j}
    \Theta^2\left(\frac{x_i}{\lambda}-z\right)Y_{m+\frac{\omega(t)}{\lambda}}(x_i-x_j)
    \Theta^2\left(\frac{x_j}{\lambda}-z\right)dz-\frac{N\omega(t)}{2\lambda}.
  \end{equation}
\end{lm}
\begin{proof} Since $\theta$ and $\Theta^2$  have the same properties it
  is enough to consider $\theta$. By rescaling we may assume that $\lambda=1$.
  We have that
  $$
  \int \gamma\theta(x-z)Y_{m+\omega}(x-y)\theta(y-z)\,dz
  =h(x-y)Y_{m+\omega}(x-y),
  $$
  where we have set $h=\gamma\theta*\theta$. We chose $\gamma$
  such that $1=h(0)=\gamma\int\theta(y)^2\,dy$.  Then $h$ satisfies all the assumptions in
  Lemma~2.1 in \cite{CLY}.  We then conclude from Lemma~2.1 in
  \cite{CLY} that the Fourier transform of the function
  $F(x)=Y_{m}(x)-h(x)Y_{m+\omega}(x)$ is non-negative, if
  $\omega$ is chosen large enough depending on $h$.
  [The detailed bounds from \cite{CLY} show that we may in fact
  choose $\omega=\const t^{-4}$, since $\omega$ has to control
  the 4th derivative of $h$.] Note, moreover, that
  $\lim_{x\to0}F(x)=\omega$.  Hence
  $$ \sum_{1\leq i<j\leq N}e_ie_jF(y_i-y_j)
  \geq -\sum_{i=1}^N\frac{e_i^2\omega}{2}=-\frac{N\omega}{2}.  $$
  The lemma follows by writing $Y_m(x) = F(x) +h(x)Y_{m+\omega}(x)$.
\end{proof}

We shall use the electrostatic decoupling (\ref{eq:Thetasliding}) when we localize into the
large cubes of size $L$ and then (\ref{eq:thetasliding}) when we localize further into
the smaller cells.  We begin with the localization into large cubes.
\begin{thm}[Localization into a large cube]\label{thm:largeloc}
Let
$$
H_{N,L}=\sum_{i=1}^N-\mfr{1}{2} \Delta_{i,\rm D}
+\sum_{1\leq i<j\leq N}e_ie_j \widetilde\gamma Y_{\frac{\omega(t)}{L}}(x_i-x_j)
$$
be a Hamiltonian acting in the space
$L^2\left(([-L/2,L/2]^3\times\{1,-1\})^N\right)$.
Here $\Delta_{\rm D}$ refers to the Dirichlet Laplacian in the cube
$[-L/2,L/2]^3$. Let
\begin{equation}\label{eq:ELN}
  E_L(N)=\inf\hbox{\rm spec} H_{N,L}.
\end{equation}
Then
$$
E(N)\geq E_L(N)-N^{7/5}\left(Ct^{-2}(N^{1/5}L)^{-2}
  +\mfr{1}{2}\omega(t)N^{-1/5}(N^{1/5}L)^{-1}\right).
$$
\end{thm}
\begin{proof}
For all
$z\in\R^3$ let $\Theta_z(x)=\Theta((x/L)-z)$. We consider
$\Theta_z$ as a multiplication operator on $L^2(\R^3)$.
We have for all $z\in\R^3$ that
$$
\sum_{q\in\Z^3}\Theta^2_{z+q}(x)=1.
$$
Using (\ref{eq:Thetasliding}) with $m=0$ and that $\int_{\R^3}
f(z) dz = \sum_{q\in\Z^3}\int_{[-1/2,1/2]^3}f(z+q) dz $ we see that
\begin{eqnarray*}
  \lefteqn{
  \sum_{1\leq i<j\leq N}\frac{e_ie_j}{|x_i-x_j|}}&&\\&\geq&
  \sum_{{\bf k}=(k_1,\ldots,k_N)\in\Z^{3N}}\int\limits_{[-1/2,1/2]^3}F_{{\mathbf
    k},z}(x_1,\ldots,x_N)^2
  \sum_{1\leq i<j\leq N}\delta_{k_ik_j}e_ie_j\widetilde\gamma Y_{\omega(t)/L}(x_i-x_j)\,dz-\frac{N\omega(t)}{2L},
\end{eqnarray*}
where
$$
F_{{\mathbf
    k},z}(x_1,\ldots,x_N)=\Theta_{k_1+z}(x_1)\cdots\Theta_{k_N+z}(x_N),\quad {\bf k}=(k_1,\ldots,k_N)\in\Z^{3N}.
$$
We localize the
kinetic energy using the formula
$$
-\Delta=\sum_{q\in\Z^3}\Theta_{z+q}(-\Delta)\Theta_{z+q} -\sum_{q\in\Z^3}(\nabla\Theta_{z+q})^2
\geq  \sum_{q\in\Z^3}\Theta_{z+q}(-\Delta)\Theta_{z+q} -\const(tL)^{-2},
$$
which gives
\begin{eqnarray*}
  \sum_{i=1}^N-\Delta_{i}\geq\sum_{{\mathbf
    k}\in\Z^{3N}}\int\limits_{[-1/2,1/2]^3}F_{{\mathbf
    k},z}(x_1,\ldots,x_N)
  \sum_{i=1}^N(-\Delta_{i})F_{{\mathbf
    k},z}(x_1,\ldots,x_N)\,dz-\const N(tL)^{-2}.
\end{eqnarray*}
The previous estimates imply that
\begin{eqnarray*}
  H_N\geq\sum_{{\mathbf
      k}\in\Z^{3N}}\int\limits_{[-1/2,1/2]^3}F_{{\mathbf
      k},z}(x_1,\ldots,x_N)\sum_{q\in\Z^3}H^{(q)}_{{\mathbf k},z}F_{{\mathbf
      k},z}(x_1,\ldots,x_N)
  \,dz-\frac{N\omega(t)}{2L}-\const N(tL)^{-2},
\end{eqnarray*}
where we have introduced the operator
$$
H^{(q)}_{{\bf k},z}=\sum_{i=1}^N-\mfr{1}{2}\delta_{k_i,q}\Delta_{i,\rm D}^{(q+z)}
+\sum_{1\leq i<j\leq N}\delta_{k_i,q}\delta_{k_j,q}
e_ie_j\widetilde\gamma Y_{\omega(t)/L}(x_i-x_j)
$$
where $\Delta_{\rm D}^{(q+z)}$ denotes the Dirichlet Laplacian in
the cube $\{(q+z)L\}+[-L/2,L/2]^3$.  Note that the operator
$H^{(q)}_{{\bf k},z}$ above acts on functions for which the space
variables are in the set $ \{(q+z)L\}+[-L/2,L/2]^3 $.
The operator $H^{(q)}_{{\mathbf k},z}$ is hence unitary equivalent to $H_{N,L}$ with $N$ replaced by
$$
N_q({\mathbf k})=\#\cN_q({\mathbf k}),\quad \cN_q(k_1,\ldots,k_N)=\{i \ |\ k_i=q\}.
$$
We therefore get the lower bound
\begin{eqnarray*}
  H_N\geq\sum_{{\mathbf
      k}\in\Z^{3N}}\int\limits_{[-1/2,1/2]^3}F_{{\mathbf
      k},z}(x_1,\ldots,x_N)^2\sum_{q\in\Z^3}E_L(N_q({\mathbf
      k})) dz-\frac{N\omega(t)}{2L}-\const N(tL)^{-2}.
\end{eqnarray*}

The theorem follows if we can prove that for fixed ${\mathbf k}\in\Z^{3N}$
\begin{equation}\label{eq:grif}
\sum_{q\in\Z^3}E_L(N_q({\mathbf k})) \geq E_L(N).
\end{equation}

To conclude (\ref{eq:grif}), let $\psi_q$ be a normalized
$N_q$-particle ground state eigenfunction of $H_{N_q,L}$, such that
$$(\psi_q,H_{N_q,L}\psi_q)=E_L(N_q).$$
(A normalized $0$-particle function we consider simply to be the number $1$.)
Define the $N$-particle eigenfunction
$$
\Psi_\tau(x_1,e_1,\ldots,x_N,e_N)=\prod_q\psi_q((x_{i},\tau_qe_{i})_{i\in\cN_q({\mathbf
  k})}),
$$
(for fixed $\mathbf k$ there are only finitely many factors different
from 1 in the above product)
where we shall choose the (finite) parameters $\tau_q\in\{1,-1\}$.
We then find that
\begin{equation}\label{eq:grtrial}
  \left(\Psi_\tau,H_{N,L}\Psi_\tau\right)=\sum_{q\in\Z^3}E_L(N_q)+\mfr{1}{2}\sum_{q\ne q'} \tau_q\tau_{q'}I_{qq'},
\end{equation}
(note that that there are again only finitely many non-zero terms in the
sums) where
$$
I_{qq'}=\Bigl(\psi_q\psi_{q'},\sum_{i}\sum_{j}\delta_{k_i,q}\delta_{k_j,q'}
e_ie_j\widetilde\gamma Y_{\omega(t)/L}(x_i-x_j)\psi_q\psi_{q'}\Bigr).
$$
It is now easy to see that by an appropriate choice of the
finite number of parameters $\tau_q$, we can make sure that the last
sum in (\ref{eq:grtrial}) is non-positive. In fact, if we average over all possible choices of $\tau_q=\pm1$
the last sum in (\ref{eq:grtrial}) averages to $0$.

The estimate
(\ref{eq:grif}) is now a consequence of the variational principle.
\end{proof}

The rest of the analysis is concerned with estimating the energy $E_L(N)$ from
(\ref{eq:ELN}).
As explained in the beginning we may define $E_L(N)$ as
$\inf\hbox{spec}H_{N,L}$ when we consider $H_{N,L}$ as an operator on
the {\it symmetric} tensor product
${\bigotimes\limits^N}_{S}L^2([-L/2,L/2]^3\times\{1,-1\})$.

In the space ${\bigotimes\limits^N}_{S}L^2(\R^3\times\{1,-1\})$ we
shall use the notation of second quantization.  If $u\in
L^2(\R^3\times\{1,-1\})$ then $a^*(u)$ denotes the creation operator
in the Fock space
$\bigoplus\limits_{N=0}^\infty{\bigotimes\limits^N}_{S}L^2(\R^3\times\{1,-1\})$.
Products of the form $a^*_0(u)\an_0(u')$ or $\an_0(u')a^*_0(u)$ are
however bounded operators on each of the fixed particle number spaces
${\bigotimes\limits^N}_{S}L^2(\R^3\times\{1,-1\})$.  We may consider
$H_{N,L}$ as acting in this space but restricted to functions with
support in $\left([-L/2,L/2]^3\times\{1,-1\}\right)^N$.

If $u\in
L^2(\R^3)$ we use the notation $a^*_{\pm}(u)$ for the creation
operator which creates the function $u$ with charge $\pm1$
respectively, i.e., the function $u$ defined on $\R^3\times\{1\}$ or
$\R^3\times\{-1\}$ respectively.

\section{The localization of the operator $H_{N,L}$}\label{sec:lochml}
We turn to the second localization into smaller cubes of size $\ell>0$, where we
shall choose $\ell$ as a function of $N$ in such a way that $\ell\leq
L$ and
$N^{2/5}\ell\to\infty$ as $N\to\infty$. As mentioned the precise choice of $\ell$
as a function of $N$ will be made at
the end of section~\ref{sec:final} when we complete the proof of
Dyson's formula.

Localizing the kinetic energy is difficult. In the paper
\cite{LSo} we used Neumann boundary conditions on the small cubes.
In this way we got a lower bound by studying independent small boxes.
In the present situation we will not get the correct answer if we
bound the kinetic energy below by Neumann energies in independent
boxes. As explained the ``kinetic energy'' between boxes is important. A lower
bound on the kinetic energy which both contains a contribution from
within each box and a contribution that connects boxes is given in
Appendix~\ref{app:kinetic}. We shall use this result now.

We  use the function $\theta$ to do the localization.
For $z\in\R^3$ we define $\upchi_z(x)=\theta((x/\ell)-z)$.
Then $\supp\upchi_z\subset
\{z\ell\}+\left[(-1+t)\ell/2,(1-t)\ell/2\right]^3$.  Let
$\eta_z=\sqrt{1-\upchi_z^2}$. Then $\|\partial^\alpha\upchi_z\|_\infty\leq \const (\ell t)^{-|\alpha|}$ and
$\|\partial^\alpha\eta_z\|_\infty\leq \const(\ell t)^{-|\alpha|}$ for
all multi-indices $\alpha$ with $|\alpha|\leq 3$.

Let $X_z$ be the characteristic function of the cube
$\{z\ell\}+[-\ell/2,\ell/2]^3$, i.e., $X_z(x)=X_0(x-\ell z)$.
Let $\cP_z$ denote the projection onto the subspace of
$L^2\left((\{z\ell\}+[-\ell/2,\ell/2]^3)\times\{1,-1\}\right)$
consisting of functions orthogonal to constants.
We shall consider $\cP_z$ as a projection in
$L^2(\R^3\times\{1,-1\})$.
We define for each $z\in\R^3$ the operator
\begin{equation}\label{eq:K}
  \cK^{(z)}=\cP_z\upchi_{z}
    \frac{(-\Delta)^2}{-\Delta
      +(\ell t^6)^{-2}}\upchi_z\cP_z.
\end{equation}
The operators $\cK^{(z)}$ shall play the role of the kinetic energy within each
box.

Let
$a^*_{0\pm}(z)$ be the creation operators
\begin{equation}\label{eq:a0}
  a^*_{0\pm}(z)=a^*_\pm(\ell^{-3/2}X_z),
\end{equation}
i.e., the operator creating the constants in the cube
$\{z\ell\}+[-\ell/2,\ell/2]^3$. We introduce the notation
\begin{equation}\label{eq:npm}
  \hn_{z}^\pm=a^*_{0\pm}(z)\an_{0\pm}(z)\quad\hbox{and}\quad
  \hn_z=\hn_z^++\hn_z^-.
\end{equation}
We refer to $\hn_z$ as the {\it number of particles in the condensate}
in the cube $\{z\ell\}+[-\ell/2,\ell/2]^3$.  The operator for the {\it
  total number of particles} in the cube
$\{z\ell\}+[-\ell/2,\ell/2]^3$ is
\begin{equation}\label{eq:nupm}
  \hnu_z=\sum_{i=1}^N X_z(x_i)
\end{equation}
The operators $\hnu^\pm_z$ for the total numbers of positively and negatively
charged particles are determined by
$$
  \hnu_z^++\hnu_z^-=\hnu_z\quad\hbox{and}\quad
  \hnu_z^+-\hnu_z^-=\sum_{i=1}^N e_iX_z(x_i).
$$
We refer to $\hnu_z-\hn_z$ as the {\it number of excited particles}.

The kinetic energy that connects boxes will be a type of lattice
Laplacian. In fact, for a map $S:\Z^3\to\R$ we define a lattice
Laplacian
\begin{equation}\label{eq:T}
  T(S)=\sum_{\sigma_1,\sigma_2\in\Z^3\atop
    |\sigma_1-\sigma_2|=\sqrt{2}}\frac{1}{12}({S}(\sigma_1)-{S}(\sigma_2))^2
  +\sum_{\sigma_1,\sigma_2\in\Z^3\atop
    |\sigma_1-\sigma_2|=\sqrt{3}}\frac{1}{24}({S}(\sigma_1)-{S}(\sigma_2))^2.
\end{equation}
This specific form of $T$ has been chosen because it is convenient
when we shall later compare with a continuum Laplacian (see Sect.~\ref{sec:lattice}).

Using Theorem~\ref{thm:kinetic} in Appendix~\ref{app:kinetic} with
$\Omega=[-(L+\ell)/(2\ell),(L+\ell)/(2\ell)]^3$  we obtain
the following result.
\begin{lm}[Kinetic energy localization]\label{lm:kineticloc}
If $\Psi\in\bigotimes^N_SL^2(\R^3\times\{1,-1\})$ is normalized and has support in
$\left([-L/2,L/2]^3\times\{1,-1\}\right)^N$ then for all $\varepsilon>0$, $0<t<1/2$
\begin{eqnarray*}
  (1+\varepsilon+Ct^3)\left(\Psi,\sum_{i=1}^N-\Delta_i\Psi\right)&\geq&\!\!\!\!
  \int\limits_{z\in[-1/2,1/2]^3}
  \left(\Psi,\sum_{\sigma\in\Z^3}\sum_{i=1}^N\left(\cK_{i}^{(z+\sigma)}
        -\varepsilon\Delta^{(z+\sigma)}_{i,\rm Neu}\right)
    \Psi\right)
      +T(S_z^\Psi)\, dz\\&& -CL^3/\ell^5,
\end{eqnarray*}
where $-\Delta_{\rm Neu}^{(z)}$ is the Neumann
Laplacian
for the cube $\{z\ell\}+[-\ell/2,\ell/2]^3$ and $S_z^\Psi:\Z^3\to\R$ is the
map
\begin{equation}\label{eq:sz}
  S_z^\Psi(\sigma)=\ell^{-1}\left(\sqrt{\langle\hn_{z+\sigma}^++\hn_{z+\sigma}^-\rangle+1}
    -1\right)=\ell^{-1}\left(\sqrt{\langle\hn_{z+\sigma}\rangle+1}
    -1\right),
\end{equation}
where
$$
\langle\hn_{z+\sigma}^++\hn_{z+\sigma}^-\rangle=\left(\Psi,
  (\hn_{z+\sigma}^++\hn_{z+\sigma}^-)\Psi\right).
$$
\end{lm}
In order to arrive at this lemma from Theorem~\ref{thm:kinetic} we
have used the inequality
$$
T(S^\Psi_z)\leq \langle T(S_z^+)+T(S_z^-)\rangle,
$$
where
$S_z^\pm(\sigma)=\ell^{-1}\sqrt{\hn_{z+\sigma}^\pm+1/2}$.
To prove this inequality first note that by the 2-dimensional triangle
inequality
\begin{equation}\label{eq:triangle}
  \left(\sqrt{S_1^2+\widetilde{S}_1^2}-\sqrt{S_2^2+\widetilde{S}_2^2}\right)^2\leq
  (S_1-S_2)^2+(\widetilde{S}_1-\widetilde{S}_2)^2,
\end{equation}
for all $S_1,S_2,\widetilde{S}_1,\widetilde{S}_2\in\R$.
The estimate (\ref{eq:triangle}) implies, in particular,
that
$$
n\mapsto T(\ell^{-1}\sqrt{n+1}-1)
$$
is a convex map on non-negative functions $n:\Z^3\to\R$. Hence
$$
T(S^\Psi_z)=T(\ell^{-1}\sqrt{\langle \hn_{z+\sigma}\rangle+1}-1)
\leq \langle T(\ell^{-1}\sqrt{\hn_{z+\sigma}+1}-1)\rangle
\leq \langle T(S_z^+)+T(S_z^-)\rangle,
$$
where the last inequality is again a consequence of (\ref{eq:triangle}).

As before we use the sliding technique to localize the electrostatic
energy. If we combine Lemma~\ref{lm:sliding} and
Lemma~\ref{lm:kineticloc} we obtain for
$\Psi\in\bigotimes^N_SL^2(\R^3\times\{1,-1\})$ normalized and
with support in $\left([-L/2,L/2]^3\times\{1,-1\}\right)^N$ that
\begin{eqnarray}
  (\Psi,H_{N,L}\Psi)&\geq&\gamma\widetilde\gamma\int\limits_{z\in[-1/2,1/2]^3}
  \left(\Psi,\left(\sum_{\sigma\in\Z^3}\wH_{z+\sigma}+\frac{\gamep}{2} T(S_z)\right)
    \Psi\right)\, dz\nonumber\\&&
  -CL^3/\ell^5-\omega(t)N\left(\frac{1}{2\ell}-\frac{1}{2L}\right)\label{eq:hmllower}
\end{eqnarray}
where
\begin{equation}\label{eq:gamma0}
  {\gamep}=(1+\varepsilon+Ct^3)^{-1}(\widetilde\gamma\gamma)^{-1}
\end{equation}
(recall that $\gamma$ and $\widetilde\gamma$ also depend on $t$)
and for $z\in\R^3$ we have introduced the operator
\begin{equation}\label{eq:Hz}
  \wH_z=\sum_{i=1}^N\left(
    \frac{{\gamep}}{2}\cK_{i}^{(z)}-\frac{{\gamep}}{2}\varepsilon\Delta^{(z)}_{i,\rm Neu}\right)
  +\sum_{1\leq i<j\leq N}e_ie_jw_{z}(x_i,x_j),
\end{equation}
with
\begin{equation}\label{eq:wz}
  w_z(x_i,x_j)=\upchi_z(x_i)Y_{\frac{\omega(t)}{\ell}}(x_i-x_j)\upchi_z(x_j).
\end{equation}
We shall choose $\varepsilon$ depending on $N$ in
Sect.~\ref{sec:final} in such a way that $\varepsilon\to0$ as
$N\to\infty$. We shall in particular assume that
$\varepsilon<1$. Hence we may assume that $\gamep$ is bounded above
and below by constants.

The expressions  $\left(\Psi,\sum_{\sigma\in\Z^3}\wH_{z+\sigma}\Psi\right)+\frac{1}{2}{\gamep}T(S_z^\Psi)$ are
equivalent for the different $z\in[-1/2,1/2]^3$. It is
therefore enough to consider $z=0$. Moreover, the operator
$\sum_{\sigma\in\Z^3}\wH_{\sigma}$ commutes with the
number operators $\hnu^\pm_\sigma$ (the operators giving the number of
positively and negatively charged particles in the cube $\{\ell\sigma\}+[-\ell/2,\ell/2]^3$).
Note that if $\Psi$ has support in the set
$\left([-L/2,L/2]^3\times\{1,-1\}\right)^N$ then for $z$ outside the set
$\Omega=[-\frac{L}{2\ell}-\frac{1}{2},\frac{L}{2\ell}+\frac{1}{2}]^3$ we have that $\Psi$ is
in the kernel of the operators $\hnu_z^\pm$.

The estimate (\ref{eq:hmllower}) then implies the following result.
\begin{thm}[Localization into small cubes]\label{thm:smallcubesloc}
With the definitions from (\ref{eq:ELN}),(\ref{eq:K}--\ref{eq:sz}),
and (\ref{eq:gamma0}--\ref{eq:wz}) we have that
$$
 E_L(N)\geq\gamma\widetilde\gamma\inf_{\Psi, \|\Psi\|=1}\left(\left(\Psi.\sum_{\sigma\in\Z^3}\wH_{\sigma}\Psi\right)
+\frac{\gamep}{2}T(S_0^\Psi)\right)
 -CL^3/\ell^5-\omega(t)N\left(\frac{1}{2\ell}-\frac{1}{2L}\right),
$$
where the infimum is over normalized functions
$$
  \Psi\in\cH_0=\Bigl\{\Psi\in{\bigotimes^N}_SL^2(\R^3\times[-1,1])\Bigl|\
  \hnu^\pm_\sigma\Psi=0\hbox{ for
    }\sigma\not\in\Z^3\cap
  \Bigl[-\frac{L}{2\ell}-\frac{1}{2},\frac{L}{2\ell}+\frac{1}{2}\Bigr]^3\Bigr\},
$$
and $0<t<1/2$ is the parameter from the beginning of Sect.~\ref{sec:loc}.
\end{thm}

In the following Sects.~\ref{sec:cutoff}--\ref{sec:energybound}
we shall study the operators $\wH_\sigma$,
$\sigma\in\Z^3$.
Since $\wH_\sigma$ commutes with $\hnu_\sigma^\pm$ we may simply restrict
to the eigenspaces of $\hnu_\sigma^\pm$. We shall therefore not think of
$\hnu_\sigma^\pm$ and $\hnu_\sigma$ as operators, but as classical parameters
$\nu_\sigma^\pm$ and $\nu_\sigma$.
Which values of these parameters that will give the
optimal energy is of course not known a-priori, but they must satisfy $\sum_\sigma\nu_\sigma=N$.
The operator $\wH_\sigma$ is in (\ref{eq:Hz}) written as an $N$-particle
operator, but it depends only on $\nu_\sigma$ particles. We shall
therefore simply think of it as a $\nu_\sigma$-particle operator.
The operators $\wH_\sigma$ for different $\sigma$ are all unitarily
equivalent. We shall in Sects.~\ref{sec:cutoff}--\ref{sec:energybound}
simply omit the subscript $\sigma$.

\section{Long and short distance cutoffs in the potential}\label{sec:cutoff}

Our aim in this section is to
replace $w$ in (\ref{eq:wz}) (omitting the index $z$)
by a function that has long and short distance
cutoffs.

We shall replace the function $w $ by
\begin{equation}\label{eq:wrR}
  w_{r,R}(x,y)=\upchi (x)V_{r,R}(x-y)\upchi (y)
\end{equation}
where
\begin{equation}\label{eq:VrR}
  V_{r,R}(x)=Y_{R^{-1}}(x)-Y_{r^{-1}}(x)=
  \frac{e^{-|x|/R}-e^{-|x|/r}}{|x|}
\end{equation}
Here $0<r\leq R\leq\omega(t)^{-1}\ell$.
Note that for $x\ll r$ then
$V_{r,R}(x)\approx r^{-1}-R^{-1}$
and for $|x|\gg R$ then
$V_{r,R}(x)\approx |x|^{-1}e^{-|x|/R}$.

We first introduce the cutoff $R$ alone, i.e., we bound the
effect of replacing $w $ by
$w_{R}(x,y)=\upchi (x)V_R(x-y)\upchi (y)$,
where $V_R(x)=|x|^{-1}e^{-|x|/R}=Y_{R^{-1}}(x)$.
Thus, since $R\leq \omega(t)^{-1}\ell$, the Fourier transforms satisfy
$$
\widehat{Y}_{\omega/\ell}(k)-\widehat{V}_R(k)
=4\pi\left(\frac{1}{k^2+(\omega(t)/\ell)^{2}}
  -\frac{1}{k^2+R^{-2}}\right)\geq0.
$$
(We use the convention that
    $\hat{f}(k)=\int f(x)e^{-ikx}\,dx$.)
Hence
$$
w (x,y)-w_{R}(x,y)=
\upchi (x)\left(Y_{\omega/\ell}-V_R\right)(x-y)\upchi (y)
$$
defines a positive semi-definite kernel.
Note, moreover, that
$\left(Y_{\omega/\ell}-V_R\right)(0)=
R^{-1}-\omega/\ell\leq R^{-1}$.
Thus,
\begin{eqnarray}
  \sum_{1\leq i<j\leq N}e_ie_j(w (x_i,x_j)-w_{R}(x_i,x_j))
  &=&\frac{1}{2}\sum_{1\leq i,j\leq N}e_ie_j(w -w_{R})(x_i,x_j)\nonumber\\
  &&-\sum_{i=1}^N\frac{e_i^2}{2}(w -w_{R})(x_i,x_i)\nonumber\\
  &\geq&  -\mfr{1}{2}\nu \left(Y_{\omega/\ell}-V_R\right)(0)
   \geq-\mfr{1}{2}\nu R^{-1}.\label{eq:R}
\end{eqnarray}
Here we have used that $\sum_{i=1}^N\upchi (x_i)^2\leq
\sum_{i=1}^NX (x_i)=\nu $.

We shall now bound the effect of replacing $w_{R}$ by $w_{r,R}$.
I.e., we are replacing
$V_R(x)=|x|^{-1}e^{-|x|/R}$ by
$|x|^{-1}\left(e^{-|x|/R}-e^{-|x|/r}\right)$. The correction is
$Y_{r^{-1}}(x)=|x|^{-1}e^{-|x|/r}$.
\begin{lm}\label{lm:r}
We have for all $\delta>0$ the operator inequality
$$
 \sum_{i=1}^N-\frac{\delta}{2}\Delta_{i,\rm Neu}^{(z)}+\sum_{1\leq i< j\leq
   N}e_ie_j\upchi (x_i)Y_{r^{-1}}(x_i-x_j)\upchi (x_j)
 \geq -\const \nu^+ \nu^- (r^2\ell^{-3}+\delta^{-3/2}r^{1/2}).
$$
\end{lm}
\begin{proof} We set $z=0$ and omit the index $z$ in this proof.
For $D>0$ we define
$$
W_D(x)=\left\{\begin{array}{ll}Y_{r^{-1}}(D),&|x|<D\\
    Y_{r^{-1}}(x), &|x|\geq D
  \end{array}\right..
$$
Then the measure
$$
\mu_D=(4\pi)^{-1}(-\Delta+r^{-2})W_D=(4\pi)^{-1}r^{-2}Y_{r^{-1}}(D){\bf 1}_{|x|<D}
-(4\pi)^{-1}Y_{r^{-1}}'(D)\delta(|x|-D)
$$
is non-negative and $W_D=\mu_D*Y_{r^{-1}}$. Here $Y_{r^{-1}}'$ refers to the radial
derivative of $Y_{r^{-1}}$ and ${\bf 1}_{|x|<D}$ is the
characteristic function of the set $\{|x|<D\}$.

Define the (signed) measure
$$
 d\mu(x)=\sum_{i=1}^Ne_i\upchi(x_i)d\mu_{D_i}(x-x_i),
$$
where
$$
  D_i=\mfr{1}{2}\min\left\{|x_i-x_j|\ | \ j=1,\ldots,N,\ e_ie_j=-1,
    \ x_i\in[-\ell/2,\ell/2]^3\right\},
$$
is half the distance from the particle at $x_i$ to the nearest
particle of opposite charge in the cube $[-\ell/2,\ell/2]^3$, i.e.,
the cube in which $\upchi$ is supported. Thus for $x_i,x_j\in
[-\ell/2,\ell/2]^3$
we have
\begin{eqnarray*}
  \iint Y_{r^{-1}}(x-y)d\mu_{D_i}(x-x_i)d\mu_{D_j}(y-x_j)&=&\int W_{D_j}(x-x_j)d\mu_{D_i}(x-x_i)
  \\&\leq&\int Y_{r^{-1}}(x-x_j)d\mu_{D_i}(x-x_i)\\&=&W_{D_i}(x_i-x_j)\leq Y_{r^{-1}}(x_i-x_j)
\end{eqnarray*}
and both inequalities become equalities if $e_ie_j=-1$. Hence
\begin{eqnarray*}
  \sum_{1\leq i<j\leq N}e_ie_j\upchi(x_i)Y_{r^{-1}}(x_i-x_j)\upchi(x_j)&\geq&
  \mfr{1}{2}\iint Y_{r^{-1}}(x-y)d\mu(x)d\mu(y)\\
  &&
  -\mfr{1}{2}\sum_{i=1}^N\upchi(x_i)^2\iint Y_{r^{-1}}(x-y)d\mu_{D_i}(x)d\mu_{D_i}(y)\\
  &\geq&-\mfr{1}{2}\sum_{i=1}^N\upchi(x_i)^2\iint Y_{r^{-1}}(x-y)d\mu_{D_i}(x)d\mu_{D_i}(y)
\end{eqnarray*}
since $Y_{r^{-1}}$ is positive type. This inequality is very similar
to Onsager's electrostatic inequality \cite{O}.

We calculate
\begin{eqnarray*}
  \iint Y_{r^{-1}}(x-y)d\mu_{D}(x)d\mu_{D}(y)&=&\int W_D(x)d\mu_D(x)=
  \mfr{1}{3}r^{-2}D^3Y_{r^{-1}}(D)^2-D^2Y_{r^{-1}}'(D)Y_{r^{-1}}(D)\\
  &=&\mfr{1}{3}\left(Dr^{-2}+3D^{-1}+3r^{-1}\right)e^{-2D/r}.
\end{eqnarray*}

We have therefore proved the operator estimate
$$
 \sum_{i=1}^N-\frac{\delta}{2}\Delta_{i,\rm Neu}+\sum_{1\leq i< j\leq N}
 e_ie_j\upchi(x_i)Y_{r^{-1}}(x_i-x_j)\upchi(x_j)
  \geq \sum_{i=1}^N-\frac{\delta}{2}\Delta_{i,\rm Neu}-v_i,
$$
where for $i=1,2,\ldots,N$
$$
  v_i(x_i)=-\mfr{1}{6}\chi(x_i)^2\left(D_ir^{-2}+3D_i^{-1}+3r^{-1}\right)e^{-2D_i/r}.
$$
Note that here $D_i$ depends on $x_i$ and the positions of all the
particles with opposite charge of $e_i$.

{F}rom the Sobolev estimate for the Neumann Laplacian in a cube of
size $\ell$ we have the general lower bound
$$
 -\Delta_{\rm Neu}-V\geq-\const\int V^{5/2}-\const\ell^{-3}\int V.
$$
This gives for $e_i=\pm1$ that
\begin{eqnarray*}
  -\frac{\delta}{2}\Delta_{i,\rm Neu}-v_i&\geq& -\const\nu^\mp
  \int_{\R^3}\delta^{-3/2}\left(\frac{|x|}{r^2}+\frac{3}{|x|}+\frac{3}{r}\right)^{5/2}e^{-5|x|/r}
  +\ell^{-3}\left(\frac{|x|}{r^2}+\frac{3}{|x|}+\frac{3}{r}\right)e^{-2|x|/r}\, dx\\
  &\geq&-\const\nu^\mp(\delta^{-3/2}r^{1/2}+\ell^{-3}r^2).
\end{eqnarray*}
When summed over $i$ this gives the result of the lemma.
\end{proof}

If we combine the bound (\ref{eq:R}) and Lemma~\ref{lm:r}
we have  the following result.
\begin{lm}[Long and short distance potential cutoffs]
  \label{lm:cutoffs}\hfill\\
  For $0<\delta<\varepsilon<1$ consider the Hamiltonian
  \begin{eqnarray}
    \wH_{r,R}^\delta&=&\sum_{i=1}^N\left(
    \frac{{\gamep}}{2}\cK_{i}^{(z)}
    -\frac{{\gamep}}{2}(\varepsilon-\delta)\Delta^{(z)}_{i,\rm Neu}\right)
  +\sum_{1\leq i<j\leq N}e_ie_jw_{r,R}(x_i,x_j)
      \label{eq:HrR}
  \end{eqnarray}
  where $w_{r,R}$ is given in (\ref{eq:wrR}) and (\ref{eq:VrR}).
  If $0<r\leq R<\omega(t)^{-1}\ell$
  then the Hamiltonian $\wH $ defined in (\ref{eq:Hz})
  obeys the lower bound
  \begin{equation}\label{eq:Hzlower}
    \wH \geq \wH^\delta_{r,R}-\mfr{1}{2}\nu  R^{-1}
    -\const\nu^+\nu^-(\delta^{-3/2}r^{1/2}+\ell^{-3}r^2).
  \end{equation}
  If $0<r\leq R=\omega(t)^{-1}\ell$
  we get
  \begin{equation}\label{eq:HzlowerRmax}
    \wH \geq \wH^\delta_{r,R}
    -\const\nu^+\nu^-(\delta^{-3/2}r^{1/2}+\ell^{-3}r^2).
  \end{equation}
\end{lm}
A similar argument gives the following result.
\begin{lm}
  \label{lm:r'R'} With the same notation as above we have for
  $0<r\leq r'\leq R'\leq R\leq\omega(t)^{-1}\ell$ that
  $$
  \wH^\delta_{r,R}\geq \wH^{2\delta}_{r',R'}-\mfr{1}{2}\nu  R'^{-1}-
  \const\nu^+\nu^-(\delta^{-3/2}r'^{1/2}+\ell^{-3}r'^2).
  $$
\end{lm}
\begin{proof}
Simply note that
$
V_{r',R'}(x)-V_{r,R}(x)=Y_{R'^{-1}}(x)-Y_{R^{-1}}(x)+
Y_{r^{-1}}(x)-Y_{r'^{-1}}(x)
$  and now use the same arguments as before.
\end{proof}

We shall not fix the cutoffs $r$ and $R$ nor the parameter $\delta$,
but rather choose them differently at different stages in the later
arguments. Moreover, we shall choose $r$ and $R$ to depend on
$\nu $. We give an example of this in the following corollary.

\begin{cl}[Simple bound on the number of excited particles]\label{cl:sn+bound}\hfill\\
  For any state such that the expectation
  $\langle \wH \rangle\leq 0$, the expectation of the number of
  excited particles satisfies
  $\langle\nu -\hn  \rangle\leq \nu
  \min\{1,
  \const \varepsilon^{-1}\ell\nu^{1/3}\}$. (Recall that $\wH $ depends
  on $\varepsilon$.)
\end{cl}
\begin{proof}
  We simply choose $r=R$ and $\delta=\varepsilon/2$ in
  Lemma~\ref{lm:cutoffs}.  Then $\wH_{r,R}^\delta\geq
  C(\varepsilon/2)(\nu -\hn ) \ell^{-2}$ since
  $(u,-\Delta_{\rm Neu}^{(z)}u)\geq \pi^2\ell^{-2}\|u\|^2$ for functions $u$
  in the cube $\{z\ell\}+[-\ell/2,\ell/2]^3$ orthogonal to constants.
  We therefore have that
  $$
  \wH \geq C(\varepsilon/2)(\nu -\hn ) \ell^{-2}
  -\mfr{1}{2}\nu  r^{-1}-
  \const\nu^+\nu^-(\varepsilon^{-3/2}r^{1/2}+\ell^{-3}r^2).
  $$
  Strictly speaking Lemma~\ref{lm:cutoffs} requires
  $r\leq R\leq\omega(t)^{-1}\ell$. If however $R=r>\omega(t)^{-1}\ell$
  we would get an even better estimate than the one stated if we set
  $r=R=\omega(t)^{-1}\ell$ and $\delta=\varepsilon/2$. Since in this
  case we may ignore the error $-\mfr{1}{2}\nu  r^{-1}$.

  We now estimate
  $\nu^+\nu^-\leq\nu^2/4$ and make the explicit choice $r=\min\{\varepsilon
  \nu^{-2/3},\ell\nu^{-1/3}\}$.
  When $\varepsilon\nu^{-2/3}\leq\ell\nu^{-1/3}$, i.e.,
  $\nu^{1/3}\geq\varepsilon\ell^{-1}$ we obtain
  $$
  \wH \geq C(\varepsilon/2)(\nu -\hn ) \ell^{-2}
  -\const\varepsilon^{-1}\nu^{5/3}
    -\const\varepsilon^2\ell^{-3}\nu^{4/3}
  \geq C(\varepsilon/2)(\nu -\hn ) \ell^{-2}
  -\const\varepsilon^{-1}\nu^{5/3}
  $$
  and when $\varepsilon\nu^{-2/3}\geq\ell\nu^{-1/3}$, i.e.,
  $\nu^{1/3}\leq\varepsilon\ell^{-1}$ we obtain
  $$
  \wH \geq C(\varepsilon/2)(\nu -\hn ) \ell^{-2}
  -\const\varepsilon^{-3/2}\ell^{1/2}\nu^{11/6}
  -\const\ell^{-1}\nu^{4/3}\geq
  C(\varepsilon/2)(\nu -\hn ) \ell^{-2}
  -\const\ell^{-1}\nu^{4/3}.
  $$
  These two bounds give that
  $$
  \langle\nu -\hn  \rangle\leq \const\nu
  \max\{(\varepsilon^{-1}\ell\nu^{1/3})^2,
  \varepsilon^{-1}\ell\nu^{1/3}\}.
  $$
  We of course clearly have that $\nu -\hn  \leq\nu $. We
  therefore get the result claimed.
\end{proof}

\section{Bound on the unimportant part of the Hamiltonian}
\label{sec:estimates}

In this section we shall bound the
Hamiltonian $\wH^\delta_{r,R}$ given in (\ref{eq:HrR}).
We emphasize that we do not necessarily have neutrality in the
cube, i.e., $\nu^+ $ and $\nu^- $ may be different.
We are simply looking for a lower bound to $\wH^\delta_{r,R}$,
that holds for all $\nu^\pm $.
The goal is to find a lower bound that will
allow us to conclude that the optimal value of $\nu^+ -\nu^- $, i.e.,
the value for which the energy of the Hamiltonian is smallest,
is indeed close to zero. We shall also conclude that the number of
excited particles $\nu -\hn $ is small. These conclusions will be
made in Sects.~\ref{sec:simple} and \ref{sec:localizingexcitations}.

In this section we shall fix $z=0$, i.e., we are working in the cube
$[-\ell/2,\ell/2]^3$. We shall simply omit the index $z$.

We shall express the Hamiltonian in second quantized language.
This is purely for convenience. We stress that we are not
in anyway changing the model by doing this and the treatment
is entirely rigorous and could have been done without
the use of second quantization.

Let $u_p$, $\ell p/\pi\in \left(\N\cup\{0\}\right)^3$ be an orthonormal
basis of  eigenfunction of the Neumann Laplacian $-\Delta_{\rm Neu}$ in the
cube $[-\ell/2,\ell/2]^3$. More precisely,
$-\Delta_{\rm Neu} u_p=|p|^2u_p$. I.e.,
$$
u_p(x_1,x_2,x_3)=c_p\ell^{-3/2}
\prod_{j=1}^3
\cos\left(p_j(x_j+\ell/2)\right),
$$
where the normalization satisfies $c_0=1$ and in general
$1\leq c_p\leq \sqrt{8}$.
The function $u_0=\ell^{-3/2}$ is the constant eigenfunction with eigenvalue $0$.
We note that for $p\ne0$ we have
\begin{equation}\label{eq:gap}
  (u_p,-\Delta_{\rm Neu} u_p)\geq \pi^2\ell^{-2}.
\end{equation}

We now express the Hamiltonian $\wH^\delta_{r,R}$ from (\ref{eq:HrR})
(omitting $z$) in terms of the creation and annihilation operators
$a^*_{p\pm}=a_\pm(u_p)^*$ and $\an_{p\pm}=a_\pm(u_p)$ .

Define
$$
\hw_{pq,\mu\nu}=\iint w_{r,R}(x,y)u_p(x)u_q(y)u_\mu(x)u_\nu(y)
\,dx\,dy.
$$

We may then express the two-body potential in $\wH^\delta_{r,R}$ as
$$
\sum_{1\leq i<j\leq n}
e_ie_jw_{r,R}(x_i,x_j)=
\mfr{1}{2}\sum_{pq,\mu\nu}\hw_{pq,\mu\nu}\left(
  a^*_{p+}a^*_{q+}\an_{\nu+} \an_{\mu+}
  +a^*_{p-}a^*_{q-}\an_{\nu-} \an_{\mu-}-2a^*_{p+}a^*_{q-}\an_{\nu-} \an_{\mu+}
\right).
$$

Motivated by Foldy's and Dyson's use of the Bogolubov
approximation it is our goal  to reduce the Hamiltonian  $\wH^\delta_{r,R}$
so that it has only what we call quadratic terms, i.e.,
terms which contain precisely two
$a^\#_{p\pm}$ with $p\ne0$.
More precisely, we want to be able to ignore all terms in the
two-body-potential containing the coefficients
\begin{itemize}
\item $\hw_{00,00}$.
\item $\hw_{p0,q0}=\hw_{0p,0q}$, where $p,q\ne0$. These terms
  are in fact quadratic, but do not appear in the Foldy Hamiltonian.
  We shall prove that they can also be ignored.
\item $\hw_{p0,00}=\hw_{0p,00}=\hw_{00,p0}=\hw_{00,0p}$,
  where  $p\ne0$.
\item
  $\hw_{pq,\mu0}=\hw_{\mu0,pq}=\hw_{qp,0\mu}=\hw_{0\mu,qp}$,
  where $p,q,\mu\ne0$.
\item $\hw_{pq,\mu\nu}$, where $p,q,\mu,\nu\ne0$.
\end{itemize}

We shall consider these cases one at a time.

\begin{lm}[Control of terms with $\hw_{00,00}$]\label{lm:0000}\hfill\\
  The sum of the terms in $\wH^\delta_{r,R}$ containing $\hw_{00,00}$
  is equal to
  $$
  \mfr{1}{2}\hw_{00,00}\left[(\hn^+-\hn^-)^2-\hn^+-\hn^-\right].
  $$
\end{lm}
\begin{proof}
  The terms containing $\hw_{00,00}$ are
  $$
  \mfr{1}{2}\hw_{00,00}\left(
    a^*_{0+}a^*_{0+}\an_{0+} \an_{0+}
    +a^*_{0-}a^*_{0-}\an_{0-} \an_{0-}-2a^*_{0+}a^*_{0-}\an_{0-} \an_{0+}
  \right),
  $$
  which gives the above result when using the commutation relation
  $[\an_p, a^*_q]=\delta_{p,q}$.
\end{proof}

\begin{lm}[Control of  terms with $\hw_{p0,q0}$]\label{lm:p0q0}\hfill\\
  The sum of the terms in $\wH^\delta_{r,R}$ containing $\hw_{p0,q0}$
  or $\hw_{0p,0q}$ with $p,q\ne0$ is bounded below by
  $$
  -4\pi R^2\ell^{-3}|\hn^+-\hn^-|(\nu-\hn).
  $$
\end{lm}
\begin{proof}
  The terms containing $\hw_{p0,q0}=\hw_{0p,0q}$ are
  \begin{eqnarray*}
    \sum_{p\ne0\atop q\ne0}\hw_{p0,q0}\left(
        a^*_{p+}a^*_{0+}\an_{0+} \an_{q+} +a^*_{p-}a^*_{0-}\an_{0-} \an_{q-}
        -a^*_{p+}a^*_{0-}\an_{0-} \an_{q+} -a^*_{p-}a^*_{0+}\an_{0+}
        \an_{q-} \right)\\
    =\sum_{p\ne0\atop q\ne0}\hw_{p0,q0}(\hn^+-\hn^-)
    (a^*_{p+}\an_{q+}-a^*_{p-}\an_{q-}).
  \end{eqnarray*}
  Note that $\hn^\pm$ commutes with
  $\sum\limits_{p\ne0\atop q\ne0}\hw_{p0,q0}a^*_{p\pm}\an_{q\pm}$.

  We have that
  $$
  \hw_{p0,q0}=\ell^{-3}\int\int w_{r,R}(x,y)\,dy u_p(x)u_q(x)\,dx.
  $$
  Hence
  \begin{eqnarray*}
    \sum\limits_{p\ne0\atop q\ne0}\hw_{p0,q0}a^*_{p\pm}\an_{q\pm}
    &=&\ell^{-3}\int\int w_{r,R}(x,y)\,dy
    \left(\sum_{p\ne0}u_p(x)a^*_{p\pm}\right)
    \left(\sum_{p\ne0}u_p(x)a^*_{p\pm}\right)^*\,dx.\\
    &\leq&\ell^{-3}\sup_{x'}\int w_{r,R}(x',y)\,dy
    \int\left(\sum_{p\ne0}u_p(x)a^*_{p\pm}\right)
    \left(\sum_{p\ne0}u_p(x)a^*_{p\pm}\right)^*\,dx.
    \\&=&\ell^{-3}\sup_{x'}\int w_{r,R}(x',y)\,dy\sum_{p\ne0}a^*_{p\pm}\an_{p\pm}.
  \end{eqnarray*}
  Since $\sum_{p\ne0}a^*_{p\pm}\an_{p\pm}=\nu^\pm-\hn^\pm$ and
  $$
  \sup_x\int w_{r,R}(x,y)\,dy\leq \int V_{r,R}(y)\,dy\leq
  4\pi R^2
  $$
  we obtain the operator inequality
  $$
  0\leq\sum_{p\ne0\atop q\ne0}\hw_{p0,q0}a^*_{p\pm}\an_{q\pm}
  \leq4\pi\ell^{-3}R^2(\nu^\pm-\hn^\pm),
  $$
  and the lemma follows.
\end{proof}

Before treating the last three types of terms
we shall need the following result on the structure
of the coefficients $\hw_{pq,\mu\nu}$.

\begin{lm}\label{lm:hw}
  For all $p',q'\in(\pi/\ell)\left(\N\cup\{0\}\right)^3$
  and $\alpha\in\N$
  there exists $J^\alpha_{p'q'}\in\R$ with
  $J^\alpha_{p'q'}=J^\alpha_{q'p'}$ such that
  for all $p,q,\mu,\nu\in(\pi/\ell)\left(\N\cup\{0\}\right)^3$
  we have
  \begin{equation}\label{eq:wpostyp}
    \hw_{pq,\mu\nu}=\sum_\alpha J^\alpha_{p\mu} J^\alpha_{q\nu}.
  \end{equation}
  Moreover we have the operator inequalities
  \begin{equation}\label{eq:p00p'}
    0\leq\sum_{p,p'\ne0}\hw_{pp',00}a^*_{p\pm}\an_{p'\pm}
    =\sum_{p,p'\ne0}\hw_{p0,0p'}a^*_{p\pm}\an_{p'\pm}
    \leq4\pi\ell^{-3}R^2(\nu^\pm-\hn^\pm)
  \end{equation}
  and
  $$
  0\leq\sum_{p,p',m\ne0}\hw_{pm,mp'}a^*_{p\pm}\an_{p'\pm}\leq r^{-1}(\nu^\pm-\hn^\pm).
  $$
\end{lm}
\begin{proof}
  The operator ${\cal A}$ with integral kernel $w_{r,R}(x,y)$ is a non-negative
  Hilbert-Schmidt
  operator on $L^2(\R^3)$ with norm less than
  $\sup_k\hat V_{r,R}(k)\leq 4\pi R^2$.
  Denote the eigenvalues of ${\cal A}$  by
  $\lambda_\alpha$, $\alpha=1,2,\ldots$ and
  corresponding orthonormal eigenfunctions by
  $\varphi_\alpha$. We may assume
  that these functions are real.
 The eigenvalues satisfy
  $0\leq\lambda_\alpha\leq {4\pi}R^2$.
  We then have
  $$
  \hw_{pq,\mu\nu}=\sum_\alpha\lambda_\alpha
  \int u_p(x)u_\mu(x)\varphi_\alpha(x)\,dx
  \int u_q(y)u_\nu(y)\varphi_\alpha(y)\,dy.
  $$
  The identity (\ref{eq:wpostyp})  thus follows with
  $J_{p\mu}^\alpha=\lambda_\alpha^{1/2}\int u_p(x)u_\mu(x)\varphi_\alpha(x)\,dx$.

  If as before $\cP$ denotes the projection orthogonal to constants we may also
  consider the operator $\cP{\cal A}\cP$.
  Denote its eigenvalues and eigenfunctions
  by $\lambda'_\alpha$ and $\varphi'_\alpha$.
  Then again $0\leq\lambda'_\alpha\leq 4\pi R^2$.
  Hence we may write
  $$
  \hw_{p0,0p'}=\ell^{-3}\sum_\alpha\lambda'_\alpha
  \int u_p(x)\varphi'_\alpha(x)\,dx
  \int u_{p'}(y)\varphi'_\alpha(y)\,dy.
  $$
  Thus since all $\varphi'_\alpha$ are orthogonal to constants
  we have
  \begin{eqnarray*}
    \lefteqn{\sum_{p,p'\ne0}\hw_{p0,0p'}a^*_{p\pm}\an_{p'\pm}}&& \\
    &=&\ell^{-3}\sum_\alpha \lambda'_\alpha
    \left(\sum_{p\ne0}\int u_p(x)\varphi'_\alpha(x)\,dx\, a^*_{p\pm}\right)
    \left(\sum_{p\ne0}\int u_p(x)\varphi'_\alpha(x)\,dx\, a^*_{p\pm}\right)^*\\
    &=&\ell^{-3}\sum_\alpha \lambda'_\alpha a_\pm^*\left(\varphi'_\alpha\right)
    \an_\pm\left(\varphi'_\alpha\right).
  \end{eqnarray*}
  The inequalities (\ref{eq:p00p'}) follow immediately from this.

  The fact that $\sum_{p,p',m\ne0}\hw_{pm,mp'}a^*_{p\pm}\an_{p'\pm}\geq 0$ follows
  from the representation (\ref{eq:wpostyp}).
  Moreover, since the kernel $w_{r,R}(x,y)$ is a continuous function
  we have that
  $w_{r,R}(x,x)=\sum_{\alpha}\lambda_\alpha\varphi_\alpha(x)^2$
  for almost all $x$ and hence
  $$
  \sum_{m\ne0}\hw_{pm,mp'}=\int u_p(x)u_{p'}(x)w_{r,R}(x,x)\,dx
  -\hw_{p0,0p'}.
  $$
  We therefore have
  \begin{eqnarray*}
    \lefteqn{\sum_{p,p',m\ne0}\hw_{pm,mp'}a^*_{p\pm}\an_{p'\pm}\leq\sum_{p,p'\ne0}
      \int u_p(x)u_{p'}(x)w_{r,R}(x,x)\,dx\,a^*_{p\pm}\an_{p'\pm}}&&\\
    &=&\int w_{r,R}(x,x)\left(\sum_{p\ne0} u_p(x)a^*_{p\pm}\right)
    \left(\sum_{p\ne0} u_p(x)a^*_{p\pm}\right)^*\,dx\\
    &\leq&\sup_{x'}w_{r,R}(x',x')\int\left(\sum_{p\ne0} u_p(x)a^*_{p\pm}\right)
    \left(\sum_{p\ne0} u_p(x)a^*_{p\pm}\right)^*\,dx
    =\sup_{x'}w_{r,R}(x',x')(\nu^\pm-\hn^\pm)
  \end{eqnarray*}
  and the lemma follows since $\sup_{x'}w_{r,R}(x',x')\leq r^{-1}$.
\end{proof}

\begin{lm}[Control of terms with $\hw_{pq,\mu\nu}$]\label{lm:pqmunu}\hfill\\
  The sum of the terms in $\wH^\delta_{r,R}$ containing $\hw_{pq,\mu\nu}$,
  with $p,q,\mu,\nu\ne0$
  is bounded below by
  $$
  \mfr{1}{2}\sum_\alpha
  \Bigl(\sum_{p\mu\ne0}J^\alpha_{p\mu}(a^*_{p+}\an_{\mu+}-a^*_{p-}\an_{\mu-})\Bigr)^2
  -\mfr{1}{2}r^{-1}(\nu-\hn).
  $$
\end{lm}
\begin{proof}
  The relevant terms are
  \begin{eqnarray*}
    \lefteqn{\mfr{1}{2}\sum_{pq,\mu\nu\ne0}\hw_{pq,\mu\nu}\left(
        a^*_{p+}a^*_{q+}\an_{\nu+} \an_{\mu+}
        +a^*_{p-}a^*_{q-}\an_{\nu-} \an_{\mu-}-2a^*_{p+}a^*_{q-}\an_{\nu-} \an_{\mu+}
      \right)}&&\\
    &=&
    \mfr{1}{2}\sum_{pq,\mu\nu\ne0}\hw_{pq,\mu\nu}\left(
      a^*_{p+}\an_{\mu+}a^*_{q+}\an_{\nu+}
      +a^*_{p-}\an_{\mu-}a^*_{q-}\an_{\nu-} -2a^*_{p+}a^*_{q-}\an_{\nu-} \an_{\mu+}
    \right)\\&&-\mfr{1}{2}\sum_{pp',m\ne0}\hw_{pm,mp'}a^*_{p+}\an_{p'+}
    -\mfr{1}{2}\sum_{pp',m\ne0}\hw_{pm,mp'}a^*_{p-}\an_{p'-}\\
    &=&\mfr{1}{2}\sum_\alpha
    \Bigl(\sum_{p\mu\ne0}J^\alpha_{p\mu}(a^*_{p+}\an_{\mu+}-a^*_{p-}\an_{\mu-})\Bigr)^2
    -\mfr{1}{2}r^{-1}(\nu-\hn),
  \end{eqnarray*}
  where we have used the last estimate in Lemma~\ref{lm:hw} and that
  $\nu^+-\hn^++\nu^--\hn^-=\nu-\hn$.
\end{proof}
\begin{lm}[Control of terms with $\hw_{p0,00}$]\label{lm:p000}\hfill\\
  The sum of the terms in $\wH^\delta_{r,R}$ containing $\hw_{p0,00}$,
  $\hw_{0p,00}$,$\hw_{00,p0}$, or $\hw_{00,0p}$, with $p\ne0$
  is, for all $\varepsilon'>0$, bounded below by
  \begin{equation}\label{eq:p000-1}
    -\varepsilon'^{-1}4\pi\ell^{-3}R^2(\hn+1)(\nu-\hn)
    -2\varepsilon'\hw_{00,00}
    \left((\hn^+-\hn^-)^2+1\right),
  \end{equation}
  or by
  \begin{eqnarray}
    &&(\nu^+-\nu^-)\sum_{p\ne0}\hw_{p0,00}\left(
    a_{p+}^*\an_{0+}-a_{p-}^*\an_{0-}
      +a_{0+}^*\an_{p+}-a_{0-}^*\an_{p-}\right)\nonumber\\
    &&-\varepsilon'^{-1}4\pi\ell^{-3}R^2(\hn+1)(\nu-\hn)
    -2\varepsilon'\hw_{00,00}\left((\nu-\hn)^2+1\right)).
    \label{eq:p000-2}
  \end{eqnarray}
\end{lm}
\begin{proof} The terms containing $\hw_{p0,00}$,
  $\hw_{0p,00}$,$\hw_{00,p0}$, or $\hw_{00,0p}$ are
  \begin{eqnarray*}
    \sum_{p\ne0}\mfr{1}{2}\hw_{p0,00}\!\!\!\!\!\!\!\!\!&\Bigl(\!\!\!\!\!\!\!\!\!&
    2a_{p+}^*a^*_{0+}\an_{0+}\an_{0+}
      +2a_{0+}^*a^*_{0+}\an_{0+}\an_{p+}+2a_{p-}^*a^*_{0-}\an_{0-}\an_{0-}
      +2a_{0-}^*a^*_{0-}\an_{0-}\an_{p-}\\&&
      -2a_{0-}^*\an_{0-}a_{0+}^*\an_{p+}-2a_{0-}^*\an_{0-}a_{p+}^*\an_{0+}
      -2a_{0+}^*\an_{0+}a_{0-}^*\an_{p-}-2a_{0+}^*\an_{0+}a_{p-}^*\an_{0-}\Bigr)\\
    &=&\sum_\alpha \sum_{p\ne0}J^\alpha_{p0}J^\alpha_{00}\left(
    (\hn^+-\hn^-)(a_{p+}^*\an_{0+}-a_{p-}^*\an_{0-})
      +(a_{0+}^*\an_{p+}-a_{0-}^*\an_{p-})(\hn^+-\hn^-)\right)\\
    &=&\sum_\alpha \sum_{p\ne0}J^\alpha_{p0}J^\alpha_{00}
    \left( a_{p+}^*\an_{0+}
      (\hn^++1-\hn^-)+ (\hn^++1-\hn^-)a_{0+}^*\an_{p+}\right)\\
    &&-\sum_\alpha \sum_{p\ne0}J^\alpha_{p0}J^\alpha_{00}
    \left(a_{p-}^*\an_{0-}(\hn^+-\hn^--1)+a_{0-}^*\an_{p-}(\hn^+-\hn^--1)\right)
    \\
    &\geq&-\varepsilon'\sum_\alpha
    \left(J^\alpha_{00}\right)^2\left(
      (\hn^+-\hn^-+1)^2+(\hn^+-\hn^--1)^2\right)
    \\&&
    -\varepsilon'^{-1}(\hn^++1)\sum_\alpha \sum_{p,p'\ne0}
    J^\alpha_{p0}J^\alpha_{p'0}a^*_{p+} \an_{p'+}
    -\varepsilon'^{-1}(\hn^-+1)\sum_\alpha \sum_{p,p'\ne0}
    J^\alpha_{p0}J^\alpha_{p'0}a^*_{p-} \an_{p'-}.
    \end{eqnarray*}
  Here we have used
  the representation (\ref{eq:wpostyp}) and in the last step a
  Cauchy-Schwarz inequality.
  We arrive at the first bound in the lemma since from (\ref{eq:p00p'})
  we have that
  $$
  \sum_\alpha \sum_{p,p'\ne0}
    J^\alpha_{p0}J^\alpha_{p'0}a^*_{p\pm}\an_{p'\pm}
    =\sum_{p,p'\ne0}\hw_{p0,0p'}a^*_{p\pm}
    \an_{p'\pm}
    \leq4\pi\ell^{-3}R^2(\nu^\pm-\hn^\pm).
  $$

  The second  bound (\ref{eq:p000-2}) follows in the same
  way if we notice that the terms containing $\hw_{p0,00}$,
  $\hw_{0p,00}$,$\hw_{00,p0}$, or $\hw_{00,0p}$  may be written as
  \begin{eqnarray*}
    (\nu^+-\nu^-)\sum_{p\ne0}\hw_{p0,00}\!\!\!\!\!\!&(\!\!\!\!\!\!&
    a_{p+}^*\an_{0+}-a_{p-}^*\an_{0-}
      +a_{0+}^*\an_{p+}-a_{0-}^*\an_{p-})\\
    +\sum_\alpha \sum_{p\ne0}J^\alpha_{p0}J^\alpha_{00}\!\!\!\!\!\!&\Bigl(\!\!\!\!\!\!&
    (\hn^+-\nu^+-\hn^-+\nu^-)(a_{p+}^*\an_{0+}-a_{p-}^*\an_{0-})\\&&
      +(a_{0+}^*\an_{p+}-a_{0-}^*\an_{p-})(\hn^+-\nu^+-\hn^-+\nu^-\Bigr).
  \end{eqnarray*}
\end{proof}

\begin{lm}[Control of terms with $\hw_{pq,m0}$]\label{lm:pqm0}\hfill\\
  The sum of the terms in $\wH^\delta_{r,R}$ containing $\hw_{pq,m0}$,
  $\hw_{qp,0m}$,$\hw_{m0,pq}$, or $\hw_{0m,qp}$, with $p,q,m\ne0$
  is bounded below by
  \begin{eqnarray*}
    -\varepsilon'^{-1}{4\pi} \ell^{-3}R^2(\hn+1)(\nu-\hn)
    -2\varepsilon'\sum_\alpha
    \Bigl(\sum_{p\mu\ne0}J^\alpha_{p\mu}(a^*_{p+}\an_{\mu+}-a^*_{p-}\an_{\mu-})\Bigr)^2,
  \end{eqnarray*}
  for all $\varepsilon'>0$.
\end{lm}
\begin{proof}The terms containing$\hw_{pq,m0}$,
  $\hw_{qp,0m}$,$\hw_{m0,pq}$, or $\hw_{0m,qp}$, with $p,q,m\ne0$  are
  \begin{eqnarray*}
    \sum_{pqm\ne0}
      \hw_{pq,m0}\!\!\!\!\!\!\!&\Bigl(\!\!\!\!\!\!\!&
      a^*_{p+}a^*_{q+}\an_{m+}\an_{0+}+a^*_{0+}a^*_{m+}\an_{q+}\an_{p+}
    +a^*_{p-}a^*_{q-}\an_{m-}\an_{0-}+a^*_{0-}a^*_{m-}\an_{q-}\an_{p-}\\
    \!\!\!\!\!\!\!&\!\!\!\!\!\!\!&
    -a^*_{p+}a^*_{q-}\an_{m+}\an_{0-}-a^*_{0-}a^*_{m+}\an_{q-}\an_{p+}
    -a^*_{p-}a^*_{q+}\an_{m-}\an_{0+}-a^*_{0+}a^*_{m-}\an_{q+}\an_{p-}\Bigr)\\
    &=&
    \sum_\alpha \sum_{q\ne0}J^\alpha_{q0}(a^*_{q+}\an_{0+}-a^*_{q-}\an_{0-})
      \sum_{pm\ne0}J^\alpha_{pm}(a^*_{p+}\an_{m+}-a^*_{p-}\an_{m-})\\&&
      +\sum_\alpha
      \sum_{pm\ne0}J^\alpha_{pm}(a^*_{p+}\an_{m+}-a^*_{p-}\an_{m-})
      \sum_{q\ne0}J^\alpha_{q0}(a^*_{q+}\an_{0+}-a^*_{q-}\an_{0-})^*
    \\
    &\geq&
    -\varepsilon'^{-1}\sum_{qq'\ne0}\hw_{q0,0q'}a^*_{q+}\an_{q'+}\an_{0+}a^*_{0+}
    -\varepsilon'^{-1}\sum_{qq'\ne0}\hw_{q0,0q'}a^*_{q-}\an_{q'-}\an_{0-}a^*_{0-}\\
    &&-2\varepsilon'\sum_\alpha\Bigl(\sum_{pm\ne0}J^\alpha_{pm}
    (a^*_{p+}\an_{m+}-a^*_{p-}\an_{m-})\Bigr)^2
  \end{eqnarray*}
  The lemma follows from (\ref{eq:p00p'}).
\end{proof}
Note that if $\varepsilon'\leq1/4$ then the last term in the estimate in Lemma~\ref{lm:pqm0} can be
controlled by the positive term in Lemma~\ref{lm:pqmunu}.

\section{Analyzing the quadratic Hamiltonian}\label{sec:quadratic}

In this section we consider the main part of the Hamiltonian
$\wH^\delta_{r,R}$, namely the ``quadratic'' Hamiltonian considered by Foldy.
As in the previous section we have fixed $z=0$, i.e., we consider the
cube $[-\ell/2,\ell/2]^3$ and we omit $z$.
The ``quadratic'' Hamiltonian
consists of the kinetic energy $\cK$ from (\ref{eq:K}) and all the two-body terms with the
coefficients $\hw_{pq,00}$, $\hw_{00,pq}$ $\hw_{p0,0q}$, and
$\hw_{0p,q0}$with $p,q\ne0$, i.e.,
\begin{eqnarray}\label{eq:Foldy}
  H_{\rm Foldy}=\sum_{i=1}^N\frac{{\gamep}}{2}\cK_{i}
  +\mfr{1}{2}\sum_{pq\ne0\atop e,e`=\pm}\hw_{pq,00}ee'\Bigl(2a^*_{pe}a^*_{0e'}\an_{qe'}\an_{0e} +
    a^*_{pe}a^*_{qe'}\an_{0e'} \an_{0e}+a^*_{0e}a^*_{0e'}\an_{pe'}\an_{qe}\Bigr).
\end{eqnarray}
In order to compute all  the bounds we found it necessary to include
the first term in (\ref{eq:p000-2}) into the ``quadratic''
Hamiltonian.
We therefore define
\begin{eqnarray}\label{eq:quadratichamiltonian}
  H_{Q}&=&H_{\rm Foldy}+(\nu^+-\nu^-)\sum_{p\ne0}\hw_{p0,00}\left(
    a_{p+}^*\an_{0+}-a_{p-}^*\an_{0-}
      +a_{0+}^*\an_{p+}-a_{0-}^*\an_{p-}\right).
\end{eqnarray}
Our goal is to give a lower bound on the ground state energy of the
Hamiltonian $H_Q$.

For any $k\in\R^3$ denote $\upchi_{k}(x)=e^{ikx}\upchi(x)$
and define the operators
\begin{equation}\label{eq:bdef}
b_{k\pm}^*=(\ell^3\nu^\pm)^{-1/2}a^*_\pm(\cP\upchi_{k})\an_{0\pm}\quad
\hbox{and}\quad
\bn_{k\pm}=(\ell^3\nu^\pm)^{-1/2}\an_\pm(\cP\upchi_{k})a_{0\pm}^*,
\end{equation}
where as before $\cP$ denotes the projection orthogonal to constants.
(On the subspace where $\nu^{+}=0$ we set $b_{k+}^*=0$ and likewise
when $\nu^{-}=0$ we set $b_{k-}^*=0$.)
These operators satisfy the commutation relations (note that
$\an_{0\pm}$ commutes with $a^*_\pm(\cP\upchi_k)$)
\begin{equation}\label{eq:bcommutation}
  [\bn_{k+},b^*_{k'-}]=[b^*_{k+},b^*_{k'-}]=[b^*_{k-},b^*_{k'-}]=0
\end{equation}
and\footnote{The corresponding equation (25) in \cite{LSo} has a typographical
error.}
$$
[\bn_{k\pm},b^*_{k'\pm}]=(\ell^3\nu^{\pm})^{-1}
a_{0\pm}^*\an_{0\pm}\left(\cP\upchi_{k},\cP\upchi_{k'}\right)
-(\ell^3\nu^{\pm})^{-1}a^*_\pm(\cP\upchi_{k'})\an_\pm(\cP\upchi_{k}).
$$
In particular, we get
\begin{equation}\label{eq:bcomk=k'}
  [\bn_{k\pm},b^*_{k\pm}]\leq(\ell^3\nu^{\pm})^{-1}
  a_{0\pm}^*\an_{0\pm}\left(\cP\upchi_{k},\cP\upchi_{k}\right)
  \leq 1.
\end{equation}
We shall now estimate the ``quadratic'' Hamiltonian
below by an operator that is really quadratic in the operators
$b^*_{k\pm}$ and $\bn_{k\pm}$.

\begin{lm}[The quadratic Hamiltonian]\label{lm:hQ}\hfill\\
  The quadratic Hamiltonian satisfies the lower bound
  \begin{equation}\label{eq:hQ}
    H_Q\geq
    \int_{\R^3}h_Q(k)\,dk
    -\sum_{pq\ne0}\hw_{pq,00}(a^*_{p+}\an_{q+}+a^*_{p-}\an_{q-}),
  \end{equation}
  where
  \begin{eqnarray}
    h_Q(k)&=&
    (2\pi)^{-3}\frac{\ell^3{\gamep}}{4}
    \frac{|k|^4}{|k|^2+(\ell t^6)^{-2}}\left(
      b_{k+}^*\bn_{k+}+b_{-k+}^*\bn_{-k+}
      +b_{k-}^*\bn_{k-}+b_{-k-}^*\bn_{-k-}\right)\nonumber\\
    &&+\frac{\hat{V}_{r,R}(k)}{2(2\pi)^3}\Bigl[
    (\nu^+-\nu^-)\ell^{-3}
    \Big(\widehat\upchi(k)(\nu^+)^{1/2}b^*_{k+}+(\nu^+)^{1/2}\bn_{-k+}
    -(\nu^-)^{1/2}b^*_{k-}-(\nu^-)^{1/2}\bn_{-k-})
      \nonumber\\&&\hspace{3.5truecm}
    +\overline{\widehat\upchi(k)}((\nu^+)^{1/2}\bn_{k+}+(\nu^+)^{1/2}b^*_{-k+}
    -(\nu^-)^{1/2}\bn_{k-}-(\nu^-)^{1/2}b^*_{-k-})
    \Big)\nonumber\\&&
    \phantom{ \frac{\hat{V}_{r,R}(k)}{2(2\pi)^3}\Bigl[}\
    +\sum_{e,e'=\pm}ee'\sqrt{\nu^e\nu^{e'}}(b^*_{ke}\bn_{ke'}+b^*_{-ke}\bn_{-ke'}+b^*_{ke}b^*_{-ke'}+
    \bn_{ke}\bn_{-ke'})\Bigr]\label{eq:hQdef}
    \end{eqnarray}
\end{lm}
\begin{proof}
  We consider first the kinetic energy
  \begin{eqnarray*}
    \frac{{\gamep}}{2} \sum_{i=1}^N\cK_{i}&=&\frac{{\gamep}}{2} \sum_{p,q}(a^*_{p+}\an_{q+}
    +a^*_{p-}\an_{q-})(u_p,\cK u_q)\\
    &=&(2\pi)^{-3}\frac{{\gamep}}{2} \int \frac{|k|^4}{|k|^2+(\ell t^6)^{-2}}
    \sum_{p,q\ne0}(a^*_{p+}\an_{q+}+a^*_{p-}
    \an_{q-})(u_p,\upchi_k)(\upchi_k,u_q)\,dk \\
    &=&(2\pi)^{-3}\frac{{\gamep}}{2}
  \int \frac{|k|^4}{|k|^2+(\ell t^6)^{-2}}
  (a^*_{+}(\cP\upchi_k)\an_{+}(\cP\upchi_k)+
  a^*_{-}(\cP\upchi_k)\an_{-}(\cP\upchi_k))\,dk.
  \end{eqnarray*}
  We get the first line in the expression for $h_Q(k)$ if we use that
  $$
  a^*_{\pm}(\cP\upchi_k)\an_{\pm}(\cP\upchi_k)\geq (\nu^\pm)^{-1}
  a^*_{\pm}(\cP\upchi_k)\an_{0\pm}a^*_{0\pm}\an_{\pm}(\cP\upchi_k)
  =\ell^3b^*_{k\pm}\bn_{k\pm}.
  $$

  Since
  $$
  w_{r,R}(x,y)=(2\pi)^{-3}\int\hat{V}_{r,R}(k)
  \upchi_{k}(x)\overline{\upchi_{k}(y)}\,dk,
  $$
  we have that
  $$
  \hw_{pq,00}=(2\pi\ell)^{-3}\int\hat{V}_{r,R}(k)
  (u_p,\cP\upchi_k)(\cP\upchi_k,u_q)\, dk=
  (2\pi\ell)^{-3}\int\hat{V}_{r,R}(k)
  (u_p,\cP\upchi_k)(u_q,\cP\upchi_{-k})\, dk.
  $$
  If we use that $\hat{V}_{r,R}(k)=\hat{V}_{r,R}(-k)$ and
  $\overline{\widehat{\upchi}(k)}={\widehat{\upchi}(-k)}$ we arrive at
  the expression for $h_Q$. The last sum in (\ref{eq:hQ}) comes from
  commuting $a^*_{0\pm}\an_{0\pm}$ to $\an_{0\pm}a^*_{0\pm}$.
\end{proof}

\begin{thm}[Particular case of Bogolubov's method]\label{thm:bogolubov}\hfill\\
For arbitrary constants $\cA,\cB_+,\cB_->0$ and $\kappa\in\C$ we
have for each fixed $k\in\R^3$
the following inequality involving the operators $\bn_{\pm k\pm}$ defined in (\ref{eq:bdef})
\begin{eqnarray}
\lefteqn{\cA(b^*_{k+}\bn_{k+}+b^*_{-k+}\bn_{-k+}+b^*_{k-}\bn_{k-}+b^*_{-k-}\bn_{-k-})}&&
\nonumber\\
&&+\kappa\left(\cB_+^{1/2}b^*_{k+}-\cB_-^{1/2}b^*_{k-}
      +\cB_+^{1/2}\bn_{-k+}-\cB_-^{1/2}\bn_{-k-}\right)\nonumber\\&&
    +\overline{\kappa}\left(\cB_+^{1/2}\bn_{k+}-\cB_-^{1/2}\bn_{k-}
      +\cB_+^{1/2}b^*_{-k+}-\cB_-^{1/2}b^*_{-k-}\right)
    \nonumber\\&&
    +\sum_{e,e'=\pm}\sqrt{\cB_e\cB_{e'}}\Bigl(b^*_{ke}\bn_{ke'}+b^*_{-ke}\bn_{-ke'}+b^*_{ke}b^*_{-ke'}+\bn_{ke}\bn_{-ke'}\Bigr)
    \nonumber\\&
    \geq&-\left(\cA+\cB_++\cB_-\right)+\sqrt{(\cA+\cB_++\cB_-)^2-(\cB_++\cB_-)^2}
    -|\kappa|^2.
\label{eq:bogolubov}
\end{eqnarray}
\end{thm}
\begin{proof}
Let us introduce
$$
d_\pm^*=(\cB_++\cB_-)^{-1/2}(\cB_+^{1/2}b^*_{\pm k+}-\cB_-^{1/2}b^*_{\pm
  k-}),\quad
c_\pm^*=(\cB_++\cB_-)^{-1/2}(\cB_-^{1/2}b^*_{\pm k+}+\cB_+^{1/2}b^*_{\pm
  k-}).
$$
These operators are analogous to the $\cn_k$ and $\dn_k$ discussed at the
end of Sect.~\ref{sec:heuristics}.
According to (\ref{eq:bcommutation}) and (\ref{eq:bcomk=k'}) these operators satisfy
$$
 [d_+^*,d_-^*]=0,\quad [c_+^*,c_-^*]=0,\quad
 [\dn_+,d_+^*]\leq1,\quad [\dn_-,d_-^*]\leq1,\quad [\cn_+,c_+^*]\leq1,\quad [\cn_-,c_-^*]\leq1.
$$
The right side of (\ref{eq:bogolubov}) can be rewritten as
\begin{eqnarray*}
  &&\cA(d_+^*\dn_++d_-^*\dn_-+c_+^*\cn_++c_-^*\cn_-)\\
  &&+(\cB_++\cB_-)\left(d_+^*\dn_++d_-^*\dn_-+d^*_+d^*_-
    +\dn_+\dn_-
    +(\cB_++\cB_-)^{-1/2}\left(\overline{\kappa}(d_+^*+\dn_-)+\kappa(\dn_++d^*_-)\right)\right).
\end{eqnarray*}
We may now complete the squares to write this as
\begin{eqnarray*}
  &&\cA(c_+^*\cn_++c_-^*\cn_-)+D(d_+^*+\alpha \dn_-+a)(d_+^*+\alpha
  \dn_-+a)^*
  +D(d^*_-+\alpha\dn_++\overline{a})(d^*_-+\alpha\dn_++\overline{a})^*\\
  &&-D\alpha^2([\dn_+,d^*_+]+[\dn_-,d^*_-])-2D|a|^2
\end{eqnarray*}
if
$$
 D(1+\alpha^2)=\cA+\cB_++\cB_-,\quad 2D\alpha=\cB_++\cB_-,\quad
 aD(1+\alpha)=\kappa(\cB_++\cB_-)^{1/2}.
$$

We choose the solution
$
\alpha=1+\frac{\cA}{\cB_++\cB_-}-\sqrt{\left(1+\frac{\cA}{(\cB_++\cB_-)}\right)^2-1}.
$
Hence
$$
D\alpha^2=(\cB_++\cB_-)\alpha/2=
\mfr{1}{2}\left(\cA+\cB_++\cB_--\sqrt{(\cA+\cB_++\cB_-)^2-(\cB_++\cB_-)^2}\right),
$$
$$
D|a|^2=\frac{|\kappa|^2(\cB_++\cB_-)}{D(1+\alpha^2+2\alpha)}=
\frac{|\kappa|^2(\cB_++\cB_-)}{\cA+2\cB_++2\cB_-}\leq\mfr{1}{2}|\kappa|^2.
$$
\end{proof}

Usually when applying Bogolubov's method the operators $b$ and $b^*$
satisfy the canonical commutation relations. In this case the lower
bound given above is indeed the lowest eigenvalue of the quadratic
operator.
In our case the operators $b$ and $b^*$ only satisfy the commutation
inequalities (\ref{eq:bcomk=k'}).

We can now apply Bogolubov's method as formulated in the theorem above
to the quadratic Hamiltonian $H_Q$.
\begin{lm}[Lower bound on the quadratic Hamiltonian]\hfill\label{lm:lbqh}\\
The quadratic Hamiltonian satisfies the lower bound
\begin{equation}\label{eq:HQlower}
H_Q\geq -\gamep^{-1/4}I_0\nu^{5/4}\ell^{-3/4}
-\mfr{1}{2}\left(\nu^+-\nu^-\right)^2\hw_{00,00}
-\const t^{-6}\nu\ell^{-1},
\end{equation}
where $I_0$ was defined in (\ref{eq:I0def}).
Likewise,
\begin{equation}\label{eq:Foldylower}
H_{\rm Foldy}\geq -\gamep^{-1/4}I_0\nu^{5/4}\ell^{-3/4}
-\const t^{-6}\nu\ell^{-1}.
\end{equation}
Moreover,
\begin{equation}\label{eq:Foldy+kin}
H_{\rm Foldy}\geq \sum_{i=1}^N\frac{{\gamep}}{4}\cK_{i}
-\const\nu^{5/4}\ell^{-3/4}-\const t^{-6}\nu\ell^{-1}.
\end{equation}
\end{lm}
\begin{proof}
We shall use (\ref{eq:hQ}). Note first that
$$
\sum_{p,q\ne0}\hw_{pq,00}(a^*_{p+}\an_{q+}+a^*_{p-}\an_{q-})
\leq 4\pi\ell^{-3}R^2 (\nu-\hn)\leq 4\pi\ell^{-1} \nu,
$$
by (\ref{eq:p00p'}) and the fact that $R\leq \ell$.

By (\ref{eq:hQdef}) and Theorem~\ref{thm:bogolubov}
with $\cB_\pm=\nu^\pm$,
$\kappa=(\nu^+-\nu^-)\widehat\upchi(k)\ell^{-3}$,
and
$$
\cA=\left(\frac{\hat{V}_{r,R}(k)}{2(2\pi)^3}\right)^{-1}
(2\pi)^{-3}\frac{\ell^3{\gamep}}{4}
    \frac{|k|^4}{|k|^2+(\ell t^6)^{-2}},
$$
we have
\begin{eqnarray*}
  h_Q(k)&\geq& \frac{\hat{V}_{r,R}(k)}{2(2\pi)^3}\left(-(\cA+\nu)+\sqrt{(\cA+\nu)^2-\nu^2})
    -(\nu^+-\nu^-)^2\left|\widehat\upchi(k)\right|^2\ell^{-6}\right)\\
  &=&-\mfr{1}{2}(2\pi)^{-3}\left(g(k)+\widetilde{f}(k)
    -\left(\left(g(k)+\widetilde{f}(k)\right)^2-g(k)^2\right)^{1/2}\right)\\&&
  -\frac{\hat{V}_{r,R}(k)}{2(2\pi)^3\ell^6}(\nu^+-\nu^-)^2
  \left|\widehat\upchi(k)\right|^2,
\end{eqnarray*}
where
$$
g(k)=\nu\hat{V}_{r,R}(k),\quad\widetilde{f}(k)=\frac{\ell^3{\gamep}}{2}
    \frac{|k|^4}{|k|^2+(\ell t^6)^{-2}}.
$$
Since
$$
g(k)=4\pi\nu\left((k^2+R^2)^{-1}-(k^2+r^2)^{-1}\right)
\leq 4\pi\nu|k|^{-2}
$$
and
the expression
$g+\widetilde{f}-((g+\widetilde{f})^2-g^2)^{1/2}$
is increasing in $g$ for fixed $\widetilde{f}$ we have
\begin{eqnarray*}
  h_Q(k)&\geq&-\mfr{1}{2}(2\pi)^{-3}\left(4\pi\nu|k|^{-2}+\widetilde{f}(k)
    -\left(\left(4\pi\nu|k|^{-2}+\widetilde{f}(k)\right)^2
      -(4\pi\nu)^2|k|^{-4}\right)^{1/2}\right)\\&&
  -\frac{\hat{V}_{r,R}(k)}{2(2\pi)^3\ell^6}(\nu^+-\nu^-)^2
  \left|\widehat\upchi(k)\right|^2.
\end{eqnarray*}
If we now replace $k$ by $\nu^{1/4}\ell^{-3/4}\gamep^{-1/4}k$, in the $k$-integration,
and
observe that
\begin{eqnarray*}
\lefteqn{\int\frac{\hat{V}_{r,R}(k)}{2(2\pi)^3\ell^6}
\left|\widehat{\upchi}_\ell(k)\right|^2\,dk}&&\\
&=&\mfr{1}{2}\ell^{-6}
\iint\upchi_\ell(x)V_{r,R}(x-y)\upchi_\ell(y)\,dx\,dy
=\mfr{1}{2}\hw_{00,00},
\end{eqnarray*}
we arrive at
\begin{eqnarray*}
H_Q&\geq& -\gamep^{-1/4}I\nu^{5/4}\ell^{-3/4}
-\mfr{1}{2}\left(\nu^+-\nu^-\right)^2\hw_{00,00}
-4\pi \nu\ell^{-1}
\end{eqnarray*}
where
\begin{equation}\label{eq:Itdef}
  I=\mfr{1}{2}(2\pi)^{-3}\int_{\R^3}
  4\pi|k|^{-2}+f(k)-\left((4\pi|k|^{-2}+f(k))^2-(4\pi)^2|k|^{-4}\right)^{1/2}\,dk
\end{equation}
with
$$
f(k)=\mfr{1}{2}\frac{|k|^4}{|k|^2+\gamep^{1/2}(\nu\ell)^{-1/2}t^{-12}}.
$$

We now study the $I$-integral in more details. Since the integrand is
monotone decreasing in $f$ we get an upper bound to $I$ if we replace
$f$ by a lower bound. Let
$a=\mfr{1}{2}\gamep^{1/2}(\nu\ell)^{-1/2}t^{-12}$. If $|k|\leq
(8a)^{1/2}$ we shall simply replace $f$ by $0$. If $|k|>(8a)^{1/2}$ we
replace $f$ by the lower bound
$$
 \mfr{1}{2}|k|^2-a.
$$
Note that
\begin{eqnarray*}
  \lefteqn{\left((4\pi|k|^{-2}+\mfr{1}{2}|k|^2-a)^2
      -(4\pi)^2|k|^{-4}\right)^{1/2}=
    \left(\mfr{1}{4}|k|^4+4\pi+a^2-(|k|^2+8\pi|k|^{-2})a\right)^{1/2}}&&\\
  &\geq&\left(\left(\mfr{1}{4}|k|^4+4\pi\right)
    \left(1-4|k|^{-2}a\right)+a^2\right)^{1/2}
  \\&\geq&\left(\mfr{1}{4}|k|^4+4\pi\right)^{1/2}\left(1-2|k|^{-2}a+\frac{a^2}{2(|k|^{4}/4+4\pi)}
    -\frac{1}{8}\left(4|k|^{-2}a-\frac{a^2}{(|k|^{4}/4+4\pi)}\right)^2\right).
\end{eqnarray*}
If $|k|>(8a)^{1/2}$ then $4|k|^{-2}a\geq{a^2}(|k|^{4}/4+4\pi)^{-1}$
and hence in this case
\begin{eqnarray*}
  \lefteqn{\left((4\pi|k|^{-2}+\mfr{1}{2}|k|^2-a)^2
      -(4\pi)^2|k|^{-4}\right)^{1/2}}&&\\
  &\geq&\left(\mfr{1}{4}|k|^4+4\pi\right)^{1/2}-2|k|^{-2}a\left(\mfr{1}{4}|k|^4+4\pi\right)^{1/2}
  +\frac{a^2}{2(|k|^{4}/4+4\pi)^{1/2}}-2|k|^{-4}a^2\left(\mfr{1}{4}|k|^4+4\pi\right)^{1/2}\\
  &\geq&  \left(\mfr{1}{4}|k|^4+4\pi\right)^{1/2} -a-\const a|k|^{-4}-\const |k|^{-6}a^2
  \geq \left(\mfr{1}{4}|k|^4+4\pi\right)^{1/2} -a-\const a|k|^{-4}.
\end{eqnarray*}
Thus
\begin{eqnarray*}
  I&\leq&\mfr{1}{2}(2\pi)^{-3}\Bigl(\int\limits_{|k|<(8a)^{1/2}}
  4\pi|k|^{-2}\, dk
  +\int\limits_{|k|>(8a)^{1/2}}4\pi|k|^{-2}+\mfr{1}{2}|k|^2
  -\left(\mfr{1}{4}|k|^4+4\pi\right)^{1/2}
  +\const|k|^{-4}a\, dk\Bigr)\\
  &\leq&\mfr{1}{2}(2\pi)^{-3}\int_{\R^3}4\pi|k|^{-2}+\mfr{1}{2}|k|^2
  -\left(\mfr{1}{4}|k|^4+4\pi\right)^{1/2}\, dk+ \const a^{1/2}
  \\&=&I_0+\const(\nu\ell)^{-1/4}t^{-6}
  \leq I_0+\const\nu^{-5/4}\ell^{3/4}
  t^{-6}\nu\ell^{-1}.
\end{eqnarray*}
This gives the stated  lower bound on $H_Q$. The proof of (\ref{eq:Foldylower})
is similar except that one should simply use
Theorem~\ref{thm:bogolubov} with $\kappa=0$. The proof of
(\ref{eq:Foldy+kin}) is the same as the proof of (\ref{eq:Foldylower})
except that we, when proving the lower bound on $H_{\rm Foldy}$,
replace $\cK$  by $\cK/2$ and simply
keep the other half $\cK/2$ in the lower bound. Note that this will of
course change the constants in the lower bound.
\end{proof}

\section{A-priori bounds on the kinetic energy, non-neutrality, and
  excitations}
\label{sec:simple}
In this section we shall estimate the kinetic energy
$\langle\sum_{i=1}^N\cK_{i}\rangle$, the non-neutrality
$\langle(\hn^+-\hn^-)^2\rangle$, and the excitations
$\langle\nu-\hn\rangle$ in a state for which the energy expectation
$\langle\wH\rangle$ is low. (Recall that $\wH$ was defined in (\ref{eq:Hz}) and that we are
omitting the subscript $z$.)
More precisely, we shall introduce cutoffs
\begin{equation}\label{eq:finalcutoffs}
  R=\omega(t)^{-1}\ell\quad\hbox{and}\quad r=\ell^{3/2}(\nu \ell)^{-1/2}.
\end{equation}
and consider a state such that
$\langle\wH^{\varepsilon/4}_{r,R}\rangle\leq0$.
For technical reasons this assumption will be more appropriate later
(see Lemma~\ref{lm:n+localization}).
The Lemmas
\ref{lm:r'R'},\ref{lm:0000},\ref{lm:p0q0},\ref{lm:pqmunu},\ref{lm:p000}
(Eq. (\ref{eq:p000-1})), and
\ref{lm:pqm0}
together with (\ref{eq:Foldy+kin}) in Lemma~\ref{lm:lbqh}
control all terms in the Hamiltonian $\wH^{\varepsilon/4}_{r,R}$ from
(\ref{eq:HrR}).
If we combine all these estimates and assume that we have further
cutoffs $0<r\leq r'\leq R'\leq R$
we obtain the following bound
\begin{eqnarray*}
  \wH^{\varepsilon/4}_{r,R}&\geq&\sum_{i=1}^N
  -\frac{{\gamep}}{4}\varepsilon\Delta_{i,\rm Neu}+
  \sum_{i=1}^N\frac{{\gamep}}{4}\cK_{i}
  -\const\nu^{5/4}\ell^{-3/4}-\const t^{-6}\nu\ell^{-1}
  \\&&
  -\mfr{1}{2}\nu R'^{-1}
  -\const\nu^+\nu^-(\varepsilon^{-3/2}r'^{1/2}+\ell^{-3}r'^2)\\&&
  +\mfr{1}{2}\hwp_{00,00}\left[(\hn^+-\hn^-)^2-\hn^+-\hn^-\right]-4\pi
  R'^2\ell^{-3}|\hn^+-\hn^-|(\nu-\hn)\\&&
  +\mfr{1}{2}\sum_\alpha
    \Bigl(\sum_{p\mu\ne0}J'^\alpha_{p\mu}(a^*_{p+}\an_{\mu+}-a^*_{p-}\an_{\mu-})\Bigr)^2
    -\mfr{1}{2}r'^{-1}(\nu-\hn)-2\varepsilon'\hwp_{00,00}\left((\hn^+-\hn^-)^2+1\right)
  \\&&
  -\varepsilon'^{-1}{8\pi} \ell^{-3}R'^2(\hn+1)(\nu-\hn)
    -2\varepsilon'\sum_\alpha
    \Bigl(\sum_{p\mu\ne0}J'^\alpha_{p\mu}(a^*_{p+}\an_{\mu+}-a^*_{p-}\an_{\mu-})\Bigr)^2.
\end{eqnarray*}
The prime on $\hwp_{00,00}$ and $J'$
  refers to the fact that these quantities are calculated with the $r'$ and $R'$ cutoffs.
If we choose $\varepsilon'=1/8$ and use $\hwp_{00,00}\leq4\pi
R'^2\ell^{-3}$
we arrive at
\begin{eqnarray}
  \wH^{\varepsilon/4}_{r,R} &\geq&
  \sum_{i=1}^N-\frac{{\gamep}}{4}\varepsilon\Delta_{i,\rm Neu}
  +\sum_{i=1}^N\frac{{\gamep}}{4}\cK_{i}
  -\const\nu^{5/4}\ell^{-3/4}-\const t^{-6}\nu\ell^{-1}
  \nonumber\\&&
  -\mfr{1}{2}\nu R'^{-1}
  -\const\nu^2(\varepsilon^{-3/2}r'^{1/2}+\ell^{-3}r'^2)
    -\mfr{1}{2} r'^{-1}(\nu-\hn)
  -\const\ell^{-3}R'^2(\hn+1)(\nu-\hn+1) \nonumber\\&&
    +\mfr{1}{4}\hwp_{00,00}(\hn^+-\hn^-)^2.
    \label{eq:l1}
\end{eqnarray}

It turns out that the quantity $\nu \ell$ is important and
that the main contribution to our energy estimate will come from
boxes where $\nu \ell$ is large (but not too large).
We shall need the control on the kinetic energy, non-neutrality, and
excitations only for boxes where $\nu\ell\geq\varepsilon\omega(t)^2$.

\begin{lm}[Bounds on non-neutrality and
  excitations]\label{lm:neutralcondensation}\hfill\\
  Let $R$ and $r$ be as in (\ref{eq:finalcutoffs}).
  There is a constant $\const_1>0$ such that if
  $\const_1
  \varepsilon\omega(t)^2\leq\nu \ell$ and $\const_1N\ell^3\leq
  \varepsilon^3$
  then for any state such that
  $\langle \wH^{\varepsilon/4}_{r,R}\rangle\leq0$ we have
   \begin{equation}\label{eq:Kbound}
     \left\langle\sum_{i=1}^N\cK_{i}\right\rangle
     \leq\const\varepsilon^{-1/2}t^{-2}\nu^{5/4}\ell^{-3/4} (\nu \ell)^{1/4}
   \end{equation}
   \begin{equation}\label{eq:nu-hn}
    \langle
    \nu -\hn \rangle\leq\const \varepsilon^{-3/2}t^{-2}(\nu \ell)^{3/2}
  \end{equation}
  and
  \begin{equation}\label{eq:nu+-nu-}
    \langle (\hn^+ -\hn^- )^2\rangle\leq
    \const\varepsilon^{-3/2}t^{-2}\nu (\nu \ell)^{3/2}.
  \end{equation}
\end{lm}
\begin{proof}
  We introduce the cutoffs
  $$
  R'=a\varepsilon^{1/2}(\nu\ell)^{-1/2}\ell\quad\hbox{and}\quad
  r'=\varepsilon^{-1/4}\ell^{3/2}(\nu\ell)^{-1/4},
  $$
  where $0<a$ shall be specified below.
  We observe first that $r\leq r'\leq R'/2\leq R/2$, which will, in particular, imply that
  $\hwp_{00,00}\geq\const^{-1} R'^2\ell^{-3}$.
  That $r\leq r'$ follows from the assumption that
  $\nu\ell\geq\const_1\varepsilon\omega(t)^2\geq\const_1\varepsilon$ if we choose $\const_1\geq1$.
  That  $R'\leq R$ follows from the assumption
  $\const_1 \varepsilon\omega(t)^2\leq\nu\ell$ if we choose
  $\const_1\geq a^2$.
  Finally, $r'\leq R'/2$ follows from the assumption that
  $\const_1N\ell^3\leq\varepsilon^3$ if $\const_1$ is chosen
  large enough depending on $a$. In fact, we get that
  $R'/r'=a\varepsilon^{3/4} (\nu\ell)^{-1/4}\ell^{-1/2}\geq
  a\varepsilon^{3/4} N^{-1/4}\ell^{-3/4}$ since $\nu\leq N$.

  We see that (\ref{eq:gap}) implies
  $$
  \sum_{i=1}^N-\frac{{\gamep}}{4}\varepsilon\Delta_{i,\rm Neu} \geq
  \mfr{1}{4}{\gamep}\varepsilon\pi^2\ell^{-2}(\nu-\hn).
  $$
  {F}rom the estimate (\ref{eq:l1})
  we hence obtain
  \begin{eqnarray*}
    \wH^{\varepsilon/4}_{r,R}
    &\geq&\sum_{i=1}^N\frac{{\gamep}}{4}\cK_{i}+
    \left(\mfr{1}{4}{\gamep}\pi^2-\varepsilon^{-3/4}\ell^{1/2}(\nu\ell)^{1/4}
      -a^2\const \right)\varepsilon\ell^{-2}(\nu-\hn)
    +\mfr{1}{4}\hwp_{00,00}(\hn^+-\hn^-)^2\\&&
    -\const\nu^{5/4}\ell^{-3/4}\Bigl(1+a^{-1}\varepsilon^{-1/2}(\nu\ell)^{1/4}
    +\varepsilon^{-1/2}(\nu\ell)^{1/4}(1
    +\varepsilon^{-9/8}\ell^{3/4}(\nu\ell)^{3/8})
    +a^2\varepsilon(\nu\ell)^{-5/4}\Bigr)\\&&
    -\const t^{-6}\nu\ell^{-1}.
  \end{eqnarray*}
  We have here used that $\hn\leq\nu$ and that we may estimate
  $\nu+1\leq2\nu$ since we  assume
  that $\nu\geq1$.
  Since $\varepsilon^{-3}\ell^{2}(\nu\ell)\leq
  \varepsilon^{-3}N\ell^3\leq\const_1^{-1}$
  we may rewrite this estimate as
  \begin{eqnarray*}
    \wH^{\varepsilon/4}_{r,R}
    &\geq&\sum_{i=1}^N\frac{{\gamep}}{4}\cK_{i}+\left(\mfr{1}{4}{\gamep}\pi^2
      -\const_1^{-1/4}
      -a^2\const \right)\varepsilon\ell^{-2}(\nu-\hn)
    +\mfr{1}{4}\hwp_{00,00}(\hn^+-\hn^-)^2\\&&
    -\const\nu^{5/4}\ell^{-3/4}\Bigl(1
    +\varepsilon^{-1/2}(\nu\ell)^{1/4}(1
    +\const_1^{-3/8}+a^{-1})
    +a^2\varepsilon(\nu\ell)^{-5/4}\Bigr)
    \\&&
    -\const(t^{-14}\nu\ell^{-1}+t^{-22}\ell^{-2}).
  \end{eqnarray*}
  If we choose $a$ and $\const_1$ appropriately we see that the second
  term on the right side above is bounded below by
  $\const^{-1}\varepsilon\ell^{-2}(\nu-\hn)$. {F}rom this and the
  assumption
  $C_1\varepsilon\leq C \varepsilon t^{-8}\leq \nu\ell$ (recall that
  $\omega(t)=Ct^{-4}$) we easily
  obtain (\ref{eq:Kbound}) and (\ref{eq:nu-hn}). We obtain (\ref{eq:nu+-nu-}) if we use
  that
  $\hwp_{00,00}\geq\const^{-1}R'^2\ell^{-3}=\const^{-1}\varepsilon
  (\nu\ell)^{-1}\ell^{-1}$.
\end{proof}

\section{Localization of the number of excited particles}\label{sec:localizingexcitations}
Although the bound on the expectation value $\langle \nu -\hn \rangle$ given in
Lemma~\ref{lm:neutralcondensation}
is sufficient for our
purposes we still need to know that
$\langle (\nu -\hn)^2\rangle\sim
\langle \nu -\hn \rangle^2$.
We shall however not prove this for a general state with negative
energy. Instead we shall show that we may
change the ground state, without changing its energy expectation
significantly, in such a way that the only occurring values of $\nu -\hn $  are
bounded by the quantity on the right side of (\ref{eq:nu-hn}).

To do this we shall use the method of localizing large matrices
in Lemma~\ref{local} of Appendix~\ref{app:local}.
Let $\Psi$ be any normalized wave function with $\nu $ particles in the
box $\{z\ell\}+[-\ell/2,\ell,2]^3$.
We may write $\Psi=\sum_{m=0}^{\nu } c_m\Psi_m$, where for all
$m=0,1,\ldots,\nu $,
$\Psi_m$, is a normalized eigenfunctions of $\hn $ with eigenvalue $m$.
We may now consider the $(\nu +1)\times(\nu +1)$ Hermitean matrix ${\cal A}$
with matrix elements ${\cal A}_{mm'}=\left(\Psi_m,\wH^{\varepsilon/4}_{r,R}\Psi_{m'}\right)$.

We shall use Lemma~\ref{local} for this matrix and the vector
$\psi=(c_0,\ldots,c_{\nu })$. We shall choose $M$ in  Lemma~\ref{local}
to be given by
\begin{equation}\label{eq:Mvalue}
  M=\hbox{ Integer part of } \varepsilon^{-3/2}t^{-2}(\nu \ell)^{3/2}.
\end{equation}
With the assumption in Lemma~\ref{lm:neutralcondensation} that
$\nu\ell\geq C_1\varepsilon\omega(t)^2$
we may assume ($C_1$ so large) that $M>2$. In particular,  we may assume that
$M\geq \const^{-1} \varepsilon^{-3/2}t^{-2}(\nu \ell)^{3/2}$.
With the notation in
Lemma~\ref{local}
we have $\lambda=(\psi,{\cal A}\psi)=(\Psi,\wH^{\varepsilon/4}_{r,R}\Psi)$.
Note also that because of the structure of $\wH^{\varepsilon/4}_{r,R}$
we have, again with  the notation in
Lemma~\ref{local}, that $d_k=0$ if $k\geq3$. Since $M>2$ this means
that the second sum in (\ref{localerror}) vanishes.
We conclude from Lemma~\ref{local} that there exists a normalized
wave function $\widetilde{\Psi}$ with the property that the only
occurring
values of $\nu -\hn $ belong to an interval of length $M$ and such that
$$
\left(\Psi,\wH^{\varepsilon/4}_{r,R}\Psi\right)
\geq \left(\widetilde{\Psi},\wH^{\varepsilon/4}_{r,R}\widetilde{\Psi}\right)
-\const M^{-2}(|d_1(\Psi)|+|d_2(\Psi)|).
$$
We shall discuss $d_1=d_1(\Psi)$ and $d_2=d_2(\Psi)$ in detail below,
but first we give the result
on the localization of $\hn $ that we shall use.
\begin{lm}[Localization of $\hn $]\label{lm:n+localization}\hfill\\
  Let $R$ and $r$ be as in (\ref{eq:finalcutoffs}) and $M$ as in (\ref{eq:Mvalue}).
  Let $\const_1>0$ be the constant in
  Lemma~\ref{lm:neutralcondensation} and assume that
  $C_1\varepsilon\leq C_1
  \varepsilon\omega(t)^2\leq\nu \ell$ and $\const_1N\ell^3\leq
  \varepsilon^3$.
  Then for any normalized wave function $\Psi$ such that
  \begin{equation}\label{eq:Psiassumption}
    \left(\Psi,\wH^{\varepsilon/4}_{r,R}\Psi\right)\leq-
    M^{-2}(|d_1(\Psi)|+|d_2(\Psi)|)
  \end{equation}
  there exists a
  normalized wave function $\widetilde{\Psi}$, which
  is a linear combination of eigenfunctions of $\nu -\hn $ with eigenvalues
  less than $\const M$ only, such that
  \begin{equation}\label{eq:hn-localization}
    \left(\Psi,\wH^{\varepsilon/4}_{r,R}\Psi\right)
    \geq \left(\widetilde{\Psi},\wH^{\varepsilon/4}_{r,R}\widetilde{\Psi}\right)
    -M^{-2} (|d_1(\Psi)|+|d_2(\Psi)|).
  \end{equation}
  Here $d_1(\Psi)$ and $d_2(\Psi)$ are given as explained
  in Lemma~\ref{local}.
\end{lm}
\begin{proof}
  We  choose $\widetilde{\Psi}$ as explained above.
  Then (\ref{eq:hn-localization}) holds.
  We also know that the possible $\hn $ values of $\widetilde{\Psi}$ range
  in an interval of length $M$. We do not know however, where this
  interval is located. The assumption (\ref{eq:Psiassumption}) will
  allow us to say more about the location of the interval.

  In fact, it follows from (\ref{eq:Psiassumption}) and
  (\ref{eq:hn-localization}) that
  $\left(\widetilde{\Psi},\wH^{\varepsilon/4}_{r,R}\widetilde{\Psi}\right)\leq 0$.
  It is then a consequence of Lemma~\ref{lm:neutralcondensation} that
  $\left(\widetilde{\Psi},(\nu -\hn) \widetilde{\Psi}\right)\leq
  \const M$. This of course establishes
  that the allowed values of $\nu -\hn $ are less than
  $(\const+1) M$ (which we of course just write as $\const M$).
\end{proof}
Our final task in this section is to bound $d_1(\Psi)$ and $d_2(\Psi)$.
We have that $d_1(\Psi)=(\Psi,\wH^{\varepsilon/4}_{r,R}(1)\Psi)$,
where $\wH^{\varepsilon/4}_{r,R}(1)$ is the part of the Hamiltonian
$\wH^{\varepsilon/4}_{r,R}$ containing all the terms with the coefficients
$\hw_{pq,\mu\nu}$ for which precisely one or three indices are 0.
These are the terms bounded in Lemmas~\ref{lm:p000} and
\ref{lm:pqm0}. These lemmas are stated as one-sided bounds.
It is clear from the proof that they could have been stated as
two sided bounds. Alternatively we may observe that
$\wH^{\varepsilon/4}_{r,R}(1)$ is transformed to
$-\wH^{\varepsilon/4}_{r,R}(1)$ by the unitary transform which maps
all operators $a^*_{p\pm}$ and $\an_{p\pm}$ with $p\ne0$ to
$-a^*_{p\pm}$ and $-\an_{p\pm}$. This unitary transform leaves the
estimates in Lemmas~\ref{lm:p000} and \ref{lm:pqm0} invariant.
We therefore immediately
get the following bound on $d_1(\Psi)$.

\begin{lm}[Control of $d_1(\Psi)$]\label{lm:d1}\hfill\\
With the assumptions in Lemma~\ref{lm:n+localization} we have for all $\varepsilon'>0$
\begin{eqnarray*}
  |d_1(\Psi)|&\leq&
  \varepsilon'^{-1}{8\pi}
  \ell^{-3}R^2(\Psi,(\hn+1)(\nu-\hn)\Psi)
  +2\varepsilon'\hw_{00,00}\left(\Psi,
    \left((\hn^+-\hn^-)^2+1\right)\Psi\right)
  \\&&
    +2\varepsilon'\sum_\alpha\left(\Psi,
    \Bigl(\sum_{p\mu\ne0}J^\alpha_{p\mu}(a^*_{p+}\an_{\mu+}-a^*_{p-}\an_{\mu-})\Bigr)^2
    \Psi\right).
\end{eqnarray*}
\end{lm}
Likewise, we  have that $d_2(\Psi)=(\Psi,\wH^{\varepsilon/4}_{r,R}(2)\Psi)$,
where $\wH^{\varepsilon/4}_{r,R}(2)$ is the part of the Hamiltonian
$\wH^{\varepsilon/4}_{r,R}$ containing all the terms with precisely two $\an_{0\pm}$
or two $a^*_{0\pm}$.
i.e., these are the terms in the Foldy Hamiltonian, which do not
commute with $\hn$.
\begin{lm}[Control of $d_2$]\label{lm:d2}\hfill\\
With the assumptions in Lemma~\ref{lm:n+localization} there exists a
constant $\const>0$ such that
$$
|d_2(\Psi)|\leq \const
\varepsilon^{-1/2}t^{-2}\nu^{5/4}\ell^{-3/4}(\nu\ell)^{1/4}+
8\pi\ell^{-3}R^2\left(\Psi, (\nu-\hn)(\hn+1)\Psi\right).
$$
\end{lm}
\begin{proof} We consider the unitary transform that replaces
all the operators  $a^*_{p\pm}$ and $\an_{p\pm}$ with
$p\ne0$ by $-ia^*_{p\pm}$ and $i\an_{p\pm}$ respectively. Under this
transformation the Foldy Hamiltonian (\ref{eq:Foldy}) changes into an
operator
that differs from the Foldy
Hamiltonian only by a change of sign on the part that
we denoted $\wH^{\varepsilon/4}_{r,R}(2)$. Since both operators satisfy
the bound in (\ref{eq:Foldylower})
we conclude that
\begin{eqnarray*}
  |d_2(\Psi)|&\leq&
  \sum_{i=1}^N\frac{{\gamep}}{2}\left(\Psi,\cK_{i}\Psi\right)
  +\mfr{1}{2}\sum_{pq\ne0}\hw_{pq,00}\biggl(\Psi,
  \Bigl(2a^*_{p+}a^*_{0+}\an_{0+} \an_{q+}+2a^*_{p-}a^*_{0-}\an_{0-} \an_{q-}\\&&
  -2a^*_{p+}a^*_{0-}\an_{q-} \an_{0+}-2a^*_{0+}a^*_{q-}\an_{p+}
  \an_{0-}
  \Bigr)\Psi\biggr)\\
  &&+\gamep^{-1/4}I_0\nu^{5/4}\ell^{-3/4}
  +\const t^{-6}\nu\ell^{-1}.
\end{eqnarray*}
If we use the representation (\ref{eq:wpostyp}) we find using a
Cauchy-Schwarz inequality that
\begin{eqnarray*}
  \lefteqn{-\sum_{pq\ne0}\hw_{pq,00}(a^*_{p+}a^*_{0-}\an_{q-} \an_{0+}+a^*_{0+}a^*_{q-}\an_{p+}
    \an_{0-})=}&&\\
  &&-\sum_{\alpha} \left(\sum_{p\ne0}J^\alpha_{p0}a^*_{p+}\an_{0+}
    \Bigl(\sum_{q\ne0}J^\alpha_{q0}a^*_{q-}\an_{0-}\Bigr)^*+
    \sum_{q\ne0}J^\alpha_{q0}a^*_{q-}\an_{0-}
    \Bigl(\sum_{p\ne0}J^\alpha_{p0}a^*_{p+}\an_{0+}\Bigr)^*\right)\\
  &\leq&\sum_{\alpha}\left( \sum_{p\ne0}J^\alpha_{p0}a^*_{p+}\an_{0+}
    \Bigl(\sum_{q\ne0}J^\alpha_{q0}a^*_{q+}\an_{0+}\Bigr)^*+
    \sum_{q\ne0}J^\alpha_{q0}a^*_{q-}\an_{0-}
    \Bigl(\sum_{p\ne0}J^\alpha_{p0}a^*_{p-}\an_{0-}\Bigr)^*\right)\\
  &=&\sum_{pq\ne0}\hw_{pq,00}
  \Bigl(a^*_{p+}(a^*_{0+}\an_{0+}+1) \an_{q+}+a^*_{p-}(a^*_{0-}\an_{0-}+1) \an_{q-}\Bigr).
\end{eqnarray*}
The lemma now follows
from  (\ref{eq:p00p'}) and from Lemma~\ref{lm:neutralcondensation}.
\end{proof}

\section{Lower bound on the energy in a small cube}\label{sec:energybound}

Our goal in this section is to prove a lower bound on the expectation
$\langle\wH\rangle=(\Psi,\wH\Psi)$ of the operator $\wH$ from (\ref{eq:Hz}). We are aiming at the following result.
\begin{thm}[Lower bound on $\wH$]\label{thm:Hzbound} There exists a
  constant $C_0>0$ such that for all $0<\varepsilon<1$ and $0<t<1/2$
  with $C_0N\ell^3\leq\varepsilon^3$, and $\varepsilon^{-1}
  t^{-4}\ell<C_0^{-1}$ we have the estimate
  \begin{eqnarray}\label{eq:Hzbound}
    \left\langle\wH\right\rangle&\geq&-\gamep^{-1/4}I_0\langle\hn\rangle^{5/4}\ell^{-3/4}
    -\const(t^{-6}\nu\ell^{-1}+\varepsilon\ell^{-2}+\varepsilon^{-1}t^{-8}\nu)\nonumber\\&&
    -K_1(\varepsilon,t,N,\ell)\langle\hn\rangle^{5/4}\ell^{-3/4}
    -\langle\hn\rangle^{5/3}\ell^{-1/3}(N^{2/5}\ell)^{-5/3}\left(\const\varepsilon^{1/6}
      +K_2(\varepsilon,t,N,\ell)\right),
  \end{eqnarray}
  where $K_1$ and $K_2$ are sums of a finite number of terms of the
  form $\const\varepsilon^{-a}t^{-b}(N^{2/5}\ell)^cN^{-d}$ with
  $d>0$.
\end{thm}
Note that the estimate is not formulated as an operator inequality.
The lower bound is not a constant nor an expectation value, but a
non-linear expression in $\langle\hn\rangle$. The
reason that we give the bound in terms of an expression in
$\langle\hn\rangle$ rather than in terms of $\nu$ is that we shall later take into
account the term $T(S_0)$, i.e., the kinetic energy between
boxes, which depends on $\langle\hn\rangle$ (see (\ref{eq:sz})).

In order to arrive at an estimate expressed in $\langle\hn\rangle$ we
shall make repeated use of the following result.

\begin{lm}\label{lm:nuhncomp}
If $\langle\wH\rangle\leq0$ we have the estimate
$$
\langle n\rangle\geq \nu(1-\const\varepsilon^{-1}(N^{2/5}\ell)N^{-1/15}).
$$
In particular, we may assume that the constant $C_1$ from
Lemma~\ref{lm:neutralcondensation}
is so large that if $C_1N\ell^3\leq\varepsilon^3$ (i.e.,
$(N^{2/5}\ell)\leq C_1^{-1}\varepsilon N^{1/15}$) then
$$
\nu\leq\const\langle\hn\rangle\quad\hbox{and}\quad
\nu^{5/4}\leq\langle\hn\rangle^{5/4}(1+\const \varepsilon^{-1}(N^{2/5}\ell)N^{-1/15}).
$$
\end{lm}
\begin{proof}
This follows from
Corollary~\ref{cl:sn+bound}
since
$\varepsilon^{-1}\ell\nu^{1/3}
\leq \varepsilon^{-1} (N^{2/5}\ell)N^{-1/15}$, where we have used that
$\nu\leq N$.
\end{proof}

We shall prove Theorem~\ref{thm:Hzbound} by dividing into several cases.
We first prove an a-priori bound on $\wH$.

\begin{lm}[A-prori bound on $\wH $]\label{lm:Hzbound1}
We may assume that the constant $C_1$ from
Lemma~\ref{lm:neutralcondensation} is large enough so that if
$C_1 N\ell^3\leq\varepsilon^3$ and $\varepsilon^{-1}t^{-4}\ell<C_1^{-1}$  there exists a constant
$\const>0$ such that
\begin{equation}\label{eq:Hzbound1}
    \wH
    \geq-\gamep^{-1/4} I_0\nu^{5/4}\ell^{-3/4}
    -\const\nu^{5/4}\ell^{-3/4}\Bigl(\varepsilon^{-1/2}(\nu \ell)^{1/4}
    +\varepsilon^{-2}(\nu \ell)^{3/4}\ell\Bigr)
    -\const(t^{-6}\nu\ell^{-1}+\varepsilon\ell^{-2}).
  \end{equation}
\end{lm}
\begin{proof}
  If $\nu=0$ there is
  nothing to prove.
  We may therefore
  assume that $\nu\geq1$.
  We now make the following choices for the cutoffs
  $$
  r=b\varepsilon^{-1}\ell^2\quad\hbox{and}\quad
  R=\min\{a\varepsilon^{1/2}(\nu\ell)^{-1/2},\omega(t)^{-1}\}\ell,
  $$
  where $0<a,b$ shall be specified below. We first note that $r\leq
  R$ if $C_1$ is sufficiently large (depending on $a$ and $b$). Indeed,
  either
  $R/r=(a/b)(\varepsilon^3\nu^{-1}\ell^{-3})^{1/2}\geq(a/b)C_1^{1/2}$
  (since $\nu\leq N$)
  or $R/r=b^{-1}\omega(t)^{-1}\varepsilon\ell^{-1}\geq\const^{-1}
  b^{-1}t^4\varepsilon\ell^{-1}\geq b^{-1}\const^{-1}C_1$.

  We proceed as in the beginning of Sect.~\ref{sec:simple}, but we now
  use Lemma~\ref{lm:cutoffs} instead of Lemma~\ref{lm:r'R'} and
  (\ref{eq:Foldylower}) instead of (\ref{eq:Foldy+kin}). We then get
  that
  \begin{eqnarray*}
    \wH &\geq&
    \sum_{i=1}^N-\frac{\gamep}{2}(\varepsilon-\delta)\Delta_{i,\rm Neu}
    -\gamep^{-1/4} I_0\nu^{5/4}\ell^{-3/4}
    -\const t^{-6}\nu\ell^{-1} -\const\nu R^{-1}\\&&
    -\const\nu^2(\delta^{-3/2}r^{1/2}+\ell^{-3}r^2)
    +\mfr{1}{2}\hw_{00,00}\left[(\hn^+-\hn^-)^2-\hn^+-\hn^-\right]-4\pi
    R^2\ell^{-3}|\hn^+-\hn^-|(\nu-\hn)\\&&
    +\mfr{1}{2}\sum_\alpha
    \Bigl(\sum_{p\mu\ne0}J^\alpha_{p\mu}(a^*_{p+}\an_{\mu+}-a^*_{p-}\an_{\mu-})\Bigr)^2
    -\mfr{1}{2}r^{-1}(\nu-\hn)-2\varepsilon'\hw_{00,00}\left((\hn^+-\hn^-)^2+1\right)
    \\&&
    -\varepsilon'^{-1}{8\pi} \ell^{-3}R^2(\hn+1)(\nu-\hn)
    -2\varepsilon'\sum_\alpha
    \Bigl(\sum_{p\mu\ne0}J^\alpha_{p\mu}(a^*_{p+}\an_{\mu+}-a^*_{p-}\an_{\mu-})\Bigr)^2.
  \end{eqnarray*}
  If we Choose $\varepsilon'=1/4$, use that $\hw_{00,00}\leq 4\pi
  R^2\ell^{-3}$, and $\hn\leq\nu$ we obtain after inserting the
  choices of $r$ and $R$ that
  \begin{eqnarray*}
    \wH
    &\geq&
    \sum_{i=1}^N-\frac{{\gamep}}{2}(\varepsilon-\delta)\Delta_{i,\rm Neu}
    -\gamep^{1/4}I_0\nu^{5/4}\ell^{-3/4}
    -\const t^{-6}\nu\ell^{-1}
    -\mfr{1}{2}\varepsilon^{-1}a^{-1}\nu \ell^{-1}(\nu\ell)^{1/2}\nonumber\\&&
    -\const\nu^2\ell(b^{1/2}\varepsilon^{-1/2}\delta^{-3/2}+\varepsilon^{-2}b^2)
    -\mfr{1}{2}\varepsilon b^{-1}(\nu-\hn)\ell^{-2}
    -a^2\varepsilon\const\ell^{-1}(\nu\ell)^{-1}\nu(\nu-\hn+1).
  \end{eqnarray*}
  We have here  inserted the upper bound  $a\varepsilon(\nu\ell)^{-1/2}\ell$
  for $R$. This is allowed, since the only term which is not monotone
  increasing in $R$, i.e., $\mfr{1}{2}\nu R^{-1}$, can
  in fact, according to (\ref{eq:HzlowerRmax}), be ignored when $R=\omega(t)^{-1}\ell$.
  We have again replaced $\nu+1$ by $2\nu$.

  If we now choose $\delta=\varepsilon/2$,
  and again use that
  $$
  \sum_{i=1}^N-\frac{{\gamep}}{2}(\varepsilon-\delta)\Delta_{i,\rm Neu} \geq
  \mfr{1}{4}{\gamep}\varepsilon\pi^2\ell^{-2}(\nu-\hn),
  $$
  we arrive at
  \begin{eqnarray*}
    \wH
    &\geq&\left(\mfr{1}{4}{\gamep}\pi^2-b^{-1}/2
      -a^2\const \right)\varepsilon\ell^{-2}(\nu-\hn)
    -\gamep^{1/4}I_0\nu^{5/4}\ell^{-3/4}
    -\const t^{-6}\nu\ell^{-1}
    \\&&
    -\const\nu^{5/4}\ell^{-3/4}\Bigl(a^{-1}\varepsilon^{-1/2}(\nu\ell)^{1/4}+
    \varepsilon^{-2}(b^{1/2}+b^2)(\nu\ell)^{3/4}\ell
    +a^2\varepsilon(\nu\ell)^{-5/4}\Bigr).
  \end{eqnarray*}
  We obtain the results of the lemma if we make appropriate
  choices of $a$ and $b$.
\end{proof}

We now use this estimate to control boxes with few or with many particles.
We shall show that these boxes do not contribute to the leading order
estimate of the energy. Thus the specific form of the first term in
(\ref{eq:Hzbound1}) is not important. We may  simply estimate it by
$\const\nu^{5/4}\ell^{-3/4}$.
\begin{lm}[Boxes with few particles]\label{lm:nuellsmall}
  \hfill\\
  Let $C_1$ be the constant in Lemma~\ref{lm:neutralcondensation} and
  assume that $\nu\ell\leq C_1\varepsilon\omega(t)^2$, $C_1
  N\ell^3\leq\varepsilon^3$, and $\varepsilon^{-1}t^{-4}\ell<C_1^{-1}$. Then there is a constant $\const>0$ such
  that
  $$
  \wH\geq
  -\const t^{-14}\nu\ell^{-1}
  -\const \varepsilon\ell^{-2}
  -\const\varepsilon^{-1}t^{-8}\nu.
  $$
\end{lm}
\begin{proof}
  The result follows immediately from Lemma~\ref{lm:Hzbound1} if we
  simply insert the assumed bound on $\nu\ell$ and use that $\omega(t)=Ct^{-4}$.
\end{proof}

\
\begin{lm}[Boxes with many particles]\label{lm:nuelllarge}
Let $C_1$ be the constant in Lemma~\ref{lm:neutralcondensation}.
Assume that
$\nu\ell\geq\varepsilon^{-4}(N^{2/5}\ell)^{10}$,
$C_1 N\ell^3\leq\varepsilon^3$, and $\varepsilon^{-1}t^{-4}\ell<C_1^{-1}$.
Then
\begin{eqnarray*}
  \langle
  \wH\rangle&\geq&-\gamep^{-1/4}I_0\langle\hn\rangle^{5/4}\ell^{-3/4}
  -\const\langle\hn\rangle^{5/4}\ell^{-3/4}\varepsilon^{-1}(N^{2/5}\ell )N^{-1/15}\\&&
  -\const\langle\hn\rangle^{5/3}\ell^{-1/3}\left(\varepsilon^{1/6}(N^{2/5}\ell)^{-5/3}
  +\varepsilon^{-2}(N^{2/5}\ell)^{4/3}N^{-3/15}\right)\\&&
  -\const(t^{-14}\nu\ell^{-1}+t^{-22}\ell^{-2}).
\end{eqnarray*}
\end{lm}
\begin{proof}
  We may of course assume that $\langle\wH\rangle\leq0$  and hence use
  the estimates in
  Lemma~\ref{lm:nuhncomp}.
  We now use Lemma~\ref{lm:Hzbound1} and insert the estimates
  \begin{eqnarray*}
    \nu^{5/4}\ell^{-3/4}(\nu\ell)^{1/4}&=&\nu^{5/3}\ell^{-1/3}(\nu\ell)^{-1/6}
    \leq\nu^{5/3}\ell^{-1/3}\varepsilon^{2/3}(N^{2/5}\ell)^{-5/3}\\
    \nu^{5/4}\ell^{-3/4}(\nu\ell)^{3/4}\ell&=&
    \nu^{5/3}\ell^{-1/3}\nu^{1/3}\ell^{4/3}\leq\nu^{5/3}\ell^{-1/3}(N^{2/5}\ell)^{4/3}N^{-3/15}.
  \end{eqnarray*}
 In the first  inequality we used the assumption on $\nu\ell$. In
  the last inequality we simply used that $\nu\leq N$.
\end{proof}

We now restrict attention to boxes with
$\varepsilon\omega(t)^2\leq\nu\ell\leq\varepsilon^{-4}(N^{2/5}\ell)^{10}$.
In the next lemma we shall prove the lower bound on  $\langle\wH\rangle$
under the restrictive assumption given in (\ref{eq:case1}) below.
Finally, Theorem~\ref{thm:Hzbound}
is proved by considering the alternative case that (\ref{eq:case1})
fails.  Let us note that, logically speaking, this could have been
done in the reverse order. I.e., we could, instead, have begun with
the case that (\ref{eq:case1}) fails.

\begin{lm}[Lower bound on $\wH$: restricted version]\label{lm:case1}
Let $R$ and $r$ be given by (\ref{eq:finalcutoffs}).
Let $C_1$ be the constant in Lemma~\ref{lm:neutralcondensation}.
Assume that $C_1\varepsilon\omega(t)^2\leq\nu\ell\leq
\varepsilon^{-4}(N^{2/5}\ell)^{10}$ and
$C_1N\ell^3\leq\varepsilon^3$.
Then there exists a constant $C_2>0$ such that if
\begin{eqnarray}
  \lefteqn{\ell^{-3}R^2\nu\langle\nu-\hn\rangle}&&\label{eq:case1}\\
  &\leq& C_2^{-1}\left\langle\hw_{00,00}(\hn^+-\hn^-)^2+\sum_\alpha
    \Bigl(\sum_{p\mu\ne0}J^\alpha_{p\mu}(a^*_{p+}\an_{\mu+}-a^*_{p-}\an_{\mu-})\Bigr)^2
    \right\rangle
      ,\nonumber
\end{eqnarray}
we have that
\begin{eqnarray*}
  \left\langle \wH\right\rangle
  &\geq& -\gamep^{-1/4}I_0\langle\hn\rangle^{5/4}\ell^{-3/4}
  -\const t^{-6}\nu\ell^{-1}\\&&
  -\const\langle\hn\rangle^{5/4}\ell^{-3/4}
  \Bigl(\varepsilon^{-1}(N^{2/5}\ell)N^{-1/15}+
  \varepsilon^{-7/2}(N^{2/5}\ell)^{23/4}N^{-3/10}\\&&
  +\varepsilon^{-9/2}t^{-2}(N^{2/5}\ell)^8N^{-1/5}
  \Bigr).
\end{eqnarray*}
\end{lm}
\begin{proof}
We may of course assume that $\langle\wH\rangle\leq0$ and hence use
Lemma~\ref{lm:nuhncomp}.
We may also assume that $\langle\wH^{\varepsilon/4}_{r,R}\rangle\leq0$
otherwise we have from (\ref{eq:HzlowerRmax}) and the choice of $r$ that
\begin{eqnarray*}
  \langle\wH\rangle&\geq&-\const\nu^2(\varepsilon^{-3/2}r^{1/2}+\ell^{-3}r^2)
  =-\const\varepsilon^{-3/2}\nu^{5/4}\ell^{-3/4}(\nu\ell)^{1/2}\ell^{3/4}
  -\const
  \nu\ell^{-1}\\&\geq&
  -\const\varepsilon^{-3/2}\langle\hn\rangle^{5/4}\ell^{-3/4}
  (\varepsilon^{-4}(N^{2/5}\ell)^{10})^{1/2}(N^{2/5}\ell)^{3/4}N^{-3/10}
  -\const \nu\ell^{-1}\\
  &=&-\const\langle\hn\rangle^{5/4}\varepsilon^{-7/2}
  \ell^{-3/4}(N^{2/5}\ell)^{23/4}N^{-3/10}
  -\const \nu\ell^{-1},
\end{eqnarray*}
which implies the stated estimate. It now follows
from Lemma~\ref{lm:neutralcondensation} that we have the estimate
(\ref{eq:nu-hn}).

We again proceed as in the beginning of Sect.~\ref{sec:simple}, but we
use (\ref{eq:Foldylower}) instead of
(\ref{eq:Foldy+kin}) and (\ref{eq:HzlowerRmax}) instead of
Lemma~\ref{lm:r'R'} or (\ref{eq:Hzlower})
(since now $R=\omega(t)^{-1}\ell$) and we choose
$\delta=\varepsilon$. We then get that
\begin{eqnarray*}
  \wH &\geq& -\gamep^{-1/4} I_0\nu^{5/4}\ell^{-3/4}
  -\const t^{-6}\nu\ell^{-1}
  -\const\nu^2(\varepsilon^{-3/2}r^{1/2}+\ell^{-3}r^2)\\&&
  +\mfr{1}{2}\hw_{00,00}\left[(\hn^+-\hn^-)^2-\hn^+-\hn^-\right]-4\pi
  R^2\ell^{-3}|\hn^+-\hn^-|(\nu-\hn)\\&&
  +\mfr{1}{2}\sum_\alpha
    \Bigl(\sum_{p\mu\ne0}J^\alpha_{p\mu}(a^*_{p+}\an_{\mu+}-a^*_{p-}\an_{\mu-})\Bigr)^2
    -\mfr{1}{2}r^{-1}(\nu-\hn)-2\varepsilon'\hw_{00,00}\left((\hn^+-\hn^-)^2+1\right)
  \\&&
  -\varepsilon'^{-1}{8\pi} \ell^{-3}R^2(\hn+1)(\nu-\hn)
    -2\varepsilon'\sum_\alpha
    \Bigl(\sum_{p\mu\ne0}J^\alpha_{p\mu}(a^*_{p+}\an_{\mu+}-a^*_{p-}\an_{\mu-})\Bigr)^2.
\end{eqnarray*}
If we  use the assumption (\ref{eq:case1}) and the facts that
$|\hn^+-\hn^-|\leq \nu$,
$\hn+1\leq2\nu$, and $\hw_{00,00}\leq4\pi R^2\ell^{-3}$
we see with appropriate choices of $\varepsilon'$ and $C_2$ that
\begin{eqnarray*}
  \langle \wH\rangle&\geq& -\gamep^{-1/4} I_0\nu^{5/4}\ell^{-3/4}
  -\const(t^{-14}\nu\ell^{-1}+t^{-22}\ell^{-2})\\&&
  -\const\varepsilon^{-3/2}\nu^{7/4}\ell^{1/2} -\const
  \nu\ell^{-1}-\const \omega(t)^{-2}\nu\ell^{-1}-\const\langle\nu-\hn\rangle
  \nu^{1/2}\ell^{-1},
\end{eqnarray*}
where we have inserted the choices of $r$ and $R$.
If we use (\ref{eq:nu-hn})
we see that
$$
\langle\nu-\hn\rangle \nu^{1/2}\ell^{-1}\leq \const
\varepsilon^{-3/2}t^{-2}\nu^{5/4}\ell^{-3/4}
(\nu\ell)^{3/4}\ell^{1/2}\leq\const
\varepsilon^{-9/2}t^{-2}\nu^{5/4}\ell^{-3/4}(N^{2/5}\ell)^{8}N^{-1/5}.
$$
Hence we arrive at the bound in the lemma if we
note, as above, that
$$\varepsilon^{-3/2}\nu^{7/4}\ell^{1/2}\leq \const
\nu^{5/4}\ell^{-3/4}\varepsilon^{-7/2}(N^{2/5}\ell)^{23/4}N^{-3/10}$$
and use the result of Lemma~\ref{lm:nuhncomp}.
\end{proof}

\begin{proof}[Proof of Theorem~\ref{thm:Hzbound}]
We again let  $r$ and $R$ be given by (\ref{eq:finalcutoffs}).
According to Lemmas~\ref{lm:nuellsmall} and \ref{lm:nuelllarge} we may assume that
$C_1\varepsilon\omega(t)^2\leq\nu\ell\leq
\varepsilon^{-4}(N^{2/5}\ell)^{10}$.
Then from Lemma~\ref{lm:case1} we may assume that
\begin{eqnarray}
  \lefteqn{\ell^{-3}R^2\nu\langle\nu-\hn\rangle}&&\label{eq:case1op}\\
  &\geq& C_2^{-1}\left\langle\hw_{00,00}(\hn^+-\hn^-)^2+\sum_\alpha
    \Bigl(\sum_{p\mu\ne0}J^\alpha_{p\mu}(a^*_{p+}\an_{\mu+}-a^*_{p-}\an_{\mu-})\Bigr)^2
    \right\rangle
      .\nonumber
\end{eqnarray}
We may also still assume that $\langle\wH\rangle\leq0$ and
$\langle\wH^{\varepsilon/4}_{r,R}\rangle\leq0$ (as in the proof of
Lemma~\ref{lm:case1}).  Hence we can use the results of
Lemmas~\ref{lm:neutralcondensation} and \ref{lm:nuhncomp}.  It is
enough to prove an estimate of the type (\ref{eq:Hzbound}) for
$\langle\wH^{\varepsilon/4}_{r,R}\rangle$ (the extra error terms
were estimated in the first paragraph of the proof of
Lemma~\ref{lm:case1}).

We begin by bounding $d_1$ and $d_2$ using
Lemmas~\ref{lm:d1} and \ref{lm:d2}. We have from
Lemmas~\ref{lm:neutralcondensation} and \ref{lm:d2} that
\begin{eqnarray*}
  |d_2|&\leq& \const
  \varepsilon^{-1/2}t^{-2}\nu^{5/4}\ell^{-3/4}(\nu\ell)^{1/4}+
  \const\ell^{-3}R^2\nu\langle\nu-\hn\rangle\\
  &\leq&\const\nu^{5/4}\ell^{-3/4}\left(\varepsilon^{-1/2}t^{-2}
    (\nu\ell)^{1/4}+t^{6}\varepsilon^{-3/2}(\nu\ell)^{5/4}\right)
  \leq \const t^{6}\varepsilon^{-3/2}\nu^{5/4}\ell^{-3/4}(\nu\ell)^{5/4},
\end{eqnarray*}
where the last estimate follows since
$\nu\ell\geq\varepsilon\omega(t)^2$.

In order to bound $d_1$ we shall use (\ref{eq:case1op}).
Together with Lemma~\ref{lm:d1} this gives (choosing $\varepsilon'=1/2$
say)
\begin{eqnarray*}
  |d_1|&\leq& \const\ell^{-3}R^2\nu
  (\langle\nu-\hn\rangle+1)\leq\const
  t^{6}\varepsilon^{-3/2}\nu^{5/4}\ell^{-3/4}(\nu\ell)^{5/4}.
\end{eqnarray*}

If the assumption (\ref{eq:Psiassumption}) fails then
$$
\langle \wH^{\varepsilon/4}_{r,R}\rangle\geq
-
\const
  t^{10}\varepsilon^{3/2}\nu^{5/4}\ell^{-3/4}(\nu\ell)^{-7/4}\geq
-\const t^{10}\varepsilon^{3/2}\nu\ell^{-1}(\varepsilon\omega(t)^2)^{-3/2}
=-\const t^{22}\nu\ell^{-1}.
$$
and we see then
that the bound (\ref{eq:Hzbound}) holds. We may therefore assume that
(\ref{eq:Psiassumption}) holds.
Thus from Lemma~\ref{lm:n+localization} it follows that we can find
a normalized wave function $\widetilde{\Psi}$,
such that
\begin{equation}\label{eq:Psitildeexpectation}
  \left(\widetilde\Psi,(\nu-\hn)\widetilde\Psi\right)\leq
  \const\varepsilon^{-3/2}t^{-2}(\nu\ell)^{3/2}
  \quad\hbox{and}
  \quad
  \left(\widetilde\Psi,(\nu-\hn)^2\widetilde\Psi\right)\leq
  \const\varepsilon^{-3}t^{-4}(\nu\ell)^{3}
\end{equation}
and such that
\begin{equation}\label{eq:PsiPsitilde}
  \left\langle\wH_{r,R}^{\varepsilon/4}\right\rangle
  \geq\left(\widetilde\Psi,\wH_{r,R}^{\varepsilon/4}\widetilde\Psi\right)
  -\const t^{22}\nu\ell^{-1}.
\end{equation}
(Here as before $\langle X\rangle$ denotes the expectation of the
operator $X$ in the state
$\Psi$.)

In order to analyze
$\left(\widetilde{\Psi},\wH_{r,R}^{\varepsilon/4}\widetilde{\Psi}\right)$
we again proceed as in the beginning of Sect.~\ref{sec:simple}, but
this time we use (\ref{eq:p000-2}) of Lemma~\ref{lm:p000} and
(\ref{eq:HQlower}) of Lemma~\ref{lm:lbqh}. We find that
\begin{eqnarray}
  \wH^{\varepsilon/4}_{r,R} &\geq& -\gamep^{-1/4} I_0\nu^{5/4}\ell^{-3/4}
  -\const t^{-6}\nu\ell^{-1}\nonumber
  \\&&
  +\mfr{1}{2}\hw_{00,00}\left[(\hn^+-\hn^-)^2-\hn^+-\hn^-\right]-4\pi
  R^2\ell^{-3}|\hn^+-\hn^-|(\nu-\hn)-\mfr{1}{2}\hw_{00,00}(\nu^+-\nu^-)^2\nonumber\\&&
  +\mfr{1}{2}\sum_\alpha
    \Bigl(\sum_{p\mu\ne0}J^\alpha_{p\mu}(a^*_{p+}\an_{\mu+}-a^*_{p-}\an_{\mu-})\Bigr)^2
    -\mfr{1}{2}r^{-1}(\nu-\hn)-2\varepsilon'\hw_{00,00}\left((\nu-\hn)^2+1\right)
  \nonumber\\&&
  -\varepsilon'^{-1}{8\pi} \ell^{-3}R^2(\hn+1)(\nu-\hn)
    -2\varepsilon'\sum_\alpha
    \Bigl(\sum_{p\mu\ne0}J^\alpha_{p\mu}(a^*_{p+}\an_{\mu+}-a^*_{p-}\an_{\mu-})\Bigr)^2.
    \label{eq:Hzfinal}
\end{eqnarray}
We shall use that for all $\varepsilon''>0$
$$
(\nu^+-\nu^-)^2\leq (1+\frac{\varepsilon''}{2})(\hn^+-\hn^-)^2+(1+\frac{1}{2\varepsilon''})(\nu-\hn)^2
$$
and
$$
|\hn^+-\hn^-|(\nu-\hn)\leq\frac{\varepsilon''}{2}(\hn^+-\hn^-)^2+\frac{1}{2\varepsilon''}(\nu-\hn)^2.
$$
According to Lemma~\ref{lm:pqmunu}
\begin{eqnarray*}
  \sum_\alpha
  \Bigl(\sum_{p\mu\ne0}J^\alpha_{p\mu}(a^*_{p+}\an_{\mu+}-a^*_{p-}\an_{\mu-})\Bigr)^2
  &\leq&\sum_{pq,\mu\nu\ne0}\hw_{pq,\mu\nu}\Bigl(
    a^*_{p+}a^*_{q+}\an_{\nu+} \an_{\mu+}
    +a^*_{p-}a^*_{q-}\an_{\nu-} \an_{\mu-}\\&&-2a^*_{p+}a^*_{q-}\an_{\nu-} \an_{\mu+}
  \Bigr)+(\nu-\hn)r^{-1}
\end{eqnarray*}
The sum on the right is simply the second quantization of the two-body operator
$\cP\otimes \cP w_{r,R} \cP\otimes \cP$, where $\cP$ again
denotes the projection onto the subspace of
$L^2\left(([-\ell/2,\ell/2]^3)\times\{1,-1\}\right)$
consisting of functions orthogonal to constants.
Since $w_{r,R}\leq r^{-1}$ this sum is bounded above by
$$
\left((\nu^+-\hn^+)^2+(\nu^--\hn^-)^2\right)r^{-1}\leq (\nu-\hn)^2r^{-1}.
$$
(This can of course also be proved by directly estimating the sum).
If we insert the above estimates into (\ref{eq:Hzfinal}) and choose
$\varepsilon'=1/\varepsilon''$
we arrive at
\begin{eqnarray*}
  \wH^{\varepsilon/4}_{r,R} &\geq& -\gamep^{-1/4} I_0\nu^{5/4}\ell^{-3/4}
  -\const t^{-6}\nu\ell^{-1}
  \\&&
  -\const
  R^2\ell^{-3}\left(\nu+\varepsilon''(\hn^+-\hn^-)^2+(1+\mfr{1}{\varepsilon''})\left((\nu-\hn)^2+1\right)
    +\varepsilon''\nu(\nu-\hn)\right)\\&&
  -\const r^{-1}(\mfr{1}{\varepsilon''}(\nu-\hn)^2+(1+\mfr{1}{\varepsilon''})(\nu-\hn)).
\end{eqnarray*}
We now use (\ref{eq:Psitildeexpectation}),
Lemma~\ref{lm:neutralcondensation} (we may of course assume that
$(\widetilde\Psi,\wH_{r,R}^{\varepsilon/4}\widetilde\Psi)\leq0$), and
the choices (\ref{eq:finalcutoffs}) of $R$ and $r$. We get
\begin{eqnarray*}
  \left(\widetilde\Psi,\wH^{\varepsilon/4}_{r,R}\widetilde\Psi\right)
  &\geq&
  -\gamep^{-1/4} I_0\nu^{5/4}\ell^{-3/4}
  -\const t^{-6}\nu\ell^{-1}
  \\&&
  -\const
  \omega(t)^{-2}\ell^{-1}\left(\nu+\varepsilon''\varepsilon^{-3/2}t^{-2}\nu(\nu\ell)^{3/2}
    +(1+\mfr{1}{\varepsilon''})
    \varepsilon^{-3}t^{-4}(\nu\ell)^{3}
    \right)\\&&
  -\const
  \ell^{-3/2}(\nu\ell)^{1/2}\left(\mfr{1}{\varepsilon''}\varepsilon^{-3}t^{-4}(\nu\ell)^{3}
  +(1+\mfr{1}{\varepsilon''})\varepsilon^{-3/2}t^{-2}(\nu\ell)^{3/2}\right).
\end{eqnarray*}
Finally we make the choice
$\varepsilon''=\varepsilon^{-3/4}t^{-5}(\nu\ell)^{1/2}\ell^{1/4}$ and arrive at
\begin{eqnarray*}
  \left(\widetilde\Psi,\wH^{\varepsilon/4}_{r,R}\widetilde\Psi\right)
  &\geq&
  -\gamep^{-1/4} I_0\nu^{5/4}\ell^{-3/4}
  -\const t^{-6}\nu\ell^{-1}
  \\&&
  -\const
  \nu^{5/4}\ell^{-3/4}\Bigl(\varepsilon^{-9/4}t^{1}(\nu\ell)^{7/4}\ell^{1/4}
    +\varepsilon^{-3}t^{4}(\nu\ell)^{7/4}\ell+\varepsilon^{9/4}t^{9}(\nu\ell)^{5/4}\ell^{3/4}\\&&
    \phantom{-\const\nu^{5/4}\ell^{-3/4}\Bigl(}+\varepsilon^{-3/2}t^{-2}(\nu\ell)^{3/4}\ell^{1/2}
    +\varepsilon^{-3/4}t^3(\nu\ell)^{1/4}\ell^{1/4}\Bigr).
\end{eqnarray*}
If we now insert the bound $\nu\ell\leq \varepsilon^{-4}(N^{2/5}\ell)^{10}$, use (\ref{eq:PsiPsitilde}) and
the result of Lemma~\ref{lm:nuhncomp}
we arrive at a bound of the form
(\ref{eq:Hzbound}) for $\langle\wH^{\varepsilon/4}_{r,R}\rangle$ .
\end{proof}

\section{The lattice approximation}\label{sec:lattice}
Given a map $S:\Z^3\to\R$ we define
a function $\phi:\R^3\to\R$ as follows. On any cube
$\{\mu\}+[-1/2,1/2]^3$ with $\mu\in\{(1/2,1/2,1/2)\}+\Z^3$
we set
\begin{equation}\label{eq:philattice}
  \phi(x)=\sum_{\tau\in\Z^3\atop|\tau|=\sqrt{3}}\lambda_\tau(x-\mu)S(\mu+\tau/2),
\end{equation}
where
\begin{equation}\label{eq:phitau}
  \lambda_\tau(x)=(1/2+\tau_1x_1)(1/2+\tau_2x_2)(1/2+\tau_3x_3).
\end{equation}
Note that the 8 points $\mu+\tau/2$ with $|\tau|=\sqrt{3}$ are the
corners of the cube $\{\mu\}+[-1/2,1/2]^3$.
The function $\phi$ is well-defined on $\R^3$ and is continuous. Moreover,
$\phi(\sigma)=S(\sigma)$ for $\sigma\in\Z^3$.
By a straightforward calculation we obtain that
\begin{equation}\label{eq:phikin}
\int\limits_{\R^3}\left(\nabla\phi\right)^2=T(S),
\end{equation}
where we have defined
$$
T(S)=\sum_{\sigma_1,\sigma_2\in\Z^3\atop
  |\sigma_1-\sigma_2|=\sqrt{2}}\frac{1}{12}({S}(\sigma_1)-{S}(\sigma_2))^2
+\sum_{\sigma_1,\sigma_2\in\Z^3\atop
  |\sigma_1-\sigma_2|=\sqrt{3}}\frac{1}{24}({S}(\sigma_1)-{S}(\sigma_2))^2.
$$
Note that there is a constant $\const>0$ such that for all maps
$S:\Z^3\to\R$ we have
$$
\const^{-1}\sum_{\sigma_1,\sigma_2\in\Z^3\atop
  |\sigma_1-\sigma_2|=1}({S}(\sigma_1)-{S}(\sigma_2))^2
\leq T(S)\leq \const\sum_{\sigma_1,\sigma_2\in\Z^3\atop
  |\sigma_1-\sigma_2|=1}({S}(\sigma_1)-{S}(\sigma_2))^2.
$$

\begin{lm}\label{lm:beta}
If $\lambda_1,\ldots,\lambda_m\geq0$ and
$\sum_{j=1}^m\lambda_j=1$ we have for all $S_1,\ldots,S_m\geq0$ and
all $\beta\geq1$ that
$$
\Bigl(\sum_{j=1}^m\lambda_jS_j\Bigr)^\beta\leq
\sum_{j=1}^m\lambda_jS_j^\beta\leq
\Bigl(\sum_{j=1}^m\lambda_jS_j\Bigr)^\beta +C_{m,\beta}
\Bigl(\sum_{1\leq i<j\leq m}(S_i-S_j)^2\Bigr)^{1/2}\sum_{j=1}^mS_j^{\beta-1}
$$
for some constant $C_{m,\beta}>0$.
\end{lm}
\begin{proof}
  The first inequality follows from an application of Jensen's
  inequality.  If we write $Y= \sum_{j=1}^m\lambda_jS_j$ then
  $\sum_{j=1}^m\lambda_jS_j^\beta-
  \Bigl(\sum_{j=1}^m\lambda_jS_j\Bigr)^\beta=\sum_{j=1}^m\lambda_j(S_j^\beta-Y^\beta)$.
  The second inequality above then follows easily from
  $|a^\beta-b^\beta|\leq\beta|a-b|(a^{\beta-1}+b^{\beta-1})$, which
  holds for all $a,b>0$, and
  $|S_i-Y|\leq\sum_{j=1}^m\lambda_j|S_i-S_j|$.
\end{proof}

\begin{lm}\label{lm:latticeineq}
If $\beta\geq2$ and if $\phi:\R^3\to\R$ is the function constructed above corresponding
to a non-negative map $S:\Z^3\to\R$ we have for all $0<\delta$ that
$$
(1-\delta)\sum_{\sigma\in\Z^3}S(\sigma)^\beta -C\delta^{-(\beta-1)}T(S)^{\beta/2}
\leq\int_{\R^3}\phi^\beta\leq \sum_{\sigma\in\Z^3}S(\sigma)^\beta
$$
and that
$$
\sum_{\sigma\in\Z^3}S(\sigma)^{5/2} -\delta
T(S)-C\delta^{-1}\Bigl(\sum_{\sigma\in\Z^3}S(\sigma)^6\Bigr)^{1/4}
\Bigl(\sum_{\sigma\in\Z^3}S(\sigma)^2\Bigr)^{3/4}
\leq\int_{\R^3}\phi^{5/2}\leq \sum_{\sigma\in\Z^3}S(\sigma)^{5/2}
$$
\end{lm}
\begin{proof}
  The functions $\lambda_\tau(x)$ in (\ref{eq:phitau}) are non-negative and satisfy
  $$
  \sum_{\tau\in\Z^3\atop|\tau|=\sqrt{3}}\lambda_\tau(x)=1
  $$
  and that $\int_{[-1/2,1/2]^3}\lambda_\tau(x)dx=1/8$ for all $\tau\in\Z^3$
  with $|\tau|=\sqrt{3}$.
  We thus get from the inequality in Lemma~\ref{lm:beta} that
  for all $\mu\in\{(1/2,1/2,1/2)\}+\Z^3$
  \begin{eqnarray*}
    \lefteqn{\frac{1}{8}\sum_{\tau\in\Z^3\atop|\tau|=\sqrt{3}}S(\mu+\tau/2)^\beta\geq
      \int\limits_{\{\mu\}+[-1/2,1/2]^3}\phi(x)^\beta dx}&&\\
    &\geq&
      \frac{1}{8}\sum_{\tau\in\Z^3\atop|\tau|=\sqrt{3}}S(\mu+\tau/2)^\beta
      -C\Bigl(\sum_{\tau_1,\tau_2\in\Z^3\atop{|\tau_1|=\sqrt{3}\atop|\tau_2|=\sqrt{3}}}
      (S(\mu+\tau_1/2)-S(\mu+\tau_2/2))^2\Bigr)^{1/2}
      \sum_{\tau\in\Z^3\atop|\tau|=\sqrt{3}}S(\mu+\tau/2)^{\beta-1}\\
    &\geq&\frac{1}{8}(1-\delta)\sum_{\tau\in\Z^3\atop|\tau|=\sqrt{3}}S(\mu+\tau/2)^\beta-
    C\delta^{-(\beta-1)}
    \Bigl(\sum_{\tau_1,\tau_2\in\Z^3\atop{|\tau_1|=\sqrt{3}\atop|\tau_2|=\sqrt{3}}}
    (S(\mu+\tau_1/2)-S(\mu+\tau_2/2))^2\Bigr)^{\beta/2}.
  \end{eqnarray*}
 If we sum these inequalities over $\mu$ (i.e., over cubes),
 use that each point in $\Z^3$ is the corner of
 8 different cubes, and
 that $\sum_\mu b_\mu^{\beta/2}\leq(\sum_\mu b_\mu)^{\beta/2}$, when
 $b_\mu\geq0$, we obtain the first inequality of the lemma.

 To arrive at the second inequality we instead  use
 the Cauchy-Schwarz estimate
 \begin{eqnarray*}
   \Bigl(\sum_{\tau_1,\tau_2\in\Z^3\atop{|\tau_1|=\sqrt{3}\atop|\tau_2|=\sqrt{3}}}
   (S(\mu+\tau_1/2)-S(\mu+\tau_2/2))^2\Bigr)^{1/2}
   \sum_{\tau\in\Z^3\atop|\tau|=\sqrt{3}}S(\mu+\tau/2)^{\beta-1}\leq\\
   \delta\sum_{\tau_1,\tau_2\in\Z^3\atop{|\tau_1|=\sqrt{3}\atop|\tau_2|=\sqrt{3}}}
   (S(\mu+\tau_1/2)-S(\mu+\tau_2/2))^2+\const\delta^{-1}
   \sum_{\tau\in\Z^3\atop|\tau|=\sqrt{3}}S(\mu+\tau/2)^{2\beta-2}.
 \end{eqnarray*}
 For $\beta=5/2$ we have $2\beta-2=3$. The second inequality in the
 lemma then follows by summing over $\mu$ as above, possibly
 redefining $\delta$, and using that
 $$
 \sum_{\sigma\in\Z^3}S(\sigma)^{3}\leq\Bigl(\sum_{\sigma\in\Z^3}S(\sigma)^6\Bigr)^{1/4}
 \Bigl(\sum_{\sigma\in\Z^3}S(\sigma)^2\Bigr)^{3/4}.
 $$
\end{proof}

\begin{thm}[The Sobolev inequality for $T$]\label{thm:sobolev}\hfill\\
  There exists a constant $C>0$ such that for all maps $S:\Z^3\to\R$
  that vanish outside some finite set we have
  $$
  \Bigl(\sum_{\sigma\in\Z^3}|S(\sigma)|^6\Bigr)^{1/3}\leq C T(S).
  $$
\end{thm}
\begin{proof}
  Since $T(S)\geq T(|S|)$ it is enough to consider only
  non-negative maps $S$. If we use the first inequality in
  Lemma~\ref{lm:latticeineq} with $\beta=6$ and $\delta=1/2$ we obtain
  that
  $$
  \sum_{\sigma\in\Z^3}|S(\sigma)|^6\leq 2\int\phi(x)^6dx+CT(S)^3.
  $$
  Since $\phi:\R^3\to\R$ is a $C^1$ function of compact support we may
  use the standard Sobolev inequality on $\R^3$
  $$
  \int\phi(x)^6dx\leq C\left(\int|\nabla\phi(x)|^2dx\right)^3
  =CT(S)^3.
  $$
\end{proof}

If $S$ vanishes outside a finite set we get from the Sobolev inequality
and the second inequality in Lemma~\ref{lm:latticeineq} that
\begin{equation}\label{eq:lattice5/2}
  \sum_{\sigma\in\Z^3}S(\sigma)^{5/2} -\delta
  T(S)-C\delta^{-7}\Bigl(\sum_{\sigma\in\Z^3}S(\sigma)^2\Bigr)^{3}
  \leq\int_{\R^3}\phi^{5/2}\leq \sum_{\sigma\in\Z^3}S(\sigma)^{5/2}.
\end{equation}

\subsection{Application of the lattice approximation to the situation in Theorem~\ref{thm:smallcubesloc}}
We shall use the above results for the map
$$
S_0(\sigma)=S_0^\Psi(\sigma)=\ell^{-1}(\sqrt{\langle n_\sigma\rangle+1}-1).
$$
The expression $T(S_0)$ is then exactly the term that appears in
Theorem~\ref{thm:smallcubesloc} on localizing into small cubes. Note that
$S_0(\sigma)\leq \ell^{-1}\sqrt{\langle n_\sigma\rangle}$.
Let $\phi:\R^3\to\R$ be the function corresponding to $S_0$ as
constructed in (\ref{eq:philattice}). Then from (\ref{eq:phikin}) and
Lemma~\ref{lm:latticeineq} we get
\begin{equation}\label{eq:phiprop}
  T(S_0)=\int(\nabla\phi)^2\quad\hbox{and}\quad  \ell^{2}\int\phi^2\leq
  \sum_{\sigma\in\Z^3}\ell^2S_0(\sigma)^2\leq\sum_{\sigma\in\Z^3}\langle n_\sigma\rangle \leq N.
\end{equation}
Moreover, we get for all $0<\delta$ that
\begin{eqnarray}
  \sum_{\sigma\in\Z^3}\langle n_\sigma\rangle ^{5/4}\ell^{-3/4}
  &\leq&(1-\delta)\sum_{\sigma\in\Z^3}S_0(\sigma)^{5/2}\ell^{7/4}+
  \const\delta^{-3/2}\ell^{-3/4}(L/\ell)^3\nonumber\\
  &\leq&(1-\delta)\ell^{7/4}\int\phi^{5/2}+\delta T(S_0)
  +\const\delta^{-7}\ell^8\Bigl(\sum_{\sigma\in\Z^3}\ell^2S_0(\sigma)^2\Bigr)^{3}
  +\const\delta^{-3/2}\ell^{-3/4}(L/\ell)^3 \nonumber\\
  &\leq&(1-\delta)\ell^{7/4}\int\phi^{5/2}+\delta T(S_0)
  +\const\delta^{-7} \ell^8 N^{3}
  +\const\delta^{-3/2}\ell^{-3/4}(L/\ell)^3,\label{eq:phi5/2}
\end{eqnarray}
where we have used firstly that $\langle n_\sigma\rangle \ne0$ for at most $(L/\ell)^3$
different points $\sigma$ and secondly the inequality  (\ref{eq:lattice5/2}) with
$\delta$ replaced by $\delta\ell^{-7/4}$. Likewise we get from the
Sobolev inequality Theorem~\ref{thm:sobolev}
\begin{eqnarray}
  \sum_{\sigma\in\Z^3}\langle n_\sigma\rangle ^{5/3}\ell^{-1/3}&\leq&
  \const\sum_{\sigma\in\Z^3}\ell^3S_0(\sigma)^{10/3}+\const\ell^{-1/3}(L/\ell)^3\nonumber\\
  &\leq&\const\Bigl(\sum_{\sigma\in\Z^3}S_0(\sigma)^6\Bigr)^{1/3}
  \Bigl(\ell^{5/2}\sum_{\sigma\in\Z^3}\ell^2S_0(\sigma)^2\Bigr)^{2/3}
  +\const\ell^{-1/3}(L/\ell)^3\nonumber\\
  &\leq&\const T(S_0)
  (N^{2/5}\ell)^{5/3}
  +\const\ell^{-1/3}(L/\ell)^3.\label{eq:n5/3}
\end{eqnarray}

Finally, note that if we define $\widetilde\Phi(x)=\ell^{-1/2}\phi(x/\ell)$ we
find that
\begin{equation}\label{eq:Phiprop}
  \int(\nabla\widetilde\Phi)^2=\int(\nabla\phi)^2,\quad
  \int\widetilde\Phi^{5/2}=\ell^{7/4}\int\phi^{5/2},\quad
  \int\widetilde\Phi^2=\ell^2\int\phi^2.
\end{equation}

\section{Completing the proof of the lower bound in Dyson's formula Theorem~\ref{thm:main}}\label{sec:final}

In this final section we shall combine all the previous results to
conclude the asymptotic lower bound in Theorem~\ref{thm:main}.  {F}rom
Theorems~\ref{thm:largeloc}, \ref{thm:smallcubesloc},
\ref{thm:Hzbound}, and (\ref{eq:phiprop}), (\ref{eq:phi5/2}),
(\ref{eq:n5/3}), and (\ref{eq:Phiprop}) we obtain, under the
assumptions that $\const_0 N\ell^3\leq\varepsilon^3$ and
$\varepsilon^{-1}t^{-4}\ell\leq \const_0^{-1}$, that
\begin{eqnarray*}
  E(N)\geq\inf_{\widetilde\Phi}\left(A
    \int\left(\nabla\widetilde\Phi\right)^2
    -B\int\widetilde\Phi^{5/2}\right)
  -D N^{7/5},
\end{eqnarray*}
where the infimum is over all functions $\widetilde\Phi\geq0$ , such that $\int\widetilde\Phi^2\leq N$ and
\begin{eqnarray*}
  A&=&(\gamma\widetilde\gamma\gamep/2-\gamma\widetilde\gamma\gamep^{-1/4} I_0\delta
  -\const\varepsilon^{1/6}-\const K_2(\varepsilon,t,N,\ell))\\
  B&=&(1-\delta)\gamma\widetilde\gamma(\gamep^{-1/4} I_0+K_1(\varepsilon,t,N,\ell))\\
  D&=&\const\Bigl(t^{-2}(N^{1/5}L)^{-2}+t^{-4}(N^{2/5}\ell)^{-1}+(N^{1/5}L)^{3}(N^{2/5}\ell)^{-5}\\&&
  +t^{-6}(N^{1/5}\ell)^{-1}N^{-1/5}+\varepsilon^{-1}t^{-8}N^{-2/5}\\&&
  +\gamma\widetilde\gamma(\gamep^{-1/4} I_0+K_1(\varepsilon,t,N,\ell))\delta^{-7}(N^{2/5}\ell)^8N^{-8/5}\\&&
  +\gamma\widetilde\gamma(\gamep^{-1/4} I_0
  +K_1(\varepsilon,t,N,\ell))\delta^{-3/2}(N^{1/5}L)^3(N^{2/5}\ell)^{-15/4}N^{-1/2}
  \\&&
  +(\varepsilon^{1/6}+K_2(\varepsilon,t,N,\ell))(N^{2/5}\ell)^{-5}(N^{1/5}L)^3N^{-2/3}
  \Bigr).
\end{eqnarray*}
We have used that $\sum_\sigma\nu_\sigma=N$ and $\sum_{\sigma,\ \nu_\sigma\ne0}\ell^{-2}\leq L^3/\ell^5$.

We shall show that as $N\to\infty$ we can choose
$\varepsilon,t,\delta\to0$ and $N^{2/5}\ell,N^{1/5}L\to\infty$ in such
a way that $A\to1/2$, $B\to I_0$, $D\to0$,
$\varepsilon^{-3}N\ell^3\to0$, $\varepsilon^{-1}t^{-4}\ell\to0$, and $\ell<L$.

Note that from (\ref{eq:gammaprop}) and (\ref{eq:gamma0}) the limits
of $A$ and $B$ will follow if we can prove that $K_1,K_2\to0$.  The
error term $D$ is of the form
$$
D=\const t^{-2}(N^{1/5}L)^{-2}+\const t^{-4}(N^{2/5}\ell)^{-1}+\const (N^{1/5}L)^{3}(N^{2/5}\ell)^{-5}+K_3
$$
where $K_3$ is a finite sum of terms of the form
\begin{equation}\label{eq:termform}
  \const\delta^{-a}\varepsilon^{-b}t^{-c}(N^{2/5}\ell)^d(N^{1/5}L)^eN^{-f},
\end{equation}
where $f>0$.  Recall that $K_1$ and $K_2$ were sums of expressions of
the same form. Note that
$\varepsilon^{-3}N\ell^3=\varepsilon^{-3}(N^{2/5}\ell)^3N^{-1/5}$,
$\varepsilon^{-1}t^{-4}\ell=\varepsilon^{-1}t^{-4}(N^{1/5}\ell)N^{-1/5}$, and
$\ell/L=(N^{2/5}\ell)(N^{1/5}L)^{-1}N^{-1/5}$ (which is required to be
less than 1)  are again all of the  form (\ref{eq:termform}).

Now first choose $N^{2/5}\ell\to\infty$ in such a way that
$(N^{2/5}\ell)^dN^{-f}\to0$ for all the occurring terms of the form (\ref{eq:termform}).  Then choose
$N^{1/5}L\to\infty$ in such a way that
$\const(N^{1/5}L)^{3}(N^{2/5}\ell)^{-5}\to0$ and
$(N^{2/5}\ell)^d(N^{1/5}L)^eN^{-f}\to0$ for all the terms of the form (\ref{eq:termform}).
Finally choose $\varepsilon,t,\delta\to0$ in such a way that all the
terms of the  form (\ref{eq:termform}) and the first two terms in $D$ still go to zero.
Hence we have achieved all the limits as claimed.

Dyson's formula now follows since by scaling we have that
$$
A\int\left(\nabla\widetilde\Phi\right)^2
    -B\int\widetilde\Phi^{5/2}=(2A)^{-3/5}(B/I_0)^{4/5}N^{7/5}\left(\mfr{1}{2}\int\left(\nabla\Phi\right)^2
    -I_0\int\Phi^{5/2}\right),
$$
where
$$
\Phi(x)=(2A I_0/B)^{6/5}N^{-8/10}\widetilde\Phi\left((2A
  I_0/B)^{4/5}N^{-1/5}x\right)
$$
satisfies $\int\Phi^2=N^{-1}\int\widetilde\Phi^2\leq 1$ if $\int\widetilde\Phi^2\leq N$.

\appendix

\section{Appendix: Localization of large matrices}\label{app:local}
\bigskip

The following theorem allows us to reduce a big Hermitean matrix, ${\cal A}$,
to a smaller principal submatrix without changing the lowest eigenvalue
very much. The proof can be found in \cite{LSo}.

\begin{thm}[Localization of large matrices]\label{local}
  Suppose that ${\cal A}$ is an $N\times N$ Hermitean matrix and let
  ${\cal A}^k$, with $k=0,1,\ldots,N-1$, denote the matrix consisting of
  the $k^{\rm th}$ supra- and infra-diagonal of ${\cal A}$.  Let $\psi
  \in {\bf C}^N$ be a normalized vector and set $d_k = (\psi , {\cal
    A}^k \psi) $ and $\lambda = (\psi , {\cal A} \psi) =
  \sum_{k=0}^{N-1} d_k$.  \ ($\psi$ need not be an eigenvector of
  ${\cal A}$.) \

  Choose some positive integer $M \leq N$.  Then, with $M$ fixed,
  there is some $n \in [0, N-M]$ and some normalized vector $ \phi \in
  {\bf C}^N$ with the property that $\phi_j =0$ unless $n+1 \leq j
  \leq n+M$ \ (i.e., $\phi $ has length $M$) and such that
  \begin{equation}\label{localerror}
    (\phi , {\cal A} \phi) \leq \lambda + \frac{C}{ M^2}
    \sum_{k=1}^{M-1} k^2 |d_k|
    +C\sum_{k=M}^{N-1} |d_k|\ ,
  \end{equation}
  where $C>0 $ is a  universal constant. (Note that the first sum starts
  with $k=1$.)
\end{thm}

\section{Appendix: Localization of the kinetic energy}\label{app:kinetic}

Our goal here is to prove a certain lower bound on the many body
kinetic energy $\sum_{i=1}^N-\Delta_i$ acting on the symmetric
tensor product space ${\bigotimes\limits^N}_{S}L^2(\R^3)$.
In order to state the bound we
need to introduce some more notation.
Let $X_0$ denote the characteristic function of the cube $[-\ell/2,\ell/2]^3$,
in $\R^3$.
Let $X_z$ be the characteristic function of the cube
$\{z\ell\}+[-\ell/2,\ell/2]^3$, i.e., $X_z(x)=X_0(x-\ell z)$.
Let $\cP_z$ denote the projection onto the subspace of
$L^2(\{z\ell\}+[-\ell/2,\ell/2]^3)$
consisting of functions orthogonal to constants.
We shall consider $\cP_z$ as a projection in $L^2(\R^3)$.
Let $-\Delta_{\rm Neu}^{(z)}$ be the Neumann
Laplacian
for the cube $\{z\ell\}+[-\ell/2,\ell/2]^3$.
Let
$a^*_0(z)$ be the creation operator
$$
a^*_0(z)=a^*(\ell^{-3/2}X_z),
$$
i.e., the operator creating the constant in the cube
$\{z\ell\}+[-\ell/2,\ell/2]^3$. Note that as before $a^*_0(z)$ acts in the Fock space
$\bigoplus\limits_{N=0}^\infty{\bigotimes\limits^N}_{S}L^2(\R^3)$.
Products of the form $a^*_0(z)\an_0(z')$ or $\an_0(z')a^*_0(z)$
are however bounded operators
on the space  ${\bigotimes\limits^N}_{S}L^2(\R^3)$.
Let for all $z\in\R^3$ the function $\upchi_z\in C^\infty_0 (\{z\ell\}+(-\ell/2,\ell/2)^3)$ be such that
$\|\partial^\alpha\upchi_z\|_\infty\leq \const(\ell t)^{-|\alpha|}$ and
$\|\partial^\alpha\sqrt{1-\upchi_z^2}\|_\infty\leq \const(\ell t)^{-|\alpha|}$ for
some $t$ with $0<t<1$  and all multi-indices $\alpha$ with $|\alpha|\leq 3$.
We can now state the operator inequality we shall prove. Observe that
that we are estimating the one-body kinetic energy operator (the
Laplacian) from below by a many-body operator.

\begin{thm}[Kinetic energy localization]\label{thm:kinetic}\hfill\\
Let $\Omega\subset\R^3$. Let $w_1,\ldots,w_r\in\R^3$ and
$\beta_1,\ldots,\beta_r>0$ be such that $X_{w_j}X_0=0$ for all
$j=1,\ldots,r$ and such that for all $v\in\R^3$ we have
that
$$
\sum_{j=1}^r\beta_j(w_j,v)^2\leq v^2.
$$
Then for all $0<s<t<1$ and all $\varepsilon>0$ we have
\begin{eqnarray}
  \lefteqn{\Bigl(1+\varepsilon+
    C(s/t)t^{-2}+Cs^{1/2}+Cs\Bigl(\sum_{j=1}^r\beta_j\Bigr)^{1/2}\Bigr)\sum_{i=1}^N-\Delta_i}&&\nonumber\\
  &\geq& \int_{\Omega}\Bigl[\sum_{i=1}^N\cP_z^{(i)}\upchi_{z}^{(i)}
    \frac{(-\Delta_i)^2}{-\Delta_i
      +(\ell s)^{-2}}\upchi^{(i)}_z\cP^{(i)}_z
    +\varepsilon(-\Delta_{i,\rm Neu}^{(z)})\label{eq:kineticmain}\\&&
    \phantom{\int_{\Omega}\Bigl[}
    +\sum_{j=1}^r\frac{\beta_j}{\ell^{2}}\left(\sqrt{a^*_0(z+w_j)\an_0(z+w_j)+1/2}-
      \sqrt{a^*_0(z)\an_0(z)+1/2}\right)^2
  \Bigr]dz-\sum_{j=1}^r\frac{\beta_j }{\ell^{2}}\hbox{\rm vol}(\Omega).\nonumber
\end{eqnarray}
Here all operators are considered in the sense of quadratic forms.
\end{thm}

\begin{proof}
By rescaling we may consider $\ell=1$.

Note first that on the quadratic form domain of the Laplacian $-\Delta$ on $\R^3$ we have
$$
\int_{\Omega}(-\Delta^{(z)}_{\rm Neu})dz\leq \int_{\R^3}(-\Delta^{(z)}_{\rm Neu})dz=-\Delta.
$$
in the sense of quadratic forms.

Let $f_s:\R^3\to\R$ be the function
${f_s}(p)=p^2/(p^2+s^{-2})$ for  some $s>0$.
Then for all $u\in L^2(\R^3)$ we have
$$
 (2\pi)^{-3}\int {f_s}(p)|\hat{u}(p)|^2dp=
 \int \overline{u(x)}\left(\frac{-\Delta}{-\Delta+s^{-2}}u\right)(x)
 dx.
$$
It follows that for all $u\in L^2(\R^3)$ we have
$$
(2\pi)^{-6}\int {f_s}*|\hat{X}|^2(p)|\hat{u}(p)|^2dp=\iint \overline{u(x)}
X_z(x) \left(\frac{-\Delta}{-\Delta+s^{-2}}(X_zu)\right)(x)dxdz.
$$
\begin{lm}
We have
$$
 \|{f_s}-(2\pi)^{-3}{f_s}*|\hat{X}|^2\|_\infty\leq Cs^{1/2}.
$$
\end{lm}
\begin{proof}
  We calculate
  $$
  |\hat{X}(p)|^2=\left(\frac{2\sin(p_1/2)}{p_1}\right)^2
  \left(\frac{2\sin(p_2/2)}{p_2}\right)^2
  \left(\frac{2\sin(p_3/2)}{p_3}\right)^2.
  $$
  In particular $\int_{|p|>r}|\hat{X}(p)|^2dp\leq Cr^{-1}$ for all $r>0$.
  Since $(2\pi)^{-3}\int|\hat{X}|^2(p)dp=1$, $\|{f_s}\|_\infty\leq 1$, and
  $\|\nabla {f_s}\|_\infty\leq \const s$ we have
  \begin{eqnarray*}
    \left|{f_s}(p)-(2\pi)^{-3}{f_s}*|\hat{X}|^2(p)\right|&
    \leq&(2\pi)^{-3}\int_{|q|<r}|{f_s}(p)-{f_s}(p-q)||\hat{X}(q)|^2dq
    \\&&{}+(2\pi)^{-3}\int_{|q|>r}|{f_s}(p)-{f_s}(p-q)||\hat{X}(q)|^2dq
    \leq\const sr+\const r^{-1} .
  \end{eqnarray*}
  The result follows from choosing $r=s^{-1/2}$.
\end{proof}

We now write
\begin{eqnarray*}
  p^2&=&(2\pi)^{-3}p^2{f_s}*|\hat{X}|^2(p)+
  p^2\left({f_s}(p)-(2\pi)^{-3}{f_s}*|\hat{X}|^2(p)\right)
  +p^2(1-{f_s}(p))\\&\geq&
  (2\pi)^{-3}p^2{f_s}*|\hat{X}|^2(p)-Cs^{1/2}p^2+p^2(1-{f_s}(p)).
\end{eqnarray*}
For all $u\in L^2(\R^3)$ we have
$$
(2\pi)^{-6}\int |\hat{u}(p)|^2p^2{f_s}*|\hat{X}|^2(p)dp
=\sum_{i=1}^3\iint \overline{\partial_i u(x)} X_z(x)
\frac{-\Delta}{-\Delta+s^{-2}}(X_z\partial_i u)(x)dxdz.
$$
In other words we have the operator inequality
\begin{equation}\label{eq:kinetic1}
  -\Delta \geq \int \cA_z dz+\frac{-\Delta s^{-2}}{-\Delta+s^{-2}}-Cs^{1/2}(-\Delta),
\end{equation}
where the operator
$$
\cA_z=\sum_{i=1}^3-\partial_i X_z\frac{-\Delta}{-\Delta+s^{-2}}
 X_z\partial_i$$
is defined as giving the positive quadratic form
$$
(u,\cA_z u)=\sum_{i=1}^3\int \overline{\partial_iu(x)} X_z(x)\frac{-\Delta}{-\Delta+s^{-2}}
 (X_z\partial_iu)(x)dx
$$
with domain $H^1(\{z\}+[-1/2,1/2]^3)$. With the similar notation the operator
$\sum_{i=1}^3-\partial_i X_z\partial_i$ is nothing but the Neumann
Laplacian $-\Delta_{\rm Neu}^{(z)}$.

\begin{lm}
Given $\theta\in C^\infty (\{z\}+[-1/2,1/2]^3)$ such that $\theta$
is constant near the boundary of the cube $\{z\}+[-1/2,1/2]^3$. Assume moreover
that $\|\partial^\alpha\theta\|_\infty\leq \const|t|^{-|\alpha|}$ for
some $t$ with $s<t$ and all multi-indices $\alpha$ with $|\alpha|\leq 3$.
Then
$$
\left[\left[\cA_z,\theta\right],\theta\right]\geq -C(s/t)(-\Delta_{\rm Neu}^{(z)})-C(s/t)t^{-2}X_z.
$$
\end{lm}
\begin{proof}
We calculate
\begin{eqnarray*}
  \left[-\partial_i X_z\frac{-\Delta}{-\Delta+s^{-2}}
    X_z\partial_i,\theta\right]&=&
  -\partial_i X_z\frac{-\Delta}{-\Delta+s^{-2}}(\partial_i\theta)
  -\partial_i X_z\left[\frac{-\Delta}{-\Delta+s^{-2}},\theta\right]
  X_z\partial_i\\&&
  -(\partial_i\theta)\frac{-\Delta}{-\Delta+s^{-2}}
  X_z\partial_i
\end{eqnarray*}
and
\begin{eqnarray*}
  \left[\left[-\partial_i X_z\frac{-\Delta}{-\Delta+s^{-2}}
      X_z\partial_i,\theta\right],\theta\right]&=&
  -2\partial_i
  X_z\left[\frac{-\Delta}{-\Delta+s^{-2}},\theta\right](\partial_i\theta)
  -2(\partial_i\theta)\left[\frac{-\Delta}{-\Delta+s^{-2}},\theta\right]
  X_z\partial_i\\&&
  -2(\partial_i\theta)
  \frac{-\Delta}{-\Delta+s^{-2}}(\partial_i\theta)
  -\partial_i X_z\left[\left[\frac{-\Delta}{-\Delta+s^{-2}},\theta\right],\theta\right]
  X_z\partial_i.
\end{eqnarray*}
We also calculate
\begin{eqnarray*}
  \left[\frac{-\Delta}{-\Delta+s^{-2}},\theta\right]&=&s^{-2}
  \frac{1}{-\Delta+s^{-2}}\left[-\Delta,\theta\right]\frac{1}{-\Delta+s^{-2}}
\end{eqnarray*}
and
\begin{eqnarray*}
  \left[\left[\frac{-\Delta}{-\Delta+s^{-2}},\theta\right],\theta\right]&=&s^{-2}
  \frac{1}{-\Delta+s^{-2}}\left[\left[-\Delta,\theta\right],\theta\right]
  \frac{1}{-\Delta+s^{-2}}\\&&+
  2s^{-2}
  \frac{1}{-\Delta+s^{-2}}\left[-\Delta,\theta\right]\frac{1}{-\Delta+s^{-2}}
  \left[\theta,-\Delta\right]\frac{1}{-\Delta+s^{-2}}
\end{eqnarray*}
Hence
$$
\left\|\left[\frac{-\Delta}{-\Delta+s^{-2}},\theta\right]\right\|
\leq Cs/t\quad\hbox{and}\quad
\left\|\left[\left[\frac{-\Delta}{-\Delta+s^{-2}},\theta\right],\theta\right]\right\|
\leq C(s/t)^2
$$
Likewise
$$
\left\|\left[\left[\frac{-\Delta}{-\Delta+s^{-2}},\partial_i\theta\right],
    \partial_i\theta\right]\right\|
\leq C(s/t)^2t^{-2}.
$$
Using these estimates we find that
\begin{eqnarray*}
  \partial_i
  \lefteqn{X_z\left[\frac{-\Delta}{-\Delta+s^{-2}},\theta\right](\partial_i\theta)
  +(\partial_i\theta)\left[\frac{-\Delta}{-\Delta+s^{-2}},\theta\right]
  X_z\partial_i}&&\\&\leq&
  -(t/s)\partial_i
  X_z\left[\frac{-\Delta}{-\Delta+s^{-2}},\theta\right]
  \left[\frac{-\Delta}{-\Delta+s^{-2}},\theta\right]^*X_z\partial_i
  +(s/t)(\partial_i\theta)^2\\&\leq&
  -C(s/t)\partial_iX_z\partial_i+C(s/t)t^{-2}X_z
\end{eqnarray*}
and
\begin{eqnarray*}
  (\partial_i\theta)
  \frac{-\Delta}{-\Delta+s^{-2}}(\partial_i\theta)&=&
  \sum_{j=1}^3-\partial_j (\partial_i\theta)
  \frac{1}{-\Delta+s^{-2}}(\partial_i\theta)\partial_j \\&&
  +\sum_{j=1}^3(\partial_j \partial_i\theta)
  \frac{\partial_j}{-\Delta+s^{-2}}(\partial_i\theta)
  -\sum_{j=1}^3(\partial_i\theta)
  \frac{\partial_j}{-\Delta+s^{-2}}(\partial_j \partial_i\theta)\\&&
   -\sum_{j=1}^3(\partial_j\partial_i\theta)
  \frac{1}{-\Delta+s^{-2}}(\partial_j \partial_i\theta)\\
  &\leq& \sum_{j=1}^3-\partial_j (\partial_i\theta)
  \frac{1}{-\Delta+s^{-2}}(\partial_i\theta)\partial_j +C(s/t)t^{-2}X_z\\
  &\leq&-C(s/t)^2\sum_{j=1}^3\partial_jX_z\partial_j+C(s/t)t^{-2}X_z.
\end{eqnarray*}
\end{proof}
With $\eta_z=\sqrt{1-\upchi_z^2}$
we may write
\begin{eqnarray*}
  \cA_z&=&
  \mfr{1}{2}(\eta_z^2+\upchi_z^2)\cA_z+
  \mfr{1}{2}\cA_z(\eta_z^2+\upchi_z^2)\\
  &=&\upchi_z\cA_z\upchi_z+\eta_z\cA_z\eta_z
  +\mfr{1}{2}\left[\left[\cA_z,\upchi_z\right],\upchi_z\right]
  +\mfr{1}{2}\left[\left[\cA_z,\eta_z\right],\eta_z\right]\\
  &\geq&\upchi_z\frac{(-\Delta)^2}{-\Delta+s^{-2}}\upchi_z
  -C(s/t)(-\Delta_{\rm Neu}^{(z)})-C(s/t)t^{-2}X_z.
\end{eqnarray*}
Here we have used that
$\upchi_z\cA_z\upchi_z=\upchi_z\frac{(-\Delta)^2}{-\Delta+s^{-2}}\upchi_z$,
where $-\Delta$ is the Laplacian on $\R^3$.

Clearly we have $\cP_z\cA_z\cP_z=\cA_z$. Thus
\begin{eqnarray*}
  \cA_z&\geq& \cP_z\upchi_z\frac{(-\Delta)^2}{-\Delta+s^{-2}}
  \upchi_z\cP_z-C(s/t)(-\Delta_{\rm Neu}^{(z)})-C(s/t)t^{-2}\cP_z
  \\&\geq&\cP_z\upchi_z
  \frac{(-\Delta)^2}{-\Delta+s^{-2}}\upchi_z\cP_z-C(s/t)t^{-2}(-\Delta_{\rm Neu}^{(z)}),
\end{eqnarray*}
where we have assumed that $t<1$ and used that
$-\Delta_{\rm Neu}^{(z)}\geq C\cP_z$.

We turn now to the term $-\Delta s^{-2}/(-\Delta+s^{-2})$ in
(\ref{eq:kinetic1}).
\begin{lm} For all $u\in L^2(\R^3)$ we have that
$$
\sum_{j=1}^r\int_{\R^3} \beta_j\left|u(x+w_j)-u(x)\right|^2dx
\leq \Bigl(u,\Bigl(\frac{-\Delta s^{-2}}{-\Delta+s^{-2}}
    +2\Bigl(\sum_{j=1}^r\beta_j\Bigr)^{1/2}s(-\Delta)\Bigr)u\Bigr)
$$
\end{lm}
\begin{proof} Write $b=2\Bigl(\sum_{j=1}^r\beta_j\Bigr)^{1/2}$.
  On one hand we may write
  \begin{eqnarray*}
    \left(u,\left(\frac{-\Delta s^{-2}}{-\Delta+s^{-2}}
        +bs(-\Delta)\right)u\right)=
    (2\pi)^{-3}\int|\hat{u}(p)|^2\left(\frac{p^2 s^{-2}}{p^2+s^{-2}}
        +bsp^2\right)dp.
  \end{eqnarray*}
  On the other hand we have
  \begin{eqnarray*}
  \sum_{j=1}^r\int \beta_j\left|u(x+w_j)-u(x)\right|^2dx
  &=&\sum_{j=1}^r(2\pi)^{-3}\int\beta_j|\hat{u}(p)|^2|\exp(i(p,w_j)/2)-\exp(-i(p,w_j)/2)|^2dp\\
  &=&\sum_{j=1}^r(2\pi)^{-3}\int\beta_j|\hat{u}(p)|^24\sin^2((p,w_j)/2)dp.
  \end{eqnarray*}
  We thus simply have to prove that
  $$
  \frac{p^2 s^{-2}}{p^2+s^{-2}}
        +bsp^2\geq\sum_{j=1}^r4\beta_j\sin^2((p,w_j)/2).
  $$
  If $2|p|<b^{1/2}s^{-1/2}$ we have
  $$
  \frac{p^2 s^{-2}}{p^2+s^{-2}}\geq
  \frac{p^2}{bs+1}\geq (1-bs)p^2\geq
  \sum_{j=1}^r4\beta_j\sin^2((p,w_j)/2)-bsp^2.
  $$
  If $|p|>b^{1/2}s^{-1/2}$ we have
  $$
  bsp^2\geq b^2\geq \sum_{j=1}^r4\beta_j\sin^2((p,w_j)/2).
  $$
\end{proof}

Since for all $y\in\R^3$, $\left(X_{y+\sigma}\right)_{\sigma\in\Z^3}$ is an orthonormal family in
$L^2(\R^3)$, we get from
Parseval's identity that
$$
\int \left|u(x+w_j)-u(x)\right|^2dx\geq \sum_{\sigma\in\Z^3}\left|\int X_{y+\sigma}(x)(u(x+w_j)-u(x))dx \right|^2.
$$
If we integrate this over $y\in[0,1]^3$ we obtain
$$
\int \left|u(x+w_j)-u(x)\right|^2dz\geq
\int \left|\int X_z (x)\left(u(x+w_j)-u(x)\right)dx\right|^2\, dz.
$$
The quadratic form on the right of this inequality corresponds to the
operator whose second quantization has the form
$$
\int \left(a^*_0(z+w_j)-a^*_0(z)\right)(\an_0(z+w_j)-\an_0(z)) dz.
$$
It is then clear that the main inequality (\ref{eq:kineticmain})
follows from the following lemma.
\begin{lm} Let $a^*_1$ and $a^*_2$ be two commuting creation
  operators.
Then
$$
(a^*_2-a^*_1)(\an_2-\an_1)\geq \left(\sqrt{a^*_2\an_2+1/2}-\sqrt{a^*_1\an_1+1/2}\right)^2-1.
$$
\end{lm}
\begin{proof}
  We calculate
  $$
  (a^*_2-a^*_1)(\an_2-\an_1)=a^*_2\an_2+a^*_1\an_1-(a^*_2\an_1+a^*_1\an_2).
  $$
  We have
  \begin{eqnarray*}
    (a^*_2\an_1+a^*_1\an_2)^2&\leq&
    (a^*_2\an_1+a^*_1\an_2)^2+(a^*_2\an_1-a^*_1\an_2)(a^*_1\an_2-a^*_2\an_1)
    \\&=&
    2a^*_2\an_1a_1^*\an_2+2a^*_1\an_2a_2^*\an_1
    =4a^*_2\an_2a_1^*\an_1+2a^*_1\an_1 +2a_2^*\an_2\\&=&
    4(a^*_2\an_2+1/2)(a_1^*\an_1+1/2)-1\leq
    4(a^*_2\an_2+1/2)(a_1^*\an_1+1/2).
  \end{eqnarray*}
  Since the square root is an operator monotone function we have that
  $$
  a^*_2\an_1+a^*_1\an_2\leq 2\sqrt{a^*_2\an_2+1/2}\sqrt{a_1^*\an_1+1/2}.
  $$
  Hence
  \begin{eqnarray*}
    (a^*_2-a^*_1)(\an_2-\an_1)&\geq&
    a^*_2\an_2+a^*_1\an_1
    - 2\sqrt{a^*_2\an_2+1/2}\sqrt{(a_1^*\an_1+1/2}\\
    &=&\left(\sqrt{a^*_2\an_2+1/2}-\sqrt{a_1^*\an_1+1/2}\right)^2-1.
  \end{eqnarray*}
\end{proof}

\end{proof}

We shall use Theorem~\ref{thm:kinetic} with $w_1,w_2,\ldots,w_r$ being
the vectors $\sigma\in\Z^3$ satisfying that $|\sigma|^2=2$ or
$|\sigma|^2=3$. Note that there is a total of 20 such vectors. We define
$$
\beta_\sigma=\left\{\begin{array}{cl}\frac{1}{12},&|\sigma|^2=2\\
 \frac{1}{24},&|\sigma|^2=3
\end{array}\right.
$$
We then have for all $v\in\R^3$ that
$$\sum_{\sigma\in\Z^3\atop
  |\sigma|=\sqrt{2}\mbox{ \tiny or }|\sigma|=\sqrt{3}}\beta_\sigma (v,\sigma)^2=
 \sum_{\sigma\in\Z^3\atop
  |\sigma|=\sqrt{2}}\frac{1}{12}(v\cdot\sigma)^2
+\sum_{\sigma\in\Z^3\atop
  |\sigma|=\sqrt{3}}\frac{1}{24}(v\cdot\sigma)^2=v^2.
$$

\end{document}